\pdfoutput=1
\documentclass[article,showpacsreprint,groupedaddress,12pt,superscriptaddress,preprintnumbers,floatfix,a4paper,nofootinbib,amsmath,amssymb]{revtex4-2}

\usepackage[english]{babel}
\usepackage[letterpaper,top=2cm,bottom=2cm,left=2.5cm,right=2.5cm,marginparwidth=1.75cm]{geometry}

\usepackage{setspace}

\usepackage{amsmath}
\usepackage{mathtools}
\usepackage{graphicx}
\usepackage[shortlabels]{enumitem}
\usepackage[colorlinks=true, allcolors=blue]{hyperref}
\usepackage{chemformula} %
\usepackage{pythonhighlight}  %
\usepackage{comment} %
\usepackage{tablefootnote} %
\usepackage{multirow}

\usepackage{amsthm}
\newtheorem{theorem}{Theorem}

\usepackage{cleveref}
\crefname{theorem}{Theorem}{Theorems}
\crefname{section}{Section}{Sections}
\crefname{equations}{Eq.}{Eqs.}

\newcommand{\tensorflow}{\textsc{TensorFlow}} %
\newcommand{\opteinsum}{\texttt{opt\_einsum}}
\newcommand{\clang}{\texttt{C}}
\newcommand{\fortran}{\texttt{Fortran}}

\usepackage{listings}
\usepackage{xcolor}

\definecolor{codegreen}{rgb}{0,0.6,0}
\definecolor{codegray}{rgb}{0.5,0.5,0.5}
\definecolor{codepurple}{rgb}{0.58,0,0.82}
\definecolor{backcolour}{rgb}{0.95,0.95,0.92}

\lstdefinestyle{mystyle}{
    backgroundcolor=\color{backcolour},   
    commentstyle=\color{codegreen},
    keywordstyle=\color{magenta},
    numberstyle=\tiny\color{codegray},
    stringstyle=\color{codepurple},
    basicstyle=\ttfamily\footnotesize,
    breakatwhitespace=false,         
    breaklines=true,                 
    captionpos=b,                    
    keepspaces=true,                 
    numbers=left,                    
    numbersep=1pt,                  
    showspaces=false,                
    showstringspaces=false,
    showtabs=false,                  
    tabsize=2
}

\newcommand{\F}{\ensuremath{\mathbb{F}}}
\newcommand{\abs}[1]{\ensuremath{|#1|}}
\usepackage{mathtools}
\newcommand{\defeq}{\vcentcolon=}

\begin{document}

\title{Nuclear correlation functions using first-principle calculations of lattice
quantum chromodynamics}

\author{Debsubhra Chakraborty}
\email{debsubhra.chakraborty@tifr.res.in}
\affiliation{Department of Theoretical Physics, Tata Institute of Fundamental Research, \\Homi Bhabha Road, Mumbai 400005, India}

\author{Piyush Srivastava}
\email{piyush.srivastava@tifr.res.in}
\affiliation{School of Technology and Computer Science, Tata Institute of Fundamental Research, \\Homi Bhabha Road, Mumbai 400005, India}

\author{Arpith Kumar}
\email{ph18013@iisermohali.ac.in}
\affiliation{Department of Physical Sciences,
IISER Mohali,
Academic Block 1, AB1-1F9,
Sector 81, Mohali, India}

\author{Nilmani Mathur}
\email{nilmani@theory.tifr.res.in}
\affiliation{Department of Theoretical Physics, Tata Institute of Fundamental Research, \\Homi Bhabha Road, Mumbai 400005, India}


\begin{abstract}
  Exploring nuclear physics through the fundamental constituents of the strong
  force -- quarks and gluons -- is a formidable challenge. While numerical
  calculations using lattice quantum chromodynamics offer the most promising
  approach for this pursuit, practical implementation is arduous, especially due
  to the uncontrollable growth of quark-combinatorics, the so-called Wick-contraction problem of nuclei.  We present here two novel methods providing a state-of-the art solution to this problem.  In the first, we exploit randomized
  algorithms inspired from computational number theory to detect and eliminate
  redundancies that arise in Wick contraction computations.  Our second method
  explores facilities for automation of tensor computations---in terms of
  efficient utilization of specialized hardware, algorithmic optimizations, as
  well as ease of programming and the potential for automatic code
  generation---that are offered by new programming models inspired by
  applications in machine learning (e.g.,~\tensorflow{}).  We demonstrate the efficacy of our methods
  by computing two-point correlation functions for Deuteron, Helium-3, Helium-4
  and Lithium-7, achieving at least an order of magnitude improvement over
  existing algorithms with efficient
  implementation on GPU-accelerators. 
  Additionally, we discover an intriguing characteristic shared by all the
  nuclei we study: specific spin-color combinations dominate the correlation
  functions,  hinting at a potential
  connection to an as-yet-unidentified symmetry in nuclei. Moreover finding them beforehand can reduce the computing time further and substantially. Our results, with the efficiency that we achieved, suggest the possibility of extending the applicability of our methods for calculating properties of light nuclei, potentially up to $A \sim 12$ and beyond. 
\end{abstract}

\maketitle

\section{Introduction}
\noindent Nuclear physics stands as one of the central pillars of 20th-century science. Theoretical developments, spanning a century, have shaped our current understanding of atomic nuclei as
quantum many-body systems composed of protons and neutrons, intriguingly bound together
by the fundamental strong force. Multiple phenomenological theories of
atomic nuclei have been developed: these include shell
models~\cite{Mayer:1948zz, PhysRev.75.1969,bhaduri}, various potential
models~\cite{Wiringa:1994wb,PhysRevLett.74.4396,PhysRevLett.84.5728,Machleidt:2000ge},
models building upon many body methods such as numerical approaches using
density functional theory (DFT) and quantum Monte Carlo
(QMC) methods~\cite{PhysRev.136.B864,PhysRev.140.A1133,dft:1991,
  LOMNITZADLER1981399,PhysRevC.36.2026, SCHMIDT199999,
  doi:10.1146/annurev.nucl.52.050102.090637,Barrett:2013nh, Hagen_2014,
  Carlson:2014vla}, as well as effective field theories
\cite{WEINBERG1990288,WEINBERG19913,WEINBERG1992114,RevModPhys.81.1773,EPELBAUM2005362}
and nuclear effective field theories \cite{meissner,PhysRevLett.106.192501}. Nonetheless, comprehension of nuclear
physics requires its understanding using strong interactions with the
fundamental degrees of freedom, quarks and gluons, and their interplay with
electroweak interactions.  
However, the role of quarks and
gluons in describing nuclear properties is still not well understood. For the case of higher nuclei, it may be argued that due to the emergence of
various possible many-body effective collective degrees of freedom, the use of
fundamental degrees of freedom may not be prudent.  Nevertheless, their
utilization is essential in understanding the origin of binding in nuclei,
nucleosynthesis of light nuclei and their properties, exploring
experimentally unknown exotic nuclei and nuclear reactions, as well as for
connecting strong force to the emergence of nuclear shell and effective degrees
of freedom in higher nuclei. While calculations involving electroweak interactions are under control, the use
of fundamental strong force to find nuclear properties is a daunting
problem. This is mainly because  Quantum Chromodynamics (QCD), the theory
underlying the strong force, is highly non-perturbative in the low-energy domain of
nuclei. With no analytical solution of QCD in this domain, as of now, it becomes
an immensely challenging problem. First principle calculations of QCD employing 3+1 dimensional space time grids, the so-called Lattice QCD (LQCD) framework, provides an excellent avenue for carrying out such 
studies.  Besides making connection to the strong force, such calculations using lattice QCD are essential for obtaining information on other important physics, for example, in searches for violations of the symmetries of the Standard Model including those by dark matter direct detection experiments, in  low-energy nuclear reactions relevant to astrophysics, equation of states in nuclear matter (which are relevant, for example, to neutron and other dense stars), in discovering a new exotic nuclear chart, as well as in providing inputs for nuclear many-body calculations that cannot be obtained experimentally. All these provide a synergy between nuclear physics, astrophysics and cosmology, high energy physics, and many body physics. Lattice QCD based calculations for nuclear physics were initiated more than a decade ago producing many impressive results \cite{Beane:2010em,NPLQCD:2010ocs,NPLQCD:2012mex,Beane:2012ey, Beane:2014ora, Beane:2015yha, 
PhysRevD.87.034506,Orginos:2015aya,Berkowitz:2015eaa,Detmold:2015daa,PhysRevD.87.114512,Savage:2016kon, Tiburzi:2017iux,Wagman:2017tmp,Detmold:2019ghl, 
Davoudi:2020ngi,Detmold:2020snb,Parreno:2021ovq,amarasinghe_variational_2023,Davoudi:2024ukx,
Ishii:2006ec,Yamazaki:2009ua,Nemura:2008sp,Yamazaki_2010,Aoki:2009ji,Inoue:2010es,Doi:2011gq,Aoki:2012tk,Gongyo:2017fjb, Francis:2018qch,
Drischler:2019xuo, Junnarkar:2019equ,green_weakly_2021,horz_two-nucleon_2021,Junnarkar:2022yak,Mathur:2022ovu}, and remain 
an active and challenging area of current research. 
Though these approaches are promising, their success remains limited compared to the single nucleon case, primarily due to algorithmic challenges, as will be discussed below.

Lattice QCD \cite{PhysRevD.10.2445} calculations  are widely used in particle physics for extracting energy levels and matrix elements, predicting new particles, studying physics at finite temperatures and/or density,  extracting the parameters of the Standard Model as well as for finding evidence for physics beyond it \cite{USQCD:2022mmc,Davoudi:2022bnl, Workman:2022ynf}. These calculations, in general, involve the following procedures~\cite{Gattringer_Lang}: $i$) generation of a large number of statistical configurations on 3+1-dimensional space-time Euclidean lattices, at multiple sizes, spacings and quark masses,  through stochastic Monte Carlo methods based on the dynamics of the QCD action, $ii$) measurement of the required correlation functions corresponding to the observables on those lattices, and $iii$) analysis of those correlation functions followed by chiral, continuum and finite volume extrapolations. These procedures lead to the extraction of many physical results with controlled systematics as has been demonstrated over many years \cite{Workman:2022ynf}.
The first principles study of nuclei using lattice QCD, with precision and controlled systematics, however is still limited to just one nucleon. That is mainly because of our current inability to compute nuclear correlation functions precisely and efficiently within lattice QCD frameworks, where two main bottlenecks are the signal-to-noise-ratio (SNR) and the exponential rise of the number of Wick-contraction terms. The SNR problem arises mainly due to poor overlap of a desired quantum state of a nuclei to the stochastically evaluated lattice correlation functions, where the noise from low-energy excitations, like multi-pion systems, inundates the large-time signals~\cite{PARISI1984203,Lepage:1989hd}. Many more studies are indeed needed to solve this SNR problem \cite{PhysRevD.90.034503,PhysRevD.93.094507,PhysRevD.100.014504,PhysRevD.102.014514}, but it  is expected that with the employment of suitably designed interpolating operator basis for nuclei, along with the availability of lattices with  larger volumes \cite{Yamazaki:2023swq, Aoyama:2024cko} in the exascale computing era, this problem can somewhat be mitigated. The second problem, the so-called Wick-contraction issue, is a combinatorial problem involving many fermions each having spin and color indices. Since a nucleus has many {\it up} ($u$) and {\it down} ($d$)-type valence quarks, the number of possible ways these quarks can combine, constrained by their fermionic properties within strong interaction dynamics, explodes rapidly with increasing number of protons and neutrons. 
For example, a naive counting procedure for a nucleus with $N_p$ protons and $N_n$ neutrons ($A = N_p+ N_n$)  yields that this number is $(2N_p+N_n)!(2N_n+N_p)!6^{2A}4^{2A}$ \cite{Doi_2013}, where the last two factors are due to spin-color combinations. For carbon ($^{12}_6C$) this count can even become  about $10^{73}$. This increment worsens when one has to deal with complicated multi-baryon operators that will be needed in future variational analyses for precision physics for nuclei as well as for three-point functions to get information on their structures. We will discuss the exponential rise of these numbers in detail in the next section. 

Much progress has been made over the years to reduce the number of Wick-contractions. For example, in Refs.~\cite{Ishii_2007, Yamazaki_2010},  permutation-symmetry of the quark fields was exploited, and a block of three quark propagators was considered 
with a zero momentum operator at the sink.
 This construction is known as the block algorithm. A further improvement has been achieved  by constructing a unified index list for the Wick-contraction with simultaneous permutations of quarks and color/spinor indices, which is known to be as the unified contraction algorithm \cite{Doi_2013}. However, finding the unified index list becomes increasingly 
 impractical with increase in atomic number. In Ref.~\cite{PhysRevD.87.114512} a matrix-determinant formulation of this problem has been devised where one calculates the correlation functions by computing determinants of a block matrix containing the elements of the propagator, and the algorithm can scale polynomially ($\mathcal{O}(n_{u}^3n_{d}^3)$) with the number of up and down quarks ({\it i.e.}, $n_u$,$n_d$). Further optimization can be achieved using rank-1 updates to relate determinants of similar matrices as shown in Ref.~\cite{Vachaspati:2014bda}. With this, a large fraction of determinants can be evaluated with only an $\mathcal{O}(n_f^2)$ cost in the number of quarks (some cases could even be  $\mathcal{O}(n_f)$). Ref.~\cite{Detmold_2021} explored the use of sparsened propagator in calculating the correlation function in coarser lattice geometry, resulting in significant speed-ups in computation. Recently Ref.~\cite{https://doi.org/10.48550/arxiv.2201.04269} presented promising improvements in the time-complexity of block algorithm by using factor-tree and tensor e-graphs. 
 Despite all these improvements, efficient evaluation of correlation functions for realistic calculations of nuclei with larger atomic number, particularly within the generalized eigenvalue framework with a large set of operators, is still a bottleneck for lattice QCD based study of nuclear physics and requires further improvements of methodologies.{\footnote{
     In the near-term calculations, for $A=2$, where the Wick-contractions problem is not severe, the solution to SNR
     problem is most urgent.}}
 {\footnote{
   Pioneering proof-of-principle calculation with a limited set of interpolating fields for higher nuclei, till A = 28, were attempted in Ref.~\cite{PhysRevD.87.114512}.}}
 
 In this work we present a state-of-the-art solution to the Wick-contraction
 problem, and show that Wick-contractions for nuclei can be performed much more efficiently, and that the evaluation of Euclidean lattice correlation functions will also be possible even for higher nuclei.
 Our approach exploits both algorithmic optimizations as well as recent
 technological advances in large scale computation that have been driven by
 applications in machine learning and artificial intelligence.
 Our first contribution is a set of new optimizations to the determinant
 approach mentioned above.  Our second contribution is the construction of
 algorithms for QCD applications that exploit new tensor-based programming
 models (in particular, \tensorflow{}~\cite{tensorflow2015-whitepaper,tensorflow_developers_2023_8118033}).  While programming models like \tensorflow{}
 have been immensely successful in training and deploying large
 scale machine learning models, we believe this is the first exploration of its
 use with accelerating hardware in the domain of lattice QCD correlation functions.

 A key component of our proposed optimizations to the determinant
 approach is a new \emph{pre-processing} step, which exploits the degeneracy in
 the system to reduce the number of \emph{distinct} determinant computations.
 The crucial aspects of this pre-processing are that i) for a nucleus it needs to be
 performed only once, and ii) it can be performed only at one time-slice on a
 lattice much smaller than the one we actually seek to study. 
 We also show that a more careful use of the
 block structure of the matrices appearing in the determinant approach can
 reduce the time complexity from $\mathcal{O}(n_u^3n_d^3)$ (as reported, e.g.,
 in \cite{PhysRevD.87.114512}) to $\mathcal{O}(n_u^3 +
 n_d^3)$. 
Our second method goes beyond the traditional algorithmic paradigm, and
 exploits more recent programming models, such as \tensorflow{}~\cite{tensorflow2015-whitepaper,tensorflow_developers_2023_8118033} and
 \textsc{PyTorch}, developed to take advantage of specialized hardware
 accelerators like GPUs.  We show that using these tools, i) a high-level ``code'' for a given interpolating operator of a nucleus can be
generated \emph{automatically}, and next ii) that ``code" can be executed
more efficiently as these tools can \emph{automatically} perform optimizations such as choosing efficient orders of contraction, and taking advantage of the parallelism afforded by hardware accelerators.
It thus
 turns out that these algorithms are not only easier to write and check (because
 of the high-level nature of the programming model used), but also found to be \emph{more efficient} than the traditional approaches especially for lighter nuclei. 
 In fact, observing the efficiency of \tensorflow{} we encourage LQCD practitioners to adapt such advantages in computing $n$-point correlation functions.

     We find that
     our optimized determinant method is about an order of magnitude faster than existing algorithms for nuclei up to $A=7$ that we tested, and the improvement is expected to be even better for higher nuclei.
         The \tensorflow{}
 method is found to be even faster for smaller nuclei like Deuteron and Helium.
 Because of the convenience of writing ``automatic code generation" and automatic optimization we propose to use \tensorflow{} method till A = 4, and the optimized determinant method for larger A. The existing block algorithm can also be accelerated easily using \tensorflow{}.  With the efficiency that we find
 the combination of \tensorflow{} and determinant algorithms stand out to be
 state-of-the-art for heavier nuclei. Moreover for higher nuclei one can use a combination of determinant and block algorithm together with the idea of using clusters of smaller nuclei (which can be analysed using \tensorflow{} based accelerated methods). With this strategy and the observed efficiency of the above methods it is possible to
 calculate both two and three-point Euclidean lattice correlation functions for
 nuclei including the higher ones. If the aforementioned SNR problem can be
 controlled, which we believe can be achieved in future, this will enable us to
 obtain precision physics for nuclei including their energy spectra and
 structures from first principles calculations.

While calculating the two-point correlation functions using our determinant/block method we find an intriguing feature of the distributions of the minimal set of determinants or the terms in blocks that enter in constructing the correlation functions. For all the nuclei that we study, their distributions follow a statistical distribution resembling the Tracy-Widom type \cite{TRACY1993115}, where only a small fraction of determinants or the block indices construct the full correlator while most others are distributed around zero and cancel each other in the correlation functions. This observation suggests that among the large number of possibilities, specific spin-color combinations dominate nuclear correlation functions, hinting at a potential connection to an as-yet unidentified symmetry. 
Finding this important set of extreme terms beforehand can reduce the computing time by at least another order of magnitude, and hence can help in performing similar calculations for  higher nuclei. The improvement observed is, so far, numerical and systematic uncertainties associated with dropping any terms need to be fully understood, and we will study that in future.

In Sec.~\ref{Sec2}, we will elaborate on the Wick-contractions problem of nuclei, and discuss its algorithmic difficulty. The existing algorithms will then be discussed showing the current status to remedy this problem. In Sec.~\ref{sec:determ-based-algor}, we will first introduce the proposed determinant algorithm, then show its implementation procedure. Next we compare its performance with other existing algorithms. In Sec.~\ref{sec4} we introduce the \tensorflow{} method for correlation functions, its automatic implementation and the results. Sec.~\ref{sec_v} will deal with the main results showing the efficiency of our proposed methods in calculating two-point correlation functions of lattice QCD for Deuteron ($^2$H), Helium-3 ($^3$He), Helium-4 ($^4$He) and Lithium-7 ($^7$Li) nuclei. We then describe our study on the distribution of determinants in constructing a two-point function with a discussion on an intriguing feature that we discover. In Sec.~\ref{sec_vi} we 
summarize the findings of this work and discuss how our proposed methods can be employed for realistic calculations in nuclear physics in future. Details on the proposed algorithms and their implementation procedures are provided in the Appendices.

\section{Difficulties in first-principles LQCD calculations for nuclei}\label{Sec2}
Using LQCD, physical observables are, in general, obtained from the large-time behavior of the Euclidean correlation functions. For example, the energy spectra of a composite hadron, like a proton, is extracted through precise evaluation of two-point correlation functions of a proton's interpolating fields, which decay exponentially with its energy excitations. Similarly, hadronic structures are generally probed 
 through their interactions with various currents, which can be obtained from three and four-point functions involving those currents. This procedure is quite standard by now and has contributed significantly in understanding the structures and interactions of subatomic composite particles \cite{Gattringer_Lang, Workman:2022ynf}.  Nevertheless,  computations of correlation functions of nuclei following these procedures still remains a daunting task: the required time to perform Wick-contraction of quark fields in a nuclear correlation function can grow exponentially with the number of quarks. Below we provide an estimate of this scaling.

We consider the following two-point correlation function of a nucleus of mass number $A$, defined by,
\begin{eqnarray}
G_{\alpha_{1}\alpha_{2}\cdots\alpha_{A};\,\alpha_{1}^{\prime}\alpha_{2}^{\prime}\cdots\alpha_{A}^{\prime}}(x; x_0)=\langle N_{\alpha_1}(x)N_{\alpha_2}(x)\cdots N_{\alpha_A}(x)\Bar{N}_{\alpha_{1}^{\prime}}(x_0)\Bar{N}_{\alpha_{2}^{\prime}}(x_0)\cdots \Bar{N}_{\alpha_{A}^{\prime}}(x_0)\rangle,
    \label{eq1.1}
\end{eqnarray}
where $N_{\alpha}(x)$ is an interpolating operator for one nucleon with spinor index $\alpha$ and coordinate index $x$. The interpolating operator for a baryon can be written in the following general expression,
\begin{equation}
    N_{\alpha}(x)=\epsilon_{abc}(C\Gamma_1)_{\beta\eta}(\Gamma_2)_{\alpha\delta}q^a_{\delta f_1}(x)q^b_{\beta f_2}(x)q^c_{\eta f_3}(x),
\end{equation}
where $a$, $b$, $c$ are the color indices, $\beta$, $\delta$, $\eta$ and
$\alpha$ are the spinor indices and $f_i$'s are the flavor indices of the quark fields.
Here $C =\gamma_4\gamma_2$ is the charge
conjugation matrix and $\Gamma_i$ are the appropriate gamma matrices for the
baryon related to the nucleus under study.

The computational cost of Wick-contractions corresponding to the operator in Eq.~(\ref{eq1.1}) increases very steeply with $A$. The number of Wick contractions for a fixed value of index set `$i$', (spin, color and spatial index) is given by $N_{perm}=n_u!n_d!n_s!$, where $n_u$, $n_d$ and $n_s$ are the number of up, down and strange quarks in a nucleus respectively.{\footnote{For a regular nucleus, there is no valence strange quark and so $N_{perm} = n_u!n_d!$}} For example, for $^2$H, $^3$He and $^4$He, corresponding numbers are $N_{perm}=36,~ 2880$ and 518400, respectively. In addition to this, we need 
to include spin and color indices. Fortunately,  $\gamma$-matrices and Levi-Civita tensor ($\epsilon$) that represent spin and color indices are sparse.  
Even so, for each nucleon we need to contract at least one $\gamma$-matrix and one  $\epsilon$ tensor at the source and also the sink which will bring $4$ and $6$ non-zero entries, respectively. Hence, naively the cost of spin-color contraction scales as $N_{loop}=6^{2A}4^{2A}$. This factor rises very fast, for example, for $^2$H, $^3$He and $^4$He, it is more than $10^5, ~ 10^8$ and $10^{11}$, respectively. In addition to this a volume factor ($N_{vol}$) enters due to the volume sum of the sink positions, along with an additional factor for the number of statistical configurations ($N_{cfg}$). This leads to an overall cost of this computation  $N_{vol}\times N_{cfg}\times 6^{2A}4^{2A}\times n_u!n_d!n_s!$. In the Table \ref{table1} below we show, for a very modest size lattice ($N_{vol} = 32^3 \times 96$) and small number of statistical gauge configurations ($N_{cfg} = 100$), 
how fast this number explodes for the first few nuclei, making it impossible to perform Wick contractions naively with the present day computing resources.

\vspace{0.1cm}
\begin{table}[h]
\begin{center}
\begin{tabular}{|l|c|c|c|}
\hline
    Nuclei ($^A$X$_{N_p}$) & $N_{loop} (6^{2A}4^{2A})$ & $N_{perm} (n_u!n_d!)$ & $N_{tot}$ \\
\hline
    Hydrogen ($^1$H$_1$)& $6^2\times4^2=576$ & $2!\times1!=2$ & $\approx4\times10^{11}$\\
\hline
    Deuteron ($^2$H$_1$) & $6^4\times4^4=331776$ & $3!\times3!=36$ & $\approx4\times10^{15}$\\
\hline
     Helium-3 ($^3$He$_2$) & $6^6\times4^6\approx 2\times 10^{8}$ & $5!\times4!=2880$ & $\approx2\times10^{20}$\\
\hline
     Helium-4 ($^4$He$_2$) & $6^8\times4^8\approx 1\times 10^{11}$ & $6!\times6!=518400$ & $\approx2\times10^{25}$\\
\hline
    Lithium-6 ($^6$Li$_3$) & $6^{12}\times4^{12}\approx 4\times 10^{16}$ & $9!\times9!\approx1\times10^{11}$ & $\approx1\times10^{36}$\\
\hline
     Lithium-7 ($^7$Li$_3$) \, & $6^{14}\times4^{14}\approx 2\times 10^{19}$ & $11!\times10!\approx1\times10^{14}$ & $\approx 9\times10^{41}$\\
\hline
    Carbon ($^{12}$C$_6$) & $6^{24}\times4^{24}\approx 1\times 10^{33}$ \, & $18!\times18!\approx4\times10^{31}$  & $\approx1\times10^{73}$ \\
\hline
\end{tabular}
\end{center}
\caption{Number of floating-point-operations that are needed, in a naive estimation, to calculate the two-point functions of the low-lying nuclei on a $32^3\times96$ lattice for $N_{cfg}=100$.}
\label{table1}
\end{table}

\subsection{Summary of existing algorithms}\label{sec2a}
To reduce the above mentioned explosion of Wick-contraction terms for nuclei, significant algorithmic developments have been made over the years.
The block algorithm is one of the widely implemented procedures 
to deal with this time-complexity by considering a hadron-block of three-quark propagators \cite{Ishii_2007, Yamazaki_2010}. Subsequently this procedure has further been improved  in Refs.~\cite{PhysRevD.87.114512, PhysRevD.87.094513, Doi_2013}. Below we briefly sketch the outline of this algorithm.  For simplicity of discussion, we  consider the following  $^2$H correlation function ($A=2$), as given by
\begin{equation}
    C_{NP}(t)=\sum_{\mathbf{x},\alpha,\beta,\alpha^{\prime},\beta^{\prime}}T_{\alpha\beta}T_{\alpha^{\prime}\beta^{\prime}}\langle p_{\alpha}(\mathbf{x},t)n_{\beta}(\mathbf{x},t)\Bar{p}_{\alpha^{\prime}}(\mathbf{0},0)\Bar{n}_{\beta^{\prime}}(\mathbf{0},0)\rangle .
    \label{eq2.1}
\end{equation}
Here, $p_{\alpha}(x)$ and $n_{\beta}(y)$ are the interpolating operators of a proton and a neutron respectively. One possible choice for such operators is given by
\begin{eqnarray}
p_{\alpha}(x)=\epsilon_{abc}u_{\alpha}^{a}(x)u_{\beta}^{b}(x)(C\gamma_{5})_{\beta\eta}d_{\eta}^{c}(x)\label{eq:1000} ,\\
n_{\alpha}(x)=\epsilon_{abc}d_{\alpha}^{a}(x)d_{\beta}^{b}(x)(C\gamma_{5})_{\beta\eta}u_{\eta}^{c}(x) , \label{eq:1001}
\end{eqnarray}
where the $\alpha$, $\beta$, $\eta$ are the spinor indices and $a$, $b$ and $c$ are the color indices.
The operator $T = \frac{1}{2}(1+\gamma_4)(1\pm i\gamma_1\gamma_2)$ projects to the positive energy and desired spin combination. With this, in terms of  individual quark propagator, $S_q(\mathbf{x},t;\mathbf{0},0)\defeq\langle q(\mathbf{x},t)\Bar{q}(\mathbf{0},0)\rangle$, blocks of three quark propagators can be tied together as below:
\begin{eqnarray}
    B^{p}_{\alpha}(\mathbf{x},t; \chi_1^{\prime}, \chi_{2}^{\prime}, \chi_3^{\prime})&=&\langle p_{\alpha}(\mathbf{x},t)\Bar{u}^{a^{\prime}}_{\alpha^{\prime}}(\mathbf{0},0)\Bar{u}^{b^{\prime}}_{\beta^{\prime}}(\mathbf{0},0)\Bar{d}^{c^{\prime}}_{\eta^{\prime}}(\mathbf{0},0)\rangle\nonumber\\
    &=&\epsilon_{abc}(C\gamma_5)_{\beta\eta} \left[(S_u)_{\alpha\alpha^{\prime}}^{aa^{\prime}}(\mathbf{x},t;\mathbf{0},0)(S_u)_{\beta\beta^{\prime}}^{bb^{\prime}}(\mathbf{x},t;\mathbf{0},0)-(a\leftrightarrow b,\alpha \leftrightarrow \beta)\right] \nonumber\\
    &&\hspace{5cm}\times\,(S_d)_{\eta\eta^{\prime}}^{cc^{\prime}}(\mathbf{x},t;\mathbf{0},0)\label{eq2.4} ,\\
    B^{n}_{\alpha}(\mathbf{x},t; \chi_1^{\prime}, \chi_{2}^{\prime}, \chi_3^{\prime})&=&\langle n_{\alpha}(\mathbf{x},t)\Bar{d}^{a^{\prime}}_{\alpha^{\prime}}(\mathbf{0},0)\Bar{d}^{b^{\prime}}_{\beta^{\prime}}(\mathbf{0},0)\Bar{u}^{c^{\prime}}_{\eta^{\prime}}(\mathbf{0},0)\rangle\nonumber\\
    &=&\epsilon_{abc}(C\gamma_5)_{\beta\eta}  \left[(S_d)_{\alpha\alpha^{\prime}}^{aa^{\prime}}(\mathbf{x},t;\mathbf{0},0)(S_d)_{\beta\beta^{\prime}}^{bb^{\prime}}(\mathbf{x},t;\mathbf{0},0)-(a\leftrightarrow b,\alpha \leftrightarrow \beta)\right]
  \nonumber\\
    &&\hspace{5cm}\times\,(S_u)_{\eta\eta^{\prime}}^{cc^{\prime}}(\mathbf{x},t;\mathbf{0},0)
     .\label{eq2.5}
\end{eqnarray}
Here, the $\chi_i^{\prime}$ are the combined indices of spin and color of the quark fields at
the source.  More explicitly, $\chi^{\prime}_1\equiv\{\alpha,a\}$. $\chi^{\prime}_2\equiv\{\beta,b\}$ and $\chi^{\prime}_3\equiv\{\eta,c\}$ where $\alpha$, $\beta$ and $\eta$ are the spinor indices of the quarks and $a,b$ and $c$ are the color indices. This notation for combined indices will be used throughout this work. Subsequently the corresponding two-point function, in terms of these blocks, reduces to, 
\begin{eqnarray}
    C_{NP}(t)&=&\sum_{\mathbf{x},\mathbf{y},\alpha,\beta,\alpha^{\prime},\beta^{\prime}}T_{\alpha\beta}T_{\alpha^{\prime}\beta^{\prime}}\sum_{\sigma}B^{p}_{\alpha}(x; \chi_{\sigma(1)}^{\prime}, \chi_{\sigma(2)}^{\prime}, \chi_{\sigma(3)}^{\prime})B^{n}_{\beta}(y; \chi_{\sigma(4)}^{\prime}, \chi_{\sigma(5)}^{\prime}, \chi_{\sigma(6)}^{\prime})\nonumber\\
    & &\hspace{5.2cm}\times\,\epsilon_{a^{\prime}b^{\prime}c^{\prime}}\epsilon_{d^{\prime}e^{\prime}f^{\prime}}(C\gamma_5)_{\gamma^{\prime}\eta^{\prime}}(C\gamma_5)_{\mu^{\prime}\delta^{\prime}}\times sgn(\sigma) .\label{eq2.6}
\end{eqnarray}
Here, $\sum_{\sigma}\equiv\sum_{\sigma_u}\sum_{\sigma_d}$ with
$\sigma_u (\sigma_d)$ representing the permutations among the up (down) quarks,
whereas the factor $sgn(\sigma)$ is the sign of the permutation $\sigma$, and
arises from the Grassmannian nature of quark fields.

There are multiple advantages of using Eq.~(\ref{eq2.6}) over the direct computation of the correlation function. The block expressions Eq.~(\ref{eq2.4}) and Eq.~(\ref{eq2.5}) are antisymmetric under the exchange of $\chi_1^{\prime}$ and $\chi_2^{\prime}$ indices. Exploiting these symmetries, the full permutation set $\sigma$ in Eq.~(\ref{eq2.6}) can be constrained within a smaller subset $\sigma_{sub}$ which is smaller than $\sigma$ by a factor of $2^A$. In addition, as the color and spin indices at the sink are contracted in Eq.~(\ref{eq2.4}) and Eq.~(\ref{eq2.5}) before the evaluation of Eq.~(\ref{eq2.6}), $N_{loop}$ is reduced from $4^{2A}6^{2A}$ to $4^A6^A$, modulo the  negligible cost of evaluating the block tensors. Hence, use of this block algorithm 
can reduce the time-complexity of the correlation function by a factor of $48^A$ compared to the direct computation.

Another efficient algorithm is the unified contraction algorithm 
which eliminates redundancies in the computation by searching for the non-zero entries of the coefficient matrices of the operator after applying the inverse permutation $\sigma^{-1}$ on it. In Ref.~\cite{PhysRevD.87.114512} a matrix-determinant formulation has been proposed where one calculates the correlation function by computing determinants of a block matrix containing the elements of the propagator which results in an asymptotic scaling of $\mathcal{O}(n_{u}^3n_{d}^3)$. We will discuss this in detail in the next section.
Ref.~\cite{Detmold_2021} presented an algorithm that uses sparsened propagator in calculating the correlation function in coarser lattice geometry resulting in significant speed-ups in computation. Recently Ref.~\cite{https://doi.org/10.48550/arxiv.2201.04269} presented some promising improvements in the time-complexity of block algorithm by using factor-tree and tensor e-graphs.

\section{Determinant based algorithms and an efficient implementation}\label{sec:determ-based-algor}
A strategy for mitigating the time-complexity of the computation of nuclear
correlation functions by exploiting the symmetries of the quark field indices is
suggested by the existence of efficient algorithms for the computation of the
determinant.  Such a determinant based algorithm was first introduced in
Ref.~\cite{PhysRevD.87.114512}, where it was shown that using this idea, the
time-complexity of evaluating the two-point correlation functions scales as
$\mathcal{O}(n_{q_1}^3n_{q_2}^3n_{q_3}^3)$, where the $n_{q_i}$ denote the
numbers of the three quark flavors in a baryon constituting a nucleus. It was
also shown in Ref.~\cite{PhysRevD.87.114512} that for simple operator choices,
LQCD study of light nuclei such as $^4$He, $^8$Be, $^{12}$C, $^{16}$O {\it
  etc.}, can in principle be performed using this algorithm. In this work (see
subsection~\ref{sec3a}) we further analyze the determinant based method to show
that it can in fact be implemented more efficiently so that its time-complexity
has the much milder scaling $\mathcal{O}(n_{q_1}^3+
n_{q_2}^3+n_{q_3}^3)$. 
 
In subsection \ref{impl} we then discuss a novel implementation which proceeds
by finding the minimal set of non-zero determinant computations that are needed
in order to compute the correlation function.  Using this idea, we show that one
can significantly reduce the redundancy in calculating the determinants and
improve the efficiency of the algorithm beyond the worst-scale analyses performed in Ref.~\cite{PhysRevD.87.114512} and subsection~\ref{sec3a}. Finally, in subsection~\ref{ss3c}, we demonstrate these improvements by
calculating the two-point correlation functions for a few low-lying nuclei.  We
also present a relative comparison of the performances of the determinant based
algorithms using the existing and the proposed methods.
\subsection{Determinant Algorithm}\label{sec3a}
By exploiting various symmetries of the 
quark field indices,  correlation functions of an atomic nucleus can be re-written in terms of suitable low-rank determinants which are dependent on quark propagators with specific spin and color indices. A generic local interpolating operator, at a space-time point ($\mathbf{x},t$), of a nucleus with $n$ number of quarks can be written in terms of its quark fields as \cite{PhysRevD.87.114512}
\begin{eqnarray}
    \mathcal{N}_{\alpha}(\mathbf{x},t)=\sum_{\boldsymbol{\chi}}C^{\chi_1\chi_2\cdots \chi_{n}}_{\alpha}q(\mathbf{x},t,\chi_1)q(\mathbf{x},t,\chi_2)\cdots q(\mathbf{x},t,\chi_{n}).\label{eq2.7}
\end{eqnarray}
Here, $\alpha$ stands for all the quantum numbers describing the nucleus, including spin, isospin,
strangeness, angular momentum etc., while the multi-dimensional
vector $\boldsymbol{\chi}$ stands for $\{\chi_1,\chi_2,\cdots \chi_{n}\}$, with
$\chi_i$ representing the  color, spin, and flavor of the quarks
combined together in one index.\footnote{For a spatially extended operator, spatial indices can also be included within $\chi$.} The tensor
$C^{\chi_1\chi_2\cdots \chi_{n}}_{\alpha}$ can be partially or fully antisymmetrized  with respect to the indices $\chi_i$'s by virtue of
the anti-commutating quark fields which are Grassmann variables. 
The corresponding zero-momentum two-point correlation function, between a source at a space-time point ($\mathbf{x_0},t_0$) to a sink-point ($\mathbf{x},t$), is given by
\begin{eqnarray}
    G_{\alpha}(t)&=&\sum_{\mathbf{x}}\langle \mathcal{N}_{\alpha}(\mathbf{x},t)\Bar{\mathcal{N}}_{\alpha}(\mathbf{x_0},t_0)\rangle \nonumber\\&=&\sum_{\mathbf{x}}\sum_{\boldsymbol{\chi}, \boldsymbol{\chi^{\prime}}}C^{\prime\chi_{1}^{\prime}\chi_{2}^{\prime}\cdots \chi_{n}^{\prime}}_{h}C^{\chi_1\chi_2\cdots \chi_{n}}_{h}\langle q(\chi_1)q(\chi_2)\cdots q(\chi_{n_q})\Bar{q}(\chi_{1}^{\prime})\Bar{q}(\chi_{2}^{\prime})\cdots \Bar{q}(\chi_{n_q}^{\prime})\rangle.
\end{eqnarray}
The evaluation of the two-point function proceeds through the path-integration
of the Grassmannian quark fields using Wick's theorem
\cite{PhysRev.80.268} (see also sec. 4.3 of \cite{Peskin:1995ev}),
\begin{eqnarray}
    \langle q(\chi_1)\cdots q(\chi_{n})\Bar{q}(\chi_{1}^{\prime})\cdots \Bar{q}(\chi_{n}^{\prime})\rangle&=&\frac{1}{\mathcal{Z}}\int\mathcal{D}U\mathcal{D}q\mathcal{D}\Bar{q}e^{-A_{QCD}}q(\chi_1)q(\chi_2)\cdots q(\chi_{n_q})\nonumber\\
    && \hspace*{1.1in} \times \,\, \Bar{q}(\chi_{1}^{\prime})\Bar{q}(\chi_{2}^{\prime})\cdots \Bar{q}(\chi_{n_q}^{\prime})\nonumber\\
    &&  \hspace*{-1.1in} = \,\frac{(-)^{n}}{\mathcal{Z}}\int\mathcal{D}U e^{-A_{eff}(U)}\sum_{\sigma\in S_{[n_q]}}sgn(\sigma)\prod_{i=1}^{n_q}S(\mathbf{x},t,\mathbf{x_0},t_0;\chi_{i},\chi^{\prime}_{\sigma(i)}) .
\end{eqnarray}
 Here, $U$ represents a gauge field in the path-integral formulation, $A$ is the QCD action,  $A_{eff}[U]=A_G[U]-\sum_{q_i}\log\det S_{q_i}[U]$, $\mathcal{Z}$ is the partition function, and $S_{[n_q]}$ denotes the permutation group on $n_q$ objects.\footnote{Note that $S_{[n_q]}$ denotes the permutation group while $S_{q_i}$ represents the quark propagator for the quark flavor $q_i$.} The matrix  $S(\mathbf{x},t,\mathbf{x_0},t_0;\chi_{i},\chi^{\prime}_{j})$
 represents the quark propagator of a particular flavor from the space time point ($\mathbf{x_0},t_0$) to ($\mathbf{x},t$) while the other indices are represented by $\chi^{\prime}$ and  $\chi$.  Since QCD is flavor-conserving, one can
separate out the Wick-contractions for individual flavors, i.e., for each flavor there will be a separate term similar to the above expression. Hence for a nucleus with $n_u$, $n_d$ and $n_s$ number of up, down and strange quarks, the above expression becomes (we will drop the overall $((-)^{n}/\mathcal{Z})e^{-A_{eff}(U)}$ factor for convenience of writing),
\begin{eqnarray}
  \langle .... \rangle \, \sim \, \eta &&\sum_{\sigma_{u}\in S_{[n_u]}}\sum_{\sigma_{d}\in S_{[n_d]}}\sum_{\sigma_{s}\in S_{[n_s]}}sgn(\sigma_{u})sgn(\sigma_{d})sgn(\sigma_{s})  \nonumber \\
  && \hspace*{0.5in}\times\,\,\prod_{i=1}^{n_u}\prod_{j=1}^{n_d}\prod_{k=1}^{n_s}S_u(\chi_{i},\chi^{\prime}_{\sigma_u(i)})S_d(\chi_{j},\chi^{\prime}_{\sigma_d(j)})S_s(\chi_{k},\chi^{\prime}_{\sigma_s(k)}).
  \label{eq2.8}
\end{eqnarray}
Here $\eta = \pm1$, which arises from grouping the quarks and anti-quark fields of the same flavors before performing the integration over the fermionic fields.

Reducing two-point functions to the form of Eq.~(\ref{eq2.8}) brings immediate benefits.
By separating the flavor, rather than combining it together, one can do the summation over $\sigma_u$, $\sigma_d$ and $\sigma_s$ independently. Hence, for any specific choice of indices $\{\chi;\chi^{\prime}\}$, the cost of performing the above sum over all possible permutations becomes $\propto \mathcal{O}(n_u!+n_d!+n_s!)$. 
(It should be noted also that for the block algorithm in subsection (\ref{sec3a}) this independent summation is not possible.) By defining a matrix for each flavor, as below,
\begin{eqnarray}
    M^q_{ij}(\mathbf{\chi},\mathbf{\chi}^{\prime})&=&S_q(\chi_i, \chi^{\prime}_j), \hspace{1.5cm} \text{for $i,j\leq n_q$ and $q\in {u,d,s}$}, 
\end{eqnarray}
one can express each of the above independent sum by a determinant as 
\begin{eqnarray}
\det(M)=\sum_{\sigma}sgn(\sigma)\prod_{i=1}^nM_{i,\sigma(i)}.
\end{eqnarray}
With this simplification, the above two-point function (Eq.~\ref{eq2.8}) can hence be expressed in terms of these determinants as
\begin{equation}
    G_{\alpha}(t)=\eta\sum_{\mathbf{x}}\sum_{\boldsymbol{\chi}, \boldsymbol{\chi^{\prime}}}C^{\prime\chi_{1}^{\prime}\chi_{2}^{\prime}\cdots \chi_{n}^{\prime}}_{\alpha}C^{\chi_1\chi_2\cdots \chi_{n}}_{\alpha}\det(M^{u}(\boldsymbol{\chi}^u,\boldsymbol{\chi}^{\prime u}))\det(M^{d}(\boldsymbol{\chi}^d,\boldsymbol{\chi}^{\prime d} ))\det(M^{s}(\boldsymbol{\chi}^s,\boldsymbol{\chi}^{\prime s})).\label{eq2.9}
\end{equation}
As an example, below we consider the carbon-12 $(^{12}C)$ nucleus which has 18 up and 18 down quarks. We consider the following interpolating operator for it:
\begin{eqnarray}
\mathcal{O}_{^{12}C}=\sum_{\boldsymbol{\chi}^u,\boldsymbol{\chi}^d}C^{\chi^u_1\cdots\chi^u_n;\chi^d_1\cdots\chi^d_n}\prod_{i=1}^{n=18} u(\chi^{u}_i)\prod_{j=1}^{n=18} d(\chi^{d}_j).
\end{eqnarray}
Here, $\chi^u_i$'s and $\chi^d_i$'s are combined index for spin, color and spatial indices for quarks with up and down flavors respectively, and $n=18$ is the number of up and down quarks in $^{12}$C. As in Eq.~(\ref{eq2.9}), the corresponding two-point function then takes the following form,
\begin{eqnarray}
    G_{^{12}C}(t)=\eta\sum_{\boldsymbol{\chi}^u,\boldsymbol{\chi}^d}\sum_{\tilde{\boldsymbol{\chi}}^{u}, \tilde{\boldsymbol{\chi}}^{d}}C^{\prime\tilde{\chi}^u_{1}\cdots\tilde{\chi}^u_{n}\tilde{\chi}^d_{1}\cdots\tilde{\chi}^d_{n}}C^{\chi^u_{1}\cdots\chi^u_{n}\chi^d_{1}\cdots\chi^d_{n}}\det(M^{u}(\boldsymbol{\chi}^u,\tilde{\boldsymbol{\chi}}^u))\det(M^{d}(\boldsymbol{\chi}^d,\tilde{\boldsymbol{\chi}}^d)).
\end{eqnarray}
It is interesting to note that these determinants are not large in size. For $^{12}$C,  $M^u$ and $M^d$  are only $18\times 18$ matrices. Of course,  the matrices $M^u$ and $M^d$ depend on the particular choices of indices $\chi$ and $\chi^{\prime}$, and hence it is necessary to calculate those determinants for all choices of $\mathbf{\chi}$ and $\mathbf{\chi}^{\prime}$ for which $C^{\chi_{1}^{\prime}\chi_{2}^ {\prime}\cdots \chi_{n}^{\prime}}_{\alpha} C^{\chi_1\chi_2\cdots \chi_{n}}_{\alpha}$ is non-zero. This number could still be very large and one needs an efficient method to calculate these large number of low-rank determinants. We find that up to $4\times 4$ matrices computation of determinants is faster via coding up the  analytical expression, while for higher sizes LU decomposition utilizing an optimized linear algebra package can efficiently be employed.

We would like to emphasize here that 
the total time to compute the correlation function using Eq.~(\ref{eq2.9}) scales as $NN^{\prime}\times \mathcal{O}(n_u^3+n_d^3+n_s^3)$, where
  $N$ and  $N^{\prime}$ are the number of non-zero entries in $C_{\alpha}$ and $C_{\alpha}^{\prime}$.
 On the contrary, in Ref.~\cite{PhysRevD.87.114512} the scaling of the determinant algorithm was given as $NN^{\prime}\times\mathcal{O}(n_u^3n_d^3n_s^3)$.\footnote{The definitions of $N$ and $N^{\prime}$ in \cite{PhysRevD.87.114512} are slightly different from the definitions we have used here. It was defined as the number of ordered list of indices $\chi_i$'s for which the entries of totally anti-symmetrized $C^{[\chi_1\chi_2\cdots \chi_{n}]}_{\alpha}$ is non-zero. However we can trivially go to the definitions used there if we had started with a completely antisymmetric a $C^{\chi_1\chi_2\cdots\chi_n}_{\alpha}$. } In the subsections below we describe how to implement the above mentioned determinant algorithm and demonstrate its efficacy by calculating two-point functions of $^2$H, $^3$He, $^4$He and $^5$Li. The vanilla method is indeed found to scale as $NN^{\prime}\times \mathcal{O}(n_u^3+n_d^3+n_s^3)$. The improved method, which uses only the independent determinants with their degeneracy factors, reduces the time-complexity substantially over the vanilla method.
\subsection{An efficient implementation of the proposed algorithm} \label{impl}
The determinant algorithm proposed in the previous subsection scales
polynomially with the number of quarks in a nucleus. However, it can be
made much more efficient if the number of nonzero entries in Eq.~(\ref{eq2.9}) are
smaller in number. In order to achieve that, it is necessary to find the redundancies, if
any, in the evaluation of Eq.~(\ref{eq2.9}). For any operator it is expected that
there will be large redundancies among these determinants when their indices are
constrained by spin and colors.
The rationale behind this are the following:
\begin{enumerate}[{\bf R.1}]
    \item \label{cause1} Suppose that for a nucleus with $n$ number of up quarks we need to calculate the determinants of a matrix $\mathbf{M^u}$ as defined in the subsection \ref{sec3a}. $\mathbf{M^u}$ can be expressed with respect to its indices as below:
    \begin{eqnarray}
  \mathbf{M^u}=\mathbf{M^u}(\chi_1, \chi_2,\cdots,\chi_n;\chi^{\prime}_1,\chi^{\prime}_2,\cdots,\chi^{\prime}_n);\hspace{0.75cm}
        M^u_{ij}=S_u(\chi_i, \chi^{\prime}_j)\hspace{0.25cm}\textrm{for} \,\,i,j\leq n.\label{eq2.10}
    \end{eqnarray}
    The first part of the above equation states that the entries of the matrix are only dependent on the spin-color indices of all $n$ up quarks at both source and sink. The determinant of this matrix can be written as,
    \begin{eqnarray}
        &&\det\mathbf{M^u}(\chi_{\sigma(1)}, \chi_{\sigma(2)},\cdots, \chi_{\sigma(n)};\chi^{\prime}_{\tau(1)}, \chi^{\prime}_{\tau(2)}, \cdots, \chi^{\prime}_{\tau(n)})\nonumber\\
        &&\hspace*{0.2in} = sgn(\sigma)sgn(\tau)\det\mathbf{M^u}(\chi_1, \chi_2,\cdots,\chi_n;\chi^{\prime}_1,\chi^{\prime}_2,\cdots,\chi^{\prime}_n),
    \end{eqnarray}
    where $\sigma,\tau\in S_{n}$. In other words, if $m$ sets of
    $\{\boldsymbol{\chi};\boldsymbol{\chi}^{\prime}\}$  are related by permutations, the determinants of various possible $\mathbf{M^u}$ matrices corresponding to these $m$ cases
    will only  differ by the product of the signs of the permutations. Hence, calculating the determinant of $\mathbf{M^{u}}$ only once for such $m$ cases will suffice provided that we keep track of the relative signs. 
  \item \label{cause2} The summation in Eq.~(\ref{eq2.9}) is over all the choices of
    $\{\chi; \chi^{\prime}\}$ for which
$C^{\chi_{1}^{\prime}\chi_{2}^{\prime}\cdots \chi_{n}^{\prime}}_{\alpha}
    C^{\chi_1\chi_2\cdots \chi_{n}}_{\alpha}$ is non-zero. Here spin, color and flavor indices are put together as
    $\{\chi\}=\{\{\chi_i\},\{\chi_j\}, ...\}$, with $\{\chi_i\}$ as the spin-color indices of $i$-th flavored quark. 
    For a flavor $f$, the matrix $\mathbf{M^{f}}$ only depends on
    $\{\chi_f;\chi^{\prime}_f\}$.  
    For a distinct set of $\{\chi\}$
    and $\{\chi^{\prime}\}$, some of the $\{\chi_f\}/\{\chi_f^{\prime}\}$ inside it could be the same, and hence that needs to be evaluated just once which saves a lot of computations.
    For example, consider two
    sets of indices $\{\chi\}^a=\{\chi_u^a, \chi_d^a, \chi_s^a\}$ for
    $a\in\{1,2\}$. As all the $\{\chi\}$'s are distinct, we know $\{\chi_f^1\}$
    cannot be equal to $\{\chi_f^2\}$ for all possible $f\in\{u,d,s\}$. However
    for $f\in H$, where $H\subset \{u,d,s\}$, $\{\chi_f\}^1=\{\chi_f\}^2$ is
    possible. If we fix $\{\chi^{\prime}\}$ to any specific choice, then the
    determinants of $\mathbf{M^{f}}$ for $f\in H$ will be equal for these
    two choices of $\{\chi\}$'s. Hence, it is sufficient to calculate them only
    once and then reuse. 
\end{enumerate}
These redundancies immediately suggest that it is advantageous to first find out
the minimal set of determinants that contribute to the sum in
Eq.~\eqref{eq2.9}.  This leads to the following two questions: how should one
\emph{compute} such a minimal set of determinants?  In fact, similar
redundancies have also been noted, and exploited, for special cases in
Ref.~\cite{pacs-cs_collaboration_helium_2010}.  Below, we propose a general
algorithmic method for detecting such redundancies, building upon techniques
from computational number theory (we provide more comparisons with the work
\cite{pacs-cs_collaboration_helium_2010} at the end of this subsection).  Next,
does this optimization actually helps in speeding up the computation for
correlation functions?  We address this second issue in subsection~\ref{ss3c},
where we show the implementation of the optimized determinant algorithm with
examples.

\subsubsection*{Computation of non-redundant minimal set of determinants}
\label{sec:computing-set-non} 
A direct method of accounting for such
    redundancies would be to manually 
    analyze the combinatorial features of the
    problem and to compute a representative set of indices.  However, that
    computation would have to be performed separately for each nucleus. Hence, instead of combinatorially accounting for all possible
    redundancies, we suggest a simpler, more automatic, algorithm inspired by
    randomized algorithms ubiquitous in computational algebra and number theory
    applications such as primality testing.  We provide the mathematical background
    for this method in Appendix~\ref{sec:append-ident-test}, and describe only the computational
    details here.

Using Eq.~(\ref{eq2.9}) we first compute the correlation function using a quark propagator on a very small-size lattice and only at one space-time point\footnote{For the case of smeared operators, we
  need to perform the calculation over any one set space-time points in lattice
  over which the operator is smeared. 
  Computation over
  the whole lattice is not necessary.
  } and get a list of the all the non-zero determinants
that are needed to calculate the correlation function. Then this list is
analysed numerically to find out the minimal set of indices
$\{\{\chi;\chi^{\prime}\}\}$ for which determinants needs to be calculated. Since
integer arithmetic is exact in computer systems, 
instead of taking an actual quark propagator, in this first step it is better to use a quark propagator containing random integers.  We describe the theoretical
underpinnings of this method in Appendix~\ref{sec:append-ident-test}.  Let us
denote this minimal set of indices as $\{\{\chi_f;\chi^{\prime}_f\}\}|_{min}$
for $f\in\{u,d,s\}$. The number of elements in these sets corresponding to
different flavors generally are different. Here, we would like to stress that
the minimal set of indices $\{\{\chi_f;\chi^{\prime}_f\}\}|_{min}$ only depends
on permutation symmetry of determinant and the specific choice of the
interpolating operator. It does not depend on the Dirac propagator
used.\footnote{Of course the actual numerical values of these determinants will
  depend on the Dirac propagator used. However the minimal set of indices
  $\{\chi;\chi^{\prime}\}$ for which we need to calculate the determinants is
  independent of the choice of propagator.} Once we calculate this minimal
set on a small-size lattice and at one space-time point that set will be the same for i) all the space-time points on
that lattice, ii) for all gauge configurations, and 
 for iii) any size of lattice. This one-time evaluation of the minimal set which is independent of the lattice-size makes this method very efficient. Once this step is performed,
we proceed to calculate the correlation function by computing only the minimal number of determinants corresponding to the indices of the minimal set.

Another way to deal with these redundancies would be to first totally
antisymmetrize the matrix $C^{\chi_1\chi_2\cdots \chi_{n}}_{\alpha}$ with
respect to the indices $\{\chi_1,\chi_2,\cdots,\chi_n\}$ and work with an ordered
list of indices for which this tensor
$C^{[\chi_1\chi_2\cdots \chi_{n}]}_{\alpha}$ is non-zero. Due to the fact that we are now working with an
ordered list of indices, the first source of redundancies (\ref{cause1} above)
is resolved (as was done in Ref.~\cite{PhysRevD.87.114512}). 
However, the second source of redundancies described above
(\ref{cause2}) still remains. To find the minimal set of determinants, one can go through a similar analysis again using randomized algorithms as mentioned above. Another way to resolve the redundancies in \ref{cause2} is to find out the
representative set of indices by inspecting the above ordered list of indices
and then arriving at the minimal set by checking their equality (a similar
inspection strategy has been described in
\cite{pacs-cs_collaboration_helium_2010}).

A comparison of the randomized method described earlier with the above method
based on inspection is in order.  Note that the inspection method is dependent
on the nucleus under investigation and needs to be modified accordingly when
analyzing different nuclei. On the other hand, the advantage of the randomized
method is that it is rather general, and independent of the nucleus under study.  In fact, it does not even rely upon \ref{cause1} and \ref{cause2}
being the only sources of redundancies. We expect that for smaller nuclei, the
two methods would be comparable in terms of time-complexity, and this is what
we find for the case of $^4$He: for this case, the methods produce the same
minimal set in roughly the same time.  However, the advantages
coming from the simplicity and generality of the randomized method become more
pronounced when studying nuclei of larger $A$ where more redundancy is expected: one can use it without worrying
about the exact sources of degeneracy.

In the following subsection we present the implementation of the proposed
determinant algorithm for a few low-lying nuclei and compare that with
Ref.~\cite{PhysRevD.87.114512}. The results below illustrate that optimization
through the minimal set of indices improves the efficiency of the algorithm
substantially.

\subsection{Implementation of the proposed algorithm with a quantitative comparative analysis}\label{ss3c}
Below we describe our optimized determinant algorithm and detail the calculations for the two-point correlation functions of $^2$H, $^3$He, $^4$He and $^7$Li. We will then make a comparative analysis of 
the efficiency of this algorithm and the existing block as well as determinant algorithms.

We use the interpolating operators for proton and neutron as given in Eq.~(\ref{eq:1000}) and Eq.~(\ref{eq:1001}) respectively. Using those we construct the following interpolating operators for $^2$H ,$^3$He, $^4$He and $^7$Li, 
\begin{eqnarray}
    \mathcal{O}_{^2{\mathrm{H}}}(x)&=&n^{T}(x)(C\gamma_5P_{+})p(x) ,\label{NP_op}\\
    \mathcal{O}^J_{^3{\mathrm{He}}}(x)&=&p^{T}_{-}(x)(C\gamma_5)n_{+}(x)p_{+}^J(x).\label{He3_op}\\
    \mathcal{O}_{^4{\mathrm{He}}} &=& \frac{1}{\sqrt{2}}(\bar{\chi}\eta - \chi\bar{\eta})
    \label{He4_op}\\
    \mathcal{O}_{^7{\mathrm{Li}}}(x) &=& \mathcal{O}_{^4{\mathrm{He}}}(x)n^{T}_{-}(x)(C\gamma_5)p_{+}(x)n_{+}^J(x).
\end{eqnarray}
Here, $\chi$ and $\bar{\chi}$ represent spin wave functions and are defined by \cite{Yamazaki:2009ua},
\begin{eqnarray}
\chi &=& \frac{1}{2}\left([\uparrow \downarrow \uparrow \downarrow] + [\downarrow \uparrow \downarrow \uparrow] - [\uparrow \downarrow \downarrow \uparrow] - [\downarrow \uparrow \uparrow \downarrow]\right) ,\\ 
\bar{\chi} &=& \frac{1}{\sqrt{12}}\left([\uparrow \downarrow \uparrow \downarrow] + [\downarrow \uparrow \downarrow \uparrow] + [\uparrow \downarrow \downarrow \uparrow] + [\downarrow \uparrow \uparrow \downarrow] - 2 [\uparrow \uparrow \downarrow \downarrow] - 2 [\downarrow \downarrow \uparrow \uparrow]\right) .
\end{eqnarray}
Here, ($\uparrow, \downarrow$)  represent the up and down spins of the nucleons respectively. Along with that, $\eta$ and $\bar{\eta}$ are isospin wave functions and can be defined by replacing 
the spin-doublet 
($\uparrow$, $\downarrow$) with isospin doublet ($p$, $n$), where $p_{\pm}^{\alpha}$ and $n_{\pm}^{\alpha}$ are,
\begin{eqnarray}
     (p_{\pm})_{\alpha}(x)=\epsilon_{abc}u_{\alpha}^{a}(x)u_{\beta}^{b}(x)(C\gamma_{5})_{\beta\eta}(P_{\pm})_{\eta\delta}d_{\delta}^{c}(x) , \label{p_int}\\
     (n_{\pm})_{\alpha}(x)=\epsilon_{abc}d_{\alpha}^{a}(x)d_{\beta}^{b}(x)(C\gamma_{5})_{\beta\eta}(P_{\pm})_{\eta\delta}u_{\delta}^{c}(x) ,
\end{eqnarray}
with the spin projection operator $P_{\pm}=\frac{1}{2}(1\pm\gamma_4)$. With these operators we proceed to calculate their zero-momentum projected two-point correlation functions. We implement the determinant algorithm in the following two ways,
\begin{enumerate}[{\bf A.}]
    \item \label{Vanl_Method}Vanilla method: Anti-symmetrize the coefficient matrix $C^{\chi_1\chi_2\cdots \chi_{n}}_{\alpha}$ for these operators and then directly use Eq.~(\ref{eq2.9}) to calculate the correlation functions. 
    This can be done by replacing $C^{\chi_1\chi_2\cdots \chi_{n}}_{\alpha}$ and $C^{\prime\chi^{\prime}_1\chi^{\prime}_2\cdots \chi^{\prime}_{n}}_{\alpha}$ in Eq.~(\ref{eq2.9}) with $C^{[\chi_1\chi_2\cdots \chi_{n}]}_{\alpha}$ and $C^{\prime[\chi^{\prime}_1\chi^{\prime}_2\cdots \chi^{\prime}_{n}]}_{\alpha}$ respectively, with a sum over only the ordered list of indices $\{\chi;\chi^{\prime}\}$, 
     where $C^{[\chi_1\chi_2\cdots \chi_{n}]}_{\alpha}$ and $C^{\prime[\chi^{\prime}_1\chi^{\prime}_2\cdots \chi^{\prime}_{n}]}_{\alpha}$ are given as,
    \begin{eqnarray}
        C^{[\chi_1\chi_2\cdots \chi_{n}]}_{\alpha}\equiv\sum_{\sigma\in S_{[n]}}sgn(\sigma)C^{\chi_{\sigma(1)}\chi_{\sigma(2)}\cdots \chi_{\sigma(n)}}_{\alpha},\\
        C^{\prime[\chi_1\chi_2\cdots \chi_{n}]}_{\alpha}\equiv\sum_{\sigma\in S_{[n]}}sgn(\sigma)C^{\prime\chi_{\sigma(1)}\chi_{\sigma(2)}\cdots \chi_{\sigma(n)}}_{\alpha}.
    \end{eqnarray}
    A similar construction was used in Ref.~\cite{PhysRevD.87.114512}.  As also
    noted above, a  scaling of $\mathcal{O}(n_u^3n_d^3n_s^3)$ was claimed in
    Ref.~\cite[eq.~(20)]{PhysRevD.87.114512}.  However, as shown above, a tight
    analysis actually gives a scaling of $\mathcal{O}(n_u^3 + n_d^3 + n_s^3)$.  We now
    move on to describe further optimizations.

    \item \label{Imp_Method}Improved method: First, anti-symmetrize the coefficient matrix $C^{\chi_1\chi_2\cdots \chi_{n}}_{\alpha}$ for these operators. Then on a single lattice point of a small size lattice calculate all the determinants, and following the procedure described in the previous section, identify the minimal set of indices $\{\{\chi_f;\chi^{\prime}_f\}\}|_{min}$ where $f\in\{u,d,s\}$ 
    to find the minimal set of determinants. Next, calculate the correlation functions by computing the determinants of these minimal set for all space-time points of the desired lattices and then for all gauge configurations. Note that the indices for this minimal set are the same at all lattice points and for all gauge configurations.
\end{enumerate}
The method \ref{Imp_Method} is our proposed optimized method of implementation of the determinant algorithm. 
In Table \ref{Table1} we present execution times for two-point function computation of $^2$H, $^3$He, $^4$He and $^7$Li using these two approaches.
The number of terms in the operator stands for the number of ordered list of indices $\{\chi\}$ for which $C^{\chi_1\chi_2\cdots \chi_{n}}_{\alpha}$ is non-zero. As we can clearly see from the execution times, calculating these correlation functions using method \ref{Imp_Method}  is more efficient compared to the method \ref{Vanl_Method}.

\begin{table}[t]
     \centering
     \begin{tabular}{|c|c|c|c|c|c|}
       \hline
       Nuclei & \multicolumn{1}{|p{2cm}|}{\centering No. of terms \\ in the operator } & \multicolumn{1}{|p{2cm}|}{\centering Execution time using \\ procedure \textbf{A}} 
              & \multicolumn{1}{|p{2cm}|}{\centering Rank of\\ Determinants}& \multicolumn{1}{|p{2cm}|}{\centering Execution time using \\ procedure \textbf{B}}& \multicolumn{1}{|p{2cm}|}{\centering Execution time ratio \textbf{A}/\textbf{B}}\\
       \hline
       $^2$H & $356$ & 148.74 & $\{3\times 3\}$ & $34.15$ & $4.35$ \\
       \hline
       $^3$He & $2\times1026$\footnote{$1026$ terms for $^3$He spin $+1/2$ and spin $-1/2$ components each.} & $20110.60 $ & $\{5\times5, 4\times4\}$\footnote{$5\times5$ matrices for up quarks and $4\times4$ matrices for down quarks.}  & $2754.16$ & $7.30$\\
       \hline
       $^4$He & $2716$ & $90547.23 $ & $\{6\times6\}$  & $3246.38$ & $27.89$\\ 
       \hline      
       $^7$Li & $2716\times1026$ & $187532.34$ & $\{6\times6,5\times5,4\times4\}$ &  $7586.86$& $24.72$\\ 
       \hline
     \end{tabular}
     \caption{Execution times (wall-clock time in seconds) for evaluating  point-sink two-point correlation functions, at a given time-slice, for a few low-lying nuclei (lattice size: $24^3\times 64$). These calculations are performed on a single core of 12-th Gen Intel(R) Core(TM) i7 processor.\label{Table1}} 
\end{table}

It is also interesting to compare the actual execution times in Table \ref{Table1} with 
the theoretical scaling $NN^{\prime}\times\mathcal{O}(n_u!+n_d!+n_s!)$ as discussed in section \ref{sec3a}. For $^2$H : $N=N^{\prime}=356$,  and $n_u=n_d=3$, and for $^3$He : $N\times N^{\prime}=2\times(1026)^2$, $n_u=5$ and $n_d=4$. One thus expects the computation time to scale as $NN^{\prime}\times\mathcal{O}(n_u!+n_d!+n_s!)$ when $n_u,n_d$ and $n_s$ are comparatively small. With these numbers, the execution times for $^3$He/$^2$H have a ratio
$$\frac{2\times(1026)^2\times(120+24)}{(356)^2\times(6+6)}\approx 200.$$ From Table \ref{Table1}, we find this number to be 135 for the vanilla method (\ref{Vanl_Method}), which is of the same order as the theoretical scaling. A similar calculation for $^4$He implies that the execution times for $^4$He/$^3$He should have a ratio of about $10$. From Table \ref{Table1}, we find this number to be $4.5$ for the vanilla method (\ref{Vanl_Method}). Of course, in the improved method it reduces further as the redundancies have been removed.

\begin{figure}
    \centering
    \includegraphics[width=0.65\textwidth, height=0.45\textwidth]{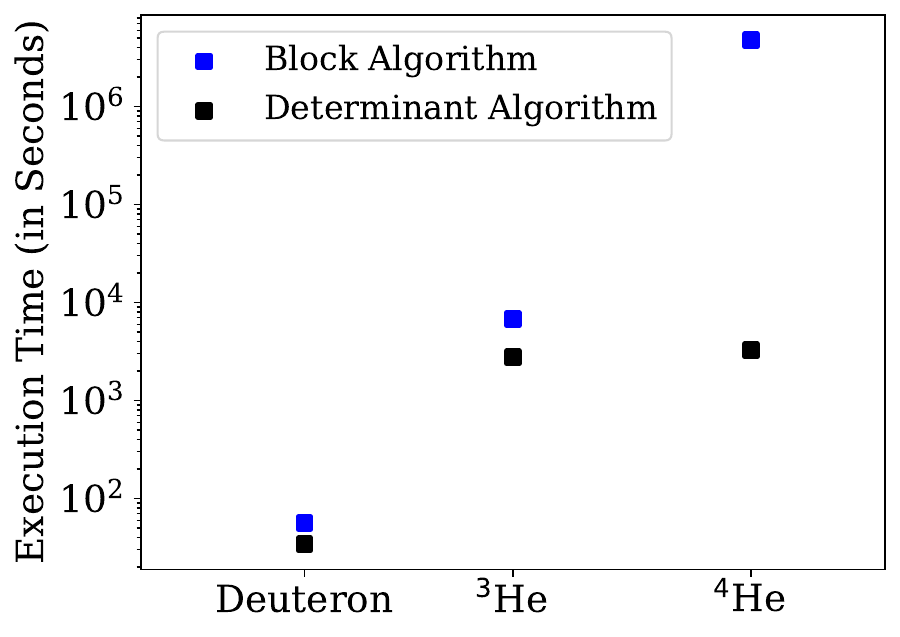}
    \caption{Execution times for calculating point-sink two-point correlation
      functions, at a given time-slice, for a few low-lying nuclei (lattice size      $24^3\times 64$). These calculations are performed on a single core of 12-th Gen Intel(R) Core(TM) i7 processor.}
    \label{fig_block_vs_det}
\end{figure}

We now discuss further the comparison of the performance of determinant and block
algorithms. It should be noted that for the block algorithm it is preferable to use a
type of sink where individual nucleons are projected to zero momentum,
whereas the determinant approach is more suitable when sink points are
correlated, {\it i.e}, for point or smeared sinks.  To compare those one thus needs
to use the same source-sink operator set up. This could be better implemented by
choosing either a point, or Gaussian (or any extended smeared sink)
operator. Here, for a comparison we use a point-source and point-sink set up for
$^2$H, $^3$He and $^4$He. 
However, in this case we need to sum over the spatial indices
  while calculating individual blocks, and hence the computational time to calculate the correlation functions by contracting the block tensors will scale with the lattice volume.
In this case, for
$^2$H, the zero-momentum projected two-point correlation function is given
by,
\begin{eqnarray}
C_{NP}(t)&=&\sum_{\mathbf{x}}\sum_{\boldsymbol{\chi^{\prime}},\alpha_1,\beta_2}\sum_{\sigma}sgn(\sigma) B^{p}_{\alpha_1}(\mathbf{x},t; \chi_{\sigma(1)}^{\prime}, \chi_{\sigma(2)}^{\prime}, \chi_{\sigma(4)}^{\prime})B^{n}_{\beta_2}(\mathbf{x},t; \chi_{\sigma(5)}^{\prime}, \chi_{\sigma(6)}^{\prime}, \chi_{\sigma(3)}^{\prime})\nonumber\\
    & &\times\epsilon_{a_{1}^{\prime}a_{2}^{\prime}b_{1}^{\prime}}\epsilon_{b_{2}^{\prime}b_{3}^{\prime}a_{3}^{\prime}}(C\gamma_5)_{\alpha_2^{\prime}\beta_1^{\prime}}(C\gamma_5)_{\beta_3^{\prime}\alpha_3^{\prime}}(C\gamma_5P_{-})_{\beta_2^{\prime}\alpha_1^{\prime}}(C\gamma_5P_{-})_{\beta_2\alpha_1} , \label{Eq3.1}
\end{eqnarray}
where $\chi^{\prime}$'s are combined spin-color indices of the quarks i.e. $\chi_i^{\prime}=\{a_i^{\prime},\alpha_i^{\prime}\}$  and $\chi_{i+3}^{\prime}=\{b_i^{\prime},\beta_i^{\prime}\}$ for $i=1,2,3$. For $^3$He and $^4$He we use a similar construction with the operator given in Eqs.~(\ref{He3_op}, \ref{He4_op}). In Fig.~(\ref{fig_block_vs_det}) we present the total execution times needed to calculate these two-point correlation functions, with the same output (equal up to the machine precision), using such block and determinant algorithms. As can be seen from the figure, determinant algorithm is about two times faster compared to block algorithm for $^2$H and $^3$He with this set up. For $^4$He, this advantage is much more, almost three order of magnitude faster.
For higher nuclei, where the number of quarks in the nuclei $(n_u,n_d)$ increases, the polynomial scaling $\mathcal{O}(n_u^3+n_d^3)$ of the determinant algorithm sets in. Hence, we expect determinant algorithm to be even more efficient while dealing with higher nuclei compared to block algorithm. 
Though we have made this comparison using a point-source and point-sink set up, the advantage of our proposed algorithm over the block algorithm
will hold for any smeared source and sink.

\subsection{GPU implementation of the proposed determinant algorithm}
In addition to the conventional data processing pipeline of CPUs, the idea of executing general-purpose parallel calculations on GPUs has rapidly become more popular within the lattice QCD community in recent years.  CUDA-C has emerged as a very effective programming languages by offering features for synchronizing threads, managing memory hierarchy, and facilitating efficient data transfers between the host (CPU) and the device (GPU). Since the determinants in our proposed algorithms are very small in size ($n_f \times n_f$, $n_f$ being the number of quarks with flavor $f$) and they do not depend on each other (thus ensuring parallelization), GPU frameworks are quite suitable in providing acceleration for this case. We detail the implementation procedure in the Appendix~\ref{GPU}. 
To see the efficacy of GPU implementation we calculate the two-point correlation functions for point-source point-sink operators for $^2$H and $^3$He. Both V100 and A100 GPUs are utilized with single and multiple nodes. In the comparison table below (Tables~\ref{tab_results1} and \ref{tab_results2}) we provide results of the optimized determinant algorithm using this programming model.

\section{Use of TensorFlow for calculating correlation functions in lattice QCD}
\label{sec4}
As mentioned previously, a major problem in the computation of lattice QCD
correlation functions for nuclei is the contraction of a large number of
color-spin indices involving Levi-Civita tensors and gamma matrices.  Thus, a
variety of methods have been used to attack the contraction problem.  These
strategies include, inter alia, use of tools such as \opteinsum{} for optimizing
the order of contractions, and also identifying common ``sub-expressions'' so as
to reduce redundant computations: see e.g. the work of Green et
al.~\cite{green_weakly_2021}, of H\"orz et
al.~\cite{horz_two-_2019,horz_two-nucleon_2021} which combines these methods and
also the work of Humphrey et
al.~\cite{https://doi.org/10.48550/arxiv.2201.04269} which explores the method
of identifying common sub-expressions.  More recent work of Amarasinghe et
al.~\cite{amarasinghe_variational_2023} develops a customized contraction
optimizer.  All these works, however, appear to use only CPUs for the final
computations, and so could not take advantages of hardware accelerators.

In this work, we propose a different approach, based on the observation
that, due to the ubiquity of tensor contractions in machine learning
computations, new programming models (such as \tensorflow{}~\cite{tensorflow2015-whitepaper,tensorflow_developers_2023_8118033})
have been developed to both simplify and speed-up tensor based computations.  In
this work, we present an efficient utilization of \tensorflow{} for calculating
two-point correlation functions in lattice QCD, which can easily be extended to
any other correlation functions.  This approach has two distinct advantages.

First, the tensor contractions in a correlation function can be implemented very
simply in a python code utilizing \tensorflow{} and \opteinsum{} packages. We
show that using these tools, the ``code'' for the relevant QCD correlation
functions simplifies significantly. In fact, it simplifies to such an extent
that one can easily write {programs} to \emph{generate the final code itself},
with a given input for an interpolating operator of a nucleus.  
  The idea of
automatically generating C++ code (using the QDP++ framework for lattice
QCD~\cite{edwards_chroma_2005}) for Wick contractions was also considered in Ref.~\cite{djukanovic_quark_2020}, in which a special purpose
\emph{Mathematica} package was developed for this purpose.  As noted in
Ref.~\cite{djukanovic_quark_2020}, automatic generation is expected to be a
simpler and less error prone method of programming Wick contraction
computations.  Our proposal instead is to generate code targeting off-the-shelf
tensor computation frameworks such as \tensorflow{} (rather than a general
purpose programming language such as C++).  A preliminary and obvious advantage
of this approach is that the code generation itself is likely to be
significantly easier (because of the easier syntax provided by these packages),
and thereby less error-prone as well.

However, targeting tools such as \tensorflow{} also provides the advantage that
these tools can \emph{automatically} optimize this \emph{automatically}
generated high-level code in two different ways.  First, they can automatically
optimize the order of tensor contractions using tools such as \opteinsum{}, in a
manner similar to the earlier work cited above.  However, in addition they can
also automatically take advantage of available hardware accelerators such as
GPUs. The performance of these tools is a result of significant technical effort
devoted towards them in the past decade, driven by the fact that tensor
contractions are ubiquitous in representing a wide class of machine learning
models.  
  We expect that the combination of a more natural syntax and easy access
to software and hardware optimizations that is available in the approach
suggested here would be especially advantageous in situations where one needs to
try out and evaluate multiple calculation strategies.  An example of this would
be the setting of carrying out comprehensive calculations involving large
operator sets in a GEVP (Generalized Eigenvalue Problem)
framework~\cite{MICHAEL198558,Luscher:1990ck}. While such calculations can be
performed using other approaches discussed above, writing the code (even using a
C++ code generator like the one in Ref.~\cite{djukanovic_quark_2020}) and then
optimizing it for all the different strategies to be evaluated would be a much
more tedious and error-prone undertaking than in the approach suggested here.

Below we provide a brief
procedure on the use of \tensorflow{} for lattice QCD correlation functions
while details are given in Appendix \ref{tf}.

In a hadronic correlation function within lattice QCD framework  a typical term involves multiplication of $(C\gamma_5)$, multiple Levi-Civita tensors $\epsilon_{abc}\epsilon_{def}...$, and multiple quark propagators ($(S_i)^{ab}_{\alpha\beta},\, i$ being the flavor) with their respective color ($a,b$) and spin ($\alpha,\beta$) indices. For example, for a single nucleon there are terms of the following form:
\begin{equation}
N = P^{pt}(C\gamma_5)^{qr}(C\gamma_5)^{hs}\epsilon_{abc}\epsilon_{def}S_d(\mathbf{x},t;\mathbf{0},0)^{ec}_{sr}S_u(\mathbf{x},t;\mathbf{0},0)^{fa}_{hp}S_u(\mathbf{x},t;\mathbf{0},0)^{db}_{tq}. \label{nucl_term}
\end{equation}
With  \tensorflow{} this 
tensor contraction can simply be
represented by
\begin{widetext}
      \begin{align}
    \texttt{N} &= \texttt{tensorflow.einsum("abc, def, qr , hs , pt, xescr ,
                 xfhap , xdtbq ->",}\nonumber\\
               &\qquad\texttt{eps , eps, C , C , P, S , S ,
                 S)} \label{tensor_N_ex}
  \end{align}
\end{widetext}
which is very compact and almost identical to the mathematical expression. In the above expression, `$eps$' is the Levi-Civita tensor, `$C$' is the $(C\gamma_5)$ matrix and `$P$' is the usual parity projection operator $P$ as in Eq.~(\ref{nucl_term}). `$S$' is the quark propagator.  Using such contractions we have implemented the block algorithm  to calculate the two-point correlation functions for $^2$H, $^3$He and $^4$He nuclei. Here we will briefly go over the procedure for $^2$H for which there are $\frac{n_u! \times n_d!}{2^A}$ = 9 independent terms in its two-point correlation function.
For a higher nucleus, the difference from the case of single nucleon only appears in the contraction part. We perform the contractions in two steps as described in the block algorithm. Similar to Eq.~(\ref{tensor_N_ex}), we first calculate the block structure (as in Eqs.~(\ref{eq2.4} and \ref{eq2.5})) in terms of the propagator tensors ($S$). Following this step, we compute the correlation function by evaluating the 9 independent terms  using the block tensors, Levi-Civita tensors and $\gamma$ matrices. We provide the snippet of the one nucleon code in the Appendix \ref{AppendixC}. Similarly, writing codes for other nuclei are also straightforward.\footnote{Due to substantially longer lengths we are not including those codes here but can be provided on request.}
In the next section we provide the results on the wall clock times for evaluating two-point functions for $^2$H, $^3$He and $^4$He nuclei using \tensorflow{} in GPU environment and compare those with corresponding serial C codes and CUDA codes.

For higher nuclei, number of terms (as in Eq.~(\ref{nucl_term})) in their correlation functions increases very rapidly and writing all these different contractions manually becomes both time-consuming and error-prone. However, \tensorflow{} being syntactic, it is straightforward to automate the ``code" generation for computing these correlation functions. The details of this procedure are provided in Appendix \ref{tf}.

We emphasize that our proposal of using such tools is not limited to the correlation function
 computations considered in this paper. In fact, any calculations on LQCD correlation functions involving tensors (for example, two-, three-, and four-point correlation functions that are commonly used for extracting observables in LQCD) can 
 benefit significantly by using tools like \tensorflow{}, and taking
 advantage of their decade long developments and optimizations in AI sector. We encourage LQCD practitioners to adapt such advantages.

\subsection{Three-point functions using \tensorflow{}}
Here we show that \tensorflow{} can also be employed efficiently to calculate three-point functions in lattice QCD. We take the example of a single nucleon and its electromagnetic form factors, which can be generalized to higher nuclei. For a proton we choose the interpolating field as given by Eq.~(\ref{p_int}) and the local vector current 
\begin{equation}
 J_{\mu} (x) = \frac{2}{3} \bar{u}(x)\gamma_\mu u(x) - \frac{1}{3} \bar{d}(x)\gamma_\mu d(x) - \frac{1}{3} \bar{s}(x)\gamma_\mu s(x) , \quad \mu = 1,  2, 3.
\end{equation}
Inserting this current at the space-time point ($\mathbf{x}_{in},t_{in}$), between the source  ($\mathbf{0},0$) and the sink point ($\mathbf{x}_f,t_f$), one obtains the expression for the corresponding three-point function where a typical term for the $u$-quark contribution to the connected diagram would be of the form:
\begin{eqnarray}
    N^{u} = P^{\alpha\alpha^{\prime}}(C\gamma_5)^{\beta\gamma}(C\gamma_5)^{\rho\sigma} \, \epsilon_{abc}\epsilon_{def} \, S_u(\mathbf{x}_f,t_f;\mathbf{0},0)^{af}_{\alpha\rho} \, 
    S_d(\mathbf{x}_f,t_f;\mathbf{0},0)^{ce}_{\gamma\sigma}\nonumber\\
    \Sigma_{u,\mu}(\mathbf{x}_f,t_f;t_{in};\mathbf{0},0)^{bd}_{\beta\alpha^{\prime}}
    \label{Proton3pt_seq} . 
\end{eqnarray}
Here, $\Sigma_{u,\mu}(\mathbf{x_f},t_f;t_{in};\mathbf{0},0)^{bd}_{\beta\alpha^{\prime}}$ %
is a sequential propagator, computed by inverting the Dirac operator with the source $\gamma_{\mu}S_{u}(\mathbf{x},t;\mathbf{0},0)\vert_{t=t_{in}}$.
With  \tensorflow{}, the above 
tensor contraction can simply be
represented by,
\begin{widetext}
      \begin{align}
    \texttt{Nu} &= \texttt{tensorflow.einsum("ab, cd, ef, ijk, lmn, xtiane,
                 xtkdmf, xtjclb  - > t ",}\nonumber\\
  &\qquad\texttt{P, C, C, eps, eps, Su, Sd, Sigma\_u\_mu)}.
  \end{align}\label{tensor_N_3pt_ex}
\end{widetext}
Similar expressions can also be easily constructed for the contributions from other quarks and also for disconnected diagrams.
Use of any other type of  propagators, as needed by a current or a source-current-sink set up,  will simply change the entries corresponding the quark propagators and the above expression can be changed accordingly. For a nucleus with higher number of protons and neutrons formulation of the above contraction terms would be straightforward and the method can also be automated.

One can also compute four-point functions in a similar way, and for higher nuclei use of determinant method will be computationally cheaper. This brings us to point out that the computation of form-factors and moments could well be possible for low-lying nuclei using our proposed methods. While such calculations have produced significant results \cite{Chang:2018uxx} for a single nucleon, those are still in early days for nuclei \cite{Parreno:2021ovq, Detmold:2020snb}. Needs for such calculations are paramount in search for new physics beyond the standard model \cite{Gonzalez-Alonso:2018omy,Engel:2016xgb,Davoudi:2022bnl}, and our proposed methods can be employed in that. 

\section{Application of the proposed algorithms to the generalized eigenvalue problem (GEVP)}

\noindent The framework of generalized eigenvalue problem \cite{MICHAEL198558, Luscher:1990ck} plays a crucial role in contemporary studies of hadron spectra using lattice QCD. It helps to extract multiple energy levels reliably, even close-by ones, from two-point correlation functions which otherwise are difficult to obtain particularly for the excited states. However, an essential part of GEVP is the inclusion of a large set of operators, preferably with the symmetric source-sink set up. For nuclei, particularly beyond A = 2, that becomes a non-trivial problem as the construction of correlation matrices with large set of operators having all possible color-spin indices requires humongous number of Wick-contraction and their coding. It becomes worse when one needs to consider the projection of individual nucleons or clusters  to zero momentum to mitigate SNR problem.

The methodology presented in this work can significantly reduce the computing time in constructing the correlation matrix.  By utilizing our randomized algorithm based common sub-expression elimination technique, each entry in the two-point correlation matrix can be computed at least an order of magnitude faster compared to conventional methods. In addition, since the quark propagators are same in each element of the correlation matrix, it is possible to eliminate common sub-expressions across different elements of the correlation matrix, resulting in additional performance improvements.

We outline this generalization by considering $N$ operators used to calculate an $N\times N$ two-point correlation matrix. Each element of this matrix is expressed as:
\begin{eqnarray}
    C_{ij}(t) = \langle \Omega \vert \mathcal{O}_i(t)\overline{\mathcal{O}}_j(0)\vert \Omega \rangle, \quad \text{for } 1\leq i,j\leq N.\label{E1}
\end{eqnarray}

Here, $\mathcal{O}_i(t)$ represents the interpolating operators. Using the same notation as in Eq.~(\ref{eq2.7}), they can be written as:
\begin{eqnarray}
    \mathcal{O}_{i}(t) = \sum_{\boldsymbol{\chi}}\sum_{\boldsymbol{x_i}} C_i^{\chi_1\chi_2\cdots \chi_n} q(\mathbf{x_1},t,\chi_1) q(\mathbf{x_2},t,\chi_2) \cdots q(\mathbf{x_n},t,\chi_n).
\end{eqnarray}

The $(i,j)$-th entry of the correlation matrix  with with the interpolating fields $\mathcal{O}_i$ and $\mathcal{O}_j$ at the sink and source respectively, can then be expressed as:
\begin{eqnarray}
    C_{ij}(t) = \eta \sum_{\mathbf{x_i}} \sum_{\mathbf{x_i'}} \sum_{\boldsymbol{\chi}, \boldsymbol{\chi'}} C_j^{\prime \chi_1' \chi_2' \cdots \chi_n'} C_i^{\chi_1 \chi_2 \cdots \chi_n} \det(M^u(\boldsymbol{\chi}^u, \boldsymbol{x}^u; \boldsymbol{\chi}'^u, \boldsymbol{x}'^u)) \nonumber \\
    \times \det(M^d(\boldsymbol{\chi}^d, \boldsymbol{x}^d; \boldsymbol{\chi}'^d, \boldsymbol{x}'^d)).\label{E3}
\end{eqnarray}

As indicated in Eq.~(\ref{E3}), all the $C_{ij}$ share the same determinant structure. Multiple $C_{ij}$ elements therefore also share common sub-expressions. By applying the randomized algorithm outlined in Sec.~\ref{impl}, a set of non-redundant sub-expressions across all matrix elements can be identified and computed only once. This results in additional performance gains over reducing redundancies for each individual element separately, and hence can really accelerate the computation of the correlation matrix. This method can be adopted for most source-sink combinations and for most types of interpolating fields.

The tensor-based programming models proposed in this work can also be employed to generate more efficient and error-resistant implementations of correlation matrix computation codes. Due to constraints on available GPU memory, at this time this can be implemented for the nuclei with $A=2, 3, 4$. For these nuclei with large set of interpolating fields the tensor-based models can be advantageous due to automatic code generation, making it error-resistant, and then taking advantage of automatic-optimization. For higher nuclei, the method of randomized preprocessing to detect determinant redundancies would be preferable due to the large potential reduction in the number of terms, particularly with the possibility of being able to use only the important terms (the latter possibility is discussed in Sec.~\ref{sec:terms-distribution} below).

Our methods can thus provide a substantial gain in computing-time in any future GEVP-based calculations for nuclei, particularly when one needs to consider repetitive calculations at multiple volumes and lattice spacings, with a large set of operators.

\section{Results}\label{sec_v}
\begin{figure}[h!]
    \centering    \includegraphics[width=0.33\textwidth]{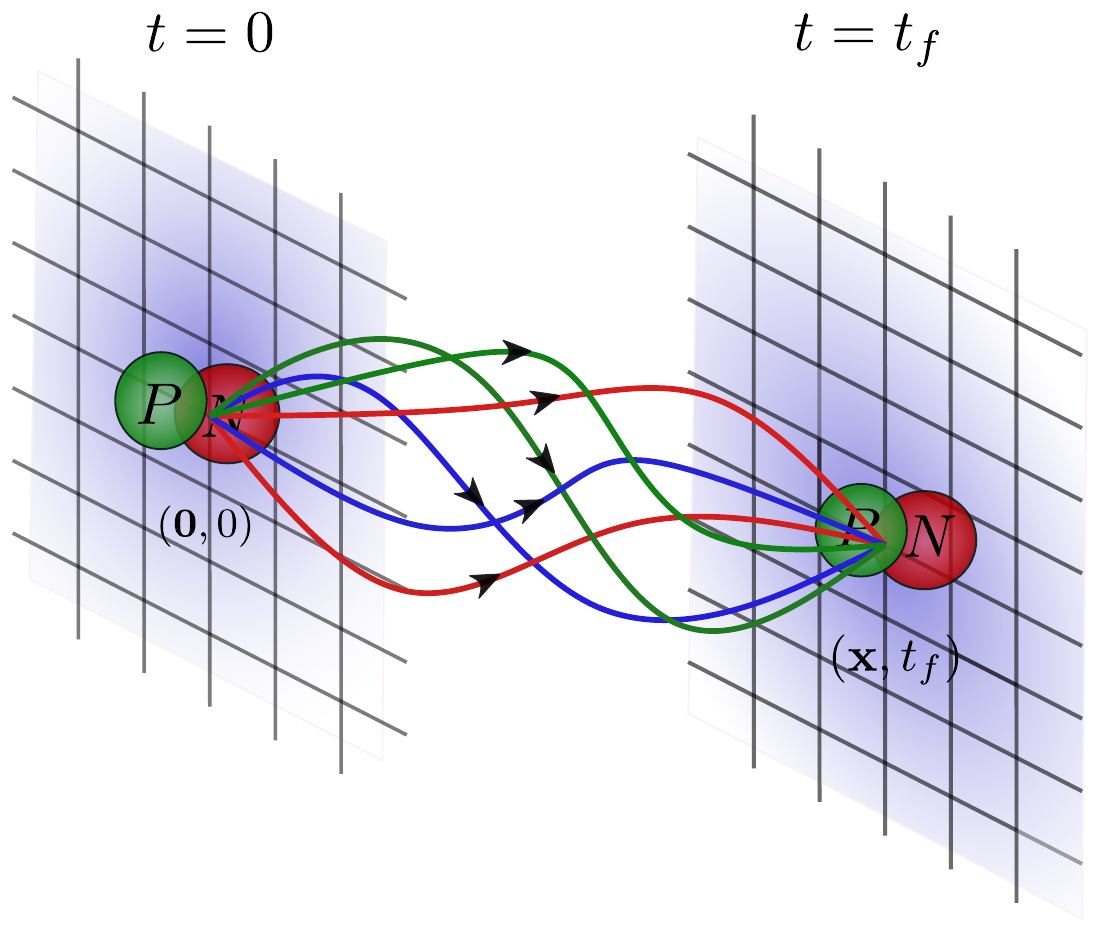} 
 \includegraphics[width=0.33\textwidth]{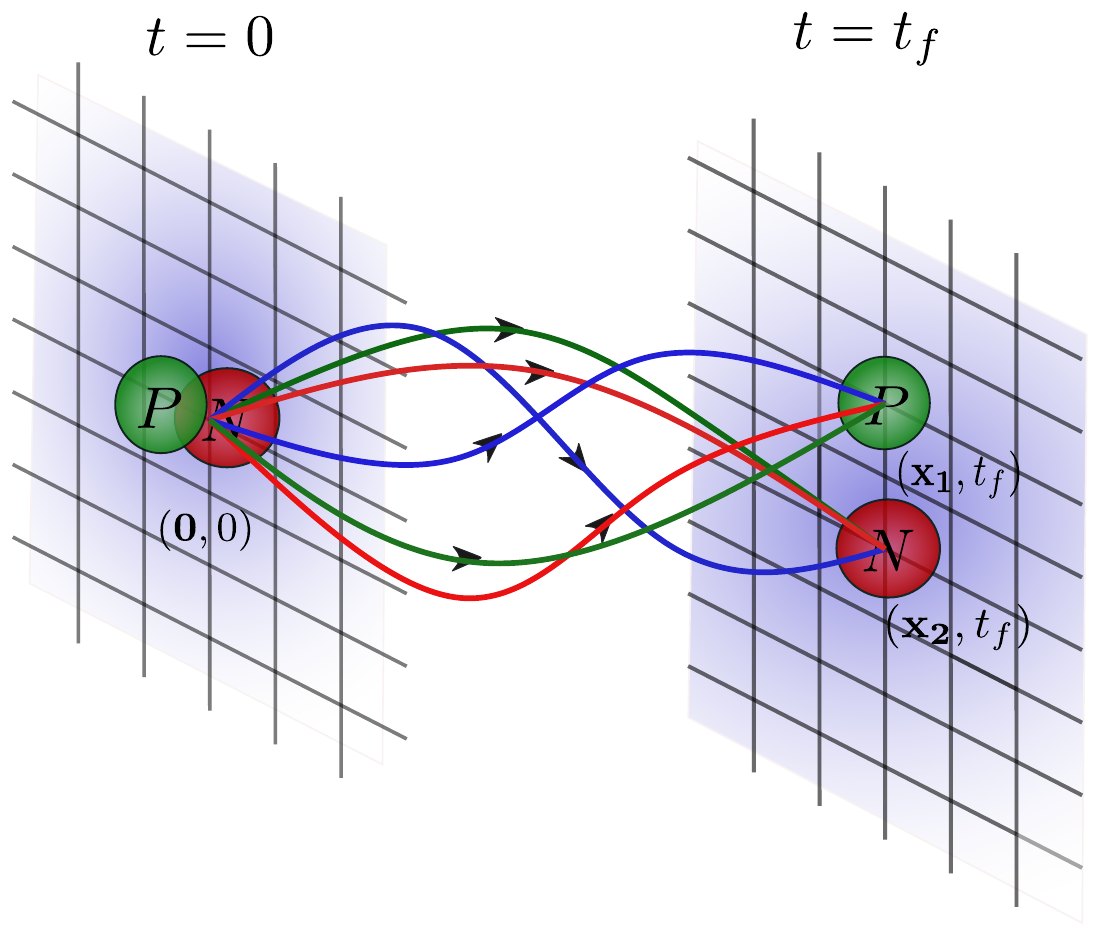} \\  
\includegraphics[width=0.33\textwidth]{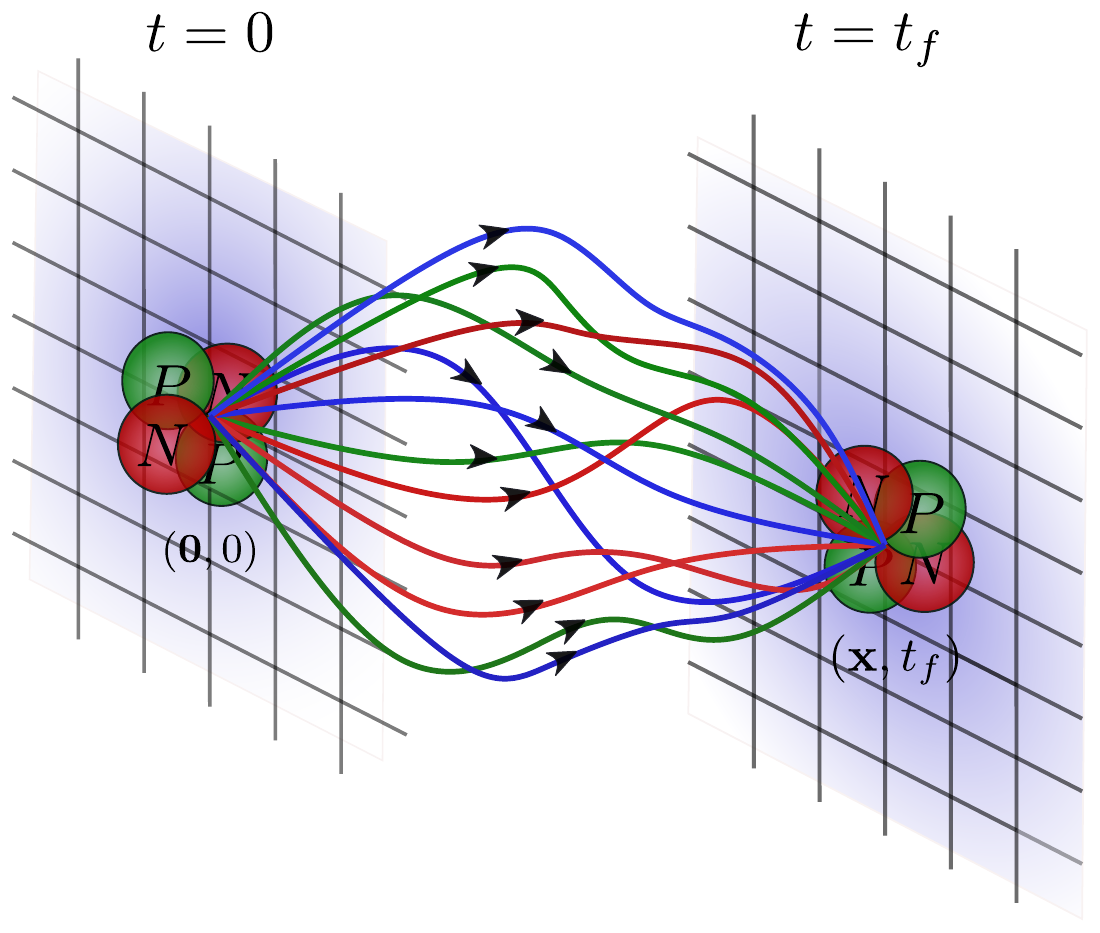} 
 \includegraphics[width=0.33\textwidth]{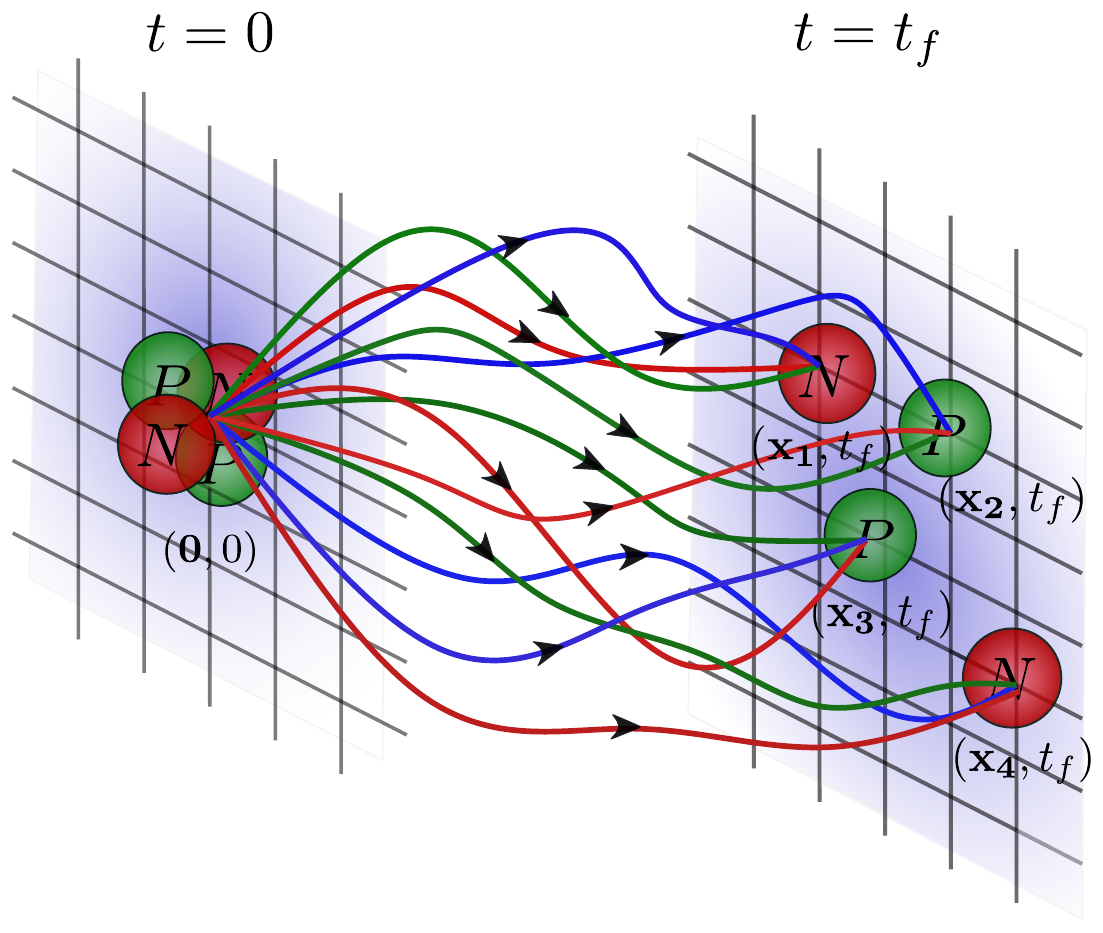}    
    \caption{Schematic sketches showing the representative Wick-contraction diagrams for the two-point correlation functions of $^2$H (top) and $^4$He (bottom) from a source at (${\bf{x_0}},0$) to a sink ({\bf{x}},$t_f$) on a space-time lattice. Left and right side figures show point-sink and extended-sink cases, respectively.}
    \label{corf_fig}
\end{figure}

After discussing our proposed algorithms and their implementation procedures, we provide here the results obtained for the two-point correlation functions for a few low-lying nuclei: $^2$H, $^3$He, $^4$He and  $^7$Li. 
In Fig.~\ref{corf_fig}, we show the representative diagrams for various types of two-point functions of some of these nuclei. The top two panes are for $^2$H: the left one where the quarks are projected to the zero-momentum at a point, while the right one with individual nucleons projected to zero-momentum separately. They are denoted by point and plane wave sink (sometimes also called as extended-sink) in Tables~\ref{tab_results1} and \ref{tab_results2}. The bottom two panes in 
Fig.~\ref{corf_fig} are for $^4$He with point and plane wave sink respectively.

\subsection{Efficiency of our proposed algorithms}
The purpose of this work is to find novel algorithms and suitable optimized programming environments 
which can enable us to compute nuclear correlation functions, including those of higher nuclei, with ease, that can eventually lead to achieving the goal of obtaining quantitative results in nuclear physics in the near future. 
For quoting the efficiency of our proposed algorithms, below we specify various hardware that we utilized to obtain our results: CPU: 
Intel(R) Xeon(R) Silver 4208 CPU\@ 2.10GHz. 
GPU V100: NVIDIA Tesla V100-PCIE-32GB
and GPU A100: NVIDIA A100-SXM4-40GB.
Calculations are performed on various lattice ensembles with sizes $24^3\times 64$, $32^3\times 96$, $48^3 \times 144$, $64^3 \times 192$ and  $96^3 \times 288$.  In Tables \ref{tab_results1} and \ref{tab_results2}, we provide our results. For a given quark propagator, the source point is the same (wall source) for all codes. At the sink we have used either a point sink, where all quarks are projected to zero momentum, or plane wave sinks where individual nucleons are separately projected to zero momentum  at any points on the lattice. For a point sink operator,  we have implemented the optimized determinant algorithm as described in the improved method (\ref{Imp_Method}) of subsection \ref{ss3c}. For a  plane-wave sink, we have used the block algorithm, as described in Eq.~(\ref{eq2.6}).
For higher nuclei, it is preferable to use a union of determinant and block algorithms. In that a cluster sink can be employed where each cluster, consisting of more than one nucleus, are projected to zero momentum separately (similar to plan wave but not with individual nucleon). The distance between the cluster points can be constrained and controlled to get the optimal overlap to a desired state. These procedures can also be helpful for carrying out the usual generalized eigenvalue problem \cite{MICHAEL198558, Luscher:1990ck} towards extracting excited states of nuclei as well as to find preferable cluster configuration for a given nucleus, which can further help in understanding its internal structures.
For $^7$Li we have used such a procedure by clustering Tritium and $^4$He together.

The notation used in the ``Program-Type'' column of Tables~\ref{tab_results1} and \ref{tab_results2} is the following: TF indicates that \tensorflow{} (version 2.4.1) is employed; SERIAL C: Serial C code on one CPU core;  
  OPENMP C : OpenMP based multi-threading parallelization of C code on $24$ CPU cores. CUDA V100: Cuda-based code with GPU type V100; CUDA A100: Cuda-based code with GPU type A100. 
The columns 4--8 represent the execution times in seconds of a code (corresponding row in the column, Program-type) for a given nucleus on different lattice volumes. Here, execution time refers to the time needed in seconds to compute a two-point correlation function at  one time-slice for a gauge configuration. On GPUs, we have also employed more than one nodes for a few cases to find the scaling with the number of GPUs. 

The main findings on the implementation of our algorithms are the following:
\begin{itemize}
\item Serial C code with our proposed method for point sink set up is more than an order of magnitude faster than the performance of similar codes with other algorithms (see Fig.~\ref{fig_block_vs_det} also). 
    \item Our algorithms scale with volume linearly for all nuclei that we study.
    \item 
      Usage of OpenMP also enables our algorithm to scale almost with the number of cores (about 17 times faster on 24 cores than that of one core).
     
    \item Porting the code to GPU with CUDA provides on-average 200 times gain over its serial version.
    \item Switching from V100 to A100 further yields a factor of two gain on-average.
    \item Multinode scaling of GPU code is almost proportional to the number of nodes. 
    \item Implementation of \tensorflow{}  is found to be even faster over A100 GPU performance for nuclei up to $^4$He.
    \item For higher nuclei, it is preferable to use a combination of block and determinant algorithms with clustering of smaller nuclei.
\end{itemize}

{\begin{small}   
\begin{table}[htbp]
	\centering
	\begin{tabular}{|c|c|c|c|c|c|c|c|} 
		\hline\hline 
		\multirow{2}{*}{Nucleus}&\multirow{2}{*}{Sink operator type}&\multirow{2}{*}{Program-type}&\multicolumn{5}{|c|}{Lattice Volume}\\\cline{4-8}
            &&&$24^3$ & $32^3$ & $48^3$ & $64^3$ & $96^3$\\\hline\hline
		\multirow{16}{*}{\ch{^2H}}&\multirow{12}{*}{Point} &\pyth|SERIAL C| &
                                                                              69.45 &  165 & 648 & 1360  & 4265\\
 \cline{3-8} &&\pyth|OPENMP C| ($24$ cores) & 4.2 & 10.0  
            &34.5 &76.1 &240.2  \\
	\cline{3-8}
		&&{\pyth|CUDA V100|} & 0.34   &  0.81   &  2.78   &  6.37   & 20.93  \\	\cline{3-8}
		&&{\pyth|CUDA V100 2-NODE|} & 0.18   &  0.41   &  1.39   &  3.19   & 10.71  \\	\cline{3-8}	
		&&{\pyth|CUDA A100|} & 0.15 &  0.35 &  1.19 &   2.80 &  11.98\\	\cline{3-8}	
		&&{\pyth|CUDA A100 2-NODE|} & 0.10 &  0.24 &  0.76 &  1.83 & 6.16\\	\cline{3-8}			
		&&{\pyth|CUDA A100 4-NODE|} & 0.06 &  0.13 &  0.40 &  0.94 & 3.16\\	\cline{3-8}			
		&& {\pyth|RATIO: V100/A100|} & 2.3 & 2.3 & 2.3 & 2.27 & 1.9 \\   \cline{3-8} 
		&&{\pyth|TF V100 |} & 0.26   & 0.66   &  2.29   & 5.47  & 18.23\\		\cline{3-8}
		&&{\pyth|TF A100|} &   0.25 &  0.33    & 1.0 & 2.34 & 7.53 \\	\cline{3-8}	
		&& {\pyth|RATIO : V100/A100|} & 1.04 & 2.00 & 2.29 & 2.33 & 2.42\\\cline{3-8}
		&&{\pyth|RATIO V100: C/CUDA|} & 204 & 204 &  233 &  214 & 203 \\	\cline{3-8}
            &&{\pyth|RATIO V100: OPENMP C/CUDA|} & 12.4 & 12.3 &  12.4 &  11.9 & 11.5 \\	\cline{3-8}
            &&\pyth|RATIO V100: CUDA/TF | & 1.31 & 1.23 &  1.21 & 1.16& 1.15\\ \cline{2-8}
        & \multirow{4}{*}{Plane-Wave} & \pyth|SERIAL C| & 53 & 108 & 335 & 783 & 2564 \\ \cline{3-8}
        && \pyth|OPENMP C| ($24$ cores) &2.8 &6.2 &18.7 &44.1 &143.5  \\  \cline{3-8}
        && {\pyth|TF V100|} & 0.15 & 0.25 &  0.61  & 1.27& 4.41  \\  \cline{3-8}
        && {\pyth|TF A100|} &   0.12 &  0.17 & 0.63   & 1.25  & 4.38 \\  \cline{3-8}
        && {\pyth|RATIO: V100/A100|} & 1.25& 1.47 & 0.96 & 1.02 & 1.01 \\\hline \hline
        \end{tabular}
	\begin{center}
		\caption{The wall clock timing of our proposed determinant and \tensorflow{} algorithms for $^2$H. Here, wall clock time refers to the time needed in seconds to compute a two-point correlation function at  one time-slice on a gauge configuration.
  The second column is for the sink-type, and the third column represents the code-types (see text above for explanation on notations). The fourth column represents the execution times  of a code in seconds (corresponding row in the third column).\label{tab_results1}
  }
\end{center}
\end{table}
\end{small}}

\begin{table}[htbp]
	\centering
	\begin{tabular}{|c|c|c|c|c|c|c|c|} 
		\hline\hline 
		\multirow{2}{*}{Nucleus}&\multirow{2}{*}{Sink operator-type}&\multirow{2}{*}{Program-type}&\multicolumn{5}{|c|}{Lattice Volume}\\\cline{4-8}
            &&&$24^3$ & $32^3$ & $48^3$ & $64^3$ & $96^3$\\\hline\hline         
		\multirow{12}{*}{\ch{^3He}}&\multirow{8}{*}{Point} &\pyth|SERIAL C| & 1196  & 2456  &8183  & 19441  &  69461 \\ \cline{3-8}
            && \pyth|OPENMP C| ($24$ cores) & 67.8  & 137.9   & 447.2   & 1062.6   & 3826.4 \\	\cline{3-8}
		&&{\pyth|CUDA V100|} & 5.24  & 12.32   & 41.50   & 98.25   & 332.75 \\	\cline{3-8}	
		&&{\pyth|CUDA V100 2-NODE|} & 2.62   & 6.20   &  21.25   & 55.37   & 196  \\		\cline{3-8}
		&&{\pyth|TF V100 |} & 8.02   & 22  &  58.8   & 141.7  & 448.72 \\	\cline{3-8}		
		&&{\pyth|TF A100 |} &   4.05  &  8.02    & 26.14 & 61.38   & 209.52 \\  \cline{3-8}			
		&& {\pyth|RATIO : V100/A100|} & 1.98 & 2.74 & 2.25 & 2.31 & 2.14 \\\cline{3-8}
		&&\pyth|RATIO V100: C/CUDA| & 228  & 199 & 200  & 198 & 208 \\\cline{3-8}
            &&\pyth|RATIO V100: OPENMP C/CUDA| & 12.9  & 11.2 & 10.8  & 10.8 & 11.5 \\\cline{3-8}
		&&\pyth|RATIO V100: CUDA/TF | & 0.65 & 0.56 &  0.71 & 0.70 & 0.74 \\ \cline{2-8}
           & \multirow{4}{*}{Plane-wave} & \pyth |SERIAL C |& 58 & 115 & 345 & 791 & 2573 \\ \cline{3-8}
           &&\pyth|OPENMP C| ($24$ cores) & 7.1   &  11.9   &  24.3   & 50.6  &148.2   \\	\cline{3-8}
           &&{\pyth|TF V100 |} & 0.23   & 0.39   &  0.89   & 2.23  & 6.39   \\	\cline{3-8}
           &&{\pyth|TF A100 |} &   0.11   &  0.15 & 0.65   & 1.23  & 3.05 \\	\cline{3-8}
          && {\pyth|RATIO : V100/A100|} & 2.09 & 2.60 & 1.36 & 1.81 & 2.09\\\hline\hline
		\multirow{6}{*}{\ch{^4He}}& \multirow{3}{*}{Point} &\pyth|SERIAL C | & 3246  &  
            7541  & 25416  &  60091  & -  \\ \cline{3-8}
            &&\pyth|OPENMP C| ($24$ cores) & 185   & 424   &  1446   & 3395  & -   \\	\cline{3-8}
  		&&{\pyth|CUDA V100|} & 24.72   &  59.05   &  198.08   &  472.83   & -  \\	\cline{3-8}
           &&\pyth|RATIO V100: C/CUDA| & 131 & 128 & 128 & 127 & - \\ \cline{3-8}
           &&\pyth|RATIO V100: OPENMP C/CUDA| & 7.48 &  7.18 & 7.30 & 7.18 & - \\ \cline{2-8} 
           &\multirow{3}{*}{Plane-wave}&\pyth|SERIAL C | & 235  & 301  & 532  & 1021  & 3017    \\ \cline{3-8}
            &&{\pyth|TF V100 |} & 2.29 & 1.84 & 1.87 & 2.41 &  7.31  \\\cline{3-8}
            &&{\pyth|RATIO V100: C/TF |} & 102 & 164 & 284 & 423 & 412 \\\hline\hline
{\ch{^7Li}}&$^4$He-Tritium cluster&\pyth|SERIAL C | & 7586 & 17876  & 60471  & 143196  &-  \\ \cline{3-8}
            &&\pyth|OPENMP C| ($24$ cores) & 432   & 1012.7   & 3379    & 8190  & -   \\ \hline \hline 
	\end{tabular}
	\begin{center}
		\caption{The wall clock times of our proposed determinant and \tensorflow{} algorithms for $^3$He, $^4$He and $^7$Li. The explanation for the table content is same as given in Table \ref{tab_results1}. \label{tab_results2}}
	\end{center}
\end{table}

\subsection{Nuclear correlation functions and their effective masses}
\begin{figure}[t!]
    \centering  
       \hspace*{-0.25in} 
\includegraphics[width=0.5\textwidth,height= 0.38\textwidth]{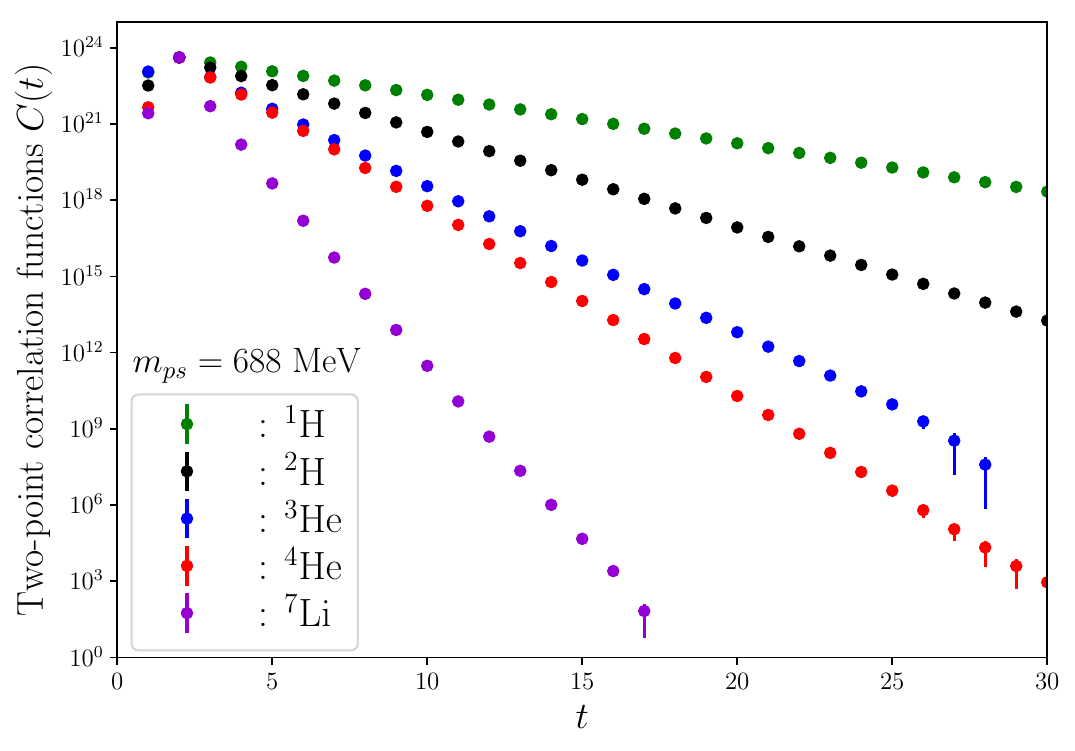}   
\includegraphics[width=0.515\textwidth,height= 0.38\textwidth]{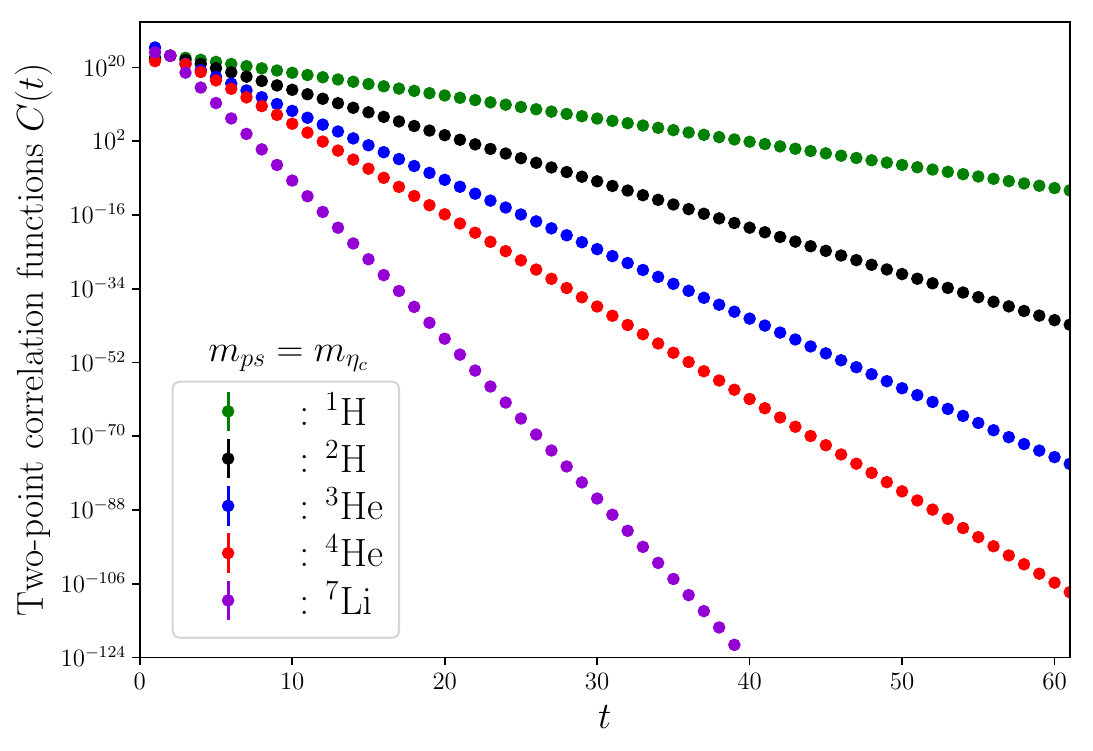} 
    \caption{The two-point correlation functions of $^1$H, $^2$H, $^3$He, $^4$He and $^7$Li on a $48^3\times 144$ lattice ensemble (wall source, lattice spacing $a$ $\sim$ 0.0582 fm with a volume $\sim$ (2.8 fm)$^3$ $\times 8.4$ fm).
    The left pane is for the case with the quark masses set at the physical strange quark mass ($m_u = m_d = m_s$; pseudoscalar meson mass, $m_{ps} = 688$ MeV, $N_\mathrm{config} = 168$)  while the right pane corresponds to that at the physical charm quark mass ($m_u = m_d = m_c$; $m_{ps} = m_{\eta_c}$, $N_\mathrm{config} = 185$). The correlation functions are normalized at the first time slice after the source point.
}
    \label{corf}
\end{figure}

\begin{figure}[t!]
    \centering  
       \hspace*{-0.25in} 
\includegraphics[width=0.5\textwidth,height= 0.38\textwidth]{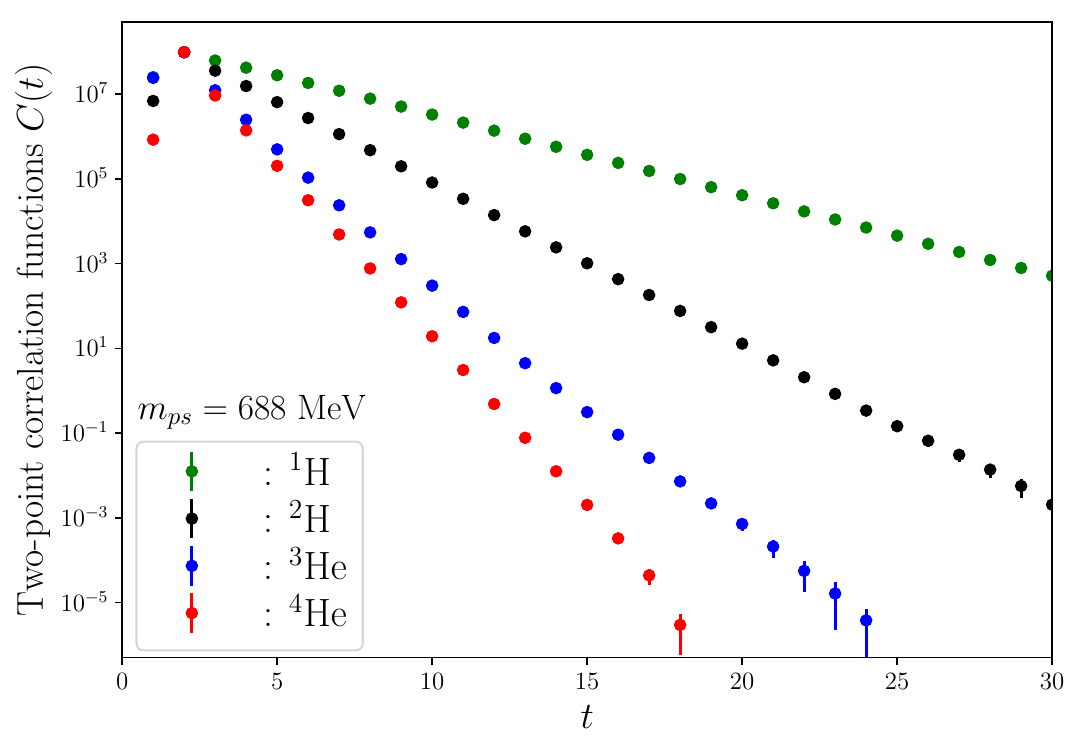}   
\includegraphics[width=0.515\textwidth,height= 0.38\textwidth]{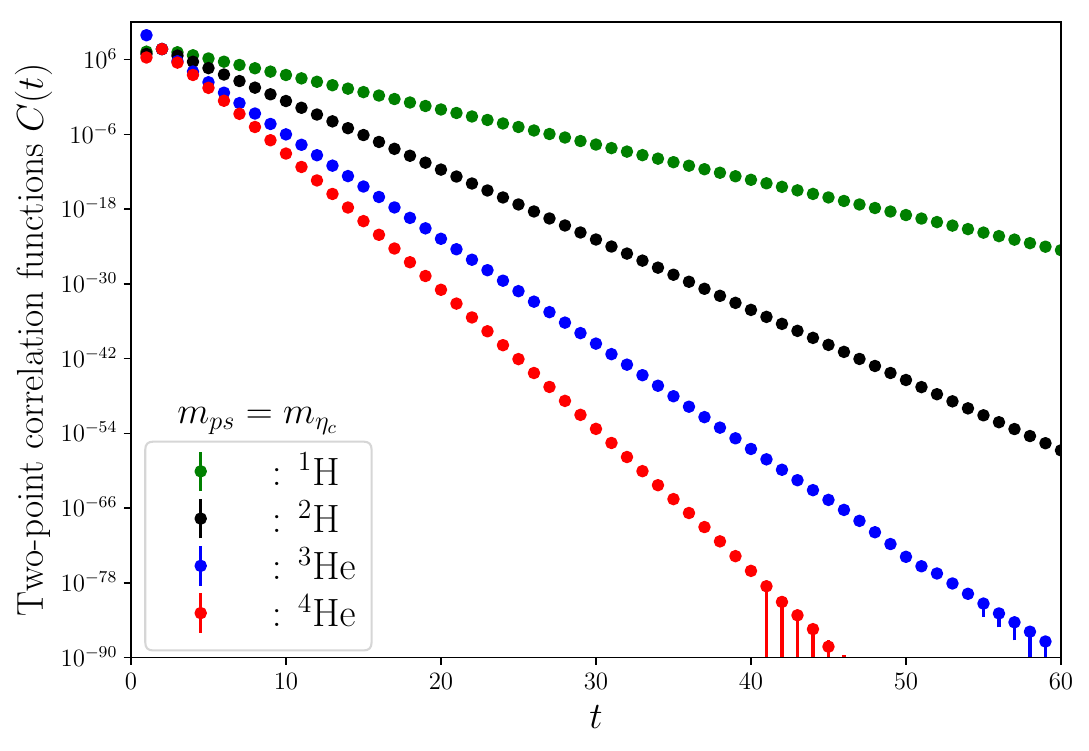} 
    \caption{Same as Fig.~\ref{corf}, but with the point sink.
}
    \label{corf_pt}
\end{figure}

In Figs.~\ref{corf} and \ref{corf_pt}, we present the two-point correlation functions 
of $^1$H, $^2$H, $^3$He, $^4$He and $^7$Li on a $48^3\times 144$ ensemble (lattice spacing $a$ $\sim$ 0.0582 fm with a total volume $\sim$ (2.8 fm)$^3$)).
The number of gauge configurations ($N_{\mathrm{config}}$) utilized are 168 at the strange quark mass and 185 at the charm quark mass. 
From hereon, all results showing correlators and effective masses are obtained 
using HISQ gauge ensembles \cite{MILC:2012znn}, and the valence quark propagators are generated with the overlap action \cite{Neuberger:1997fp, NEUBERGER1998141}.
To depict their relative fall-off with time, these correlation functions are normalized at the first time slice after source, i.e., $t = 2$. The interpolating fields for these nuclei are given  in subsection \ref{ss3c}. We have employed a wall-source to generate the valence quark propagators. For $^2$H, $^3$He and $^4$He, we have utilized Eq.~(\ref{eq2.6}) within the block algorithm with a sink where each individual nucleon is projected to zero momentum. However, we used \tensorflow{} to compute their correlation functions. We have also used Eq.~(\ref{eq2.9}) within the determinant algorithm with a point sink. For $^7$Li, we use two clusters with  $^4$He and Tritium and each of them are projected to the zero-momentum. Results for plane-wave and point sinks are shown in Fig.~\ref{corf} and \ref{corf_pt}, respectively.  The valence quark mass is chosen to be at the physical strange (left pane) and the charm quark (right pane) masses. As will be discussed later and also in the Appendix \ref{distribution},
the utilization of {\it most important} terms, extracted from the analysis of distribution of terms, can further reduce the computing time for these correlators significantly, often by more than an order of magnitude over the above-mentioned {\it improved} determinant/block method.
\begin{figure}[htbp]
    \centering    
    \hspace*{-0.25in}
    \includegraphics[width=0.51\textwidth,height= 0.4\textwidth]
{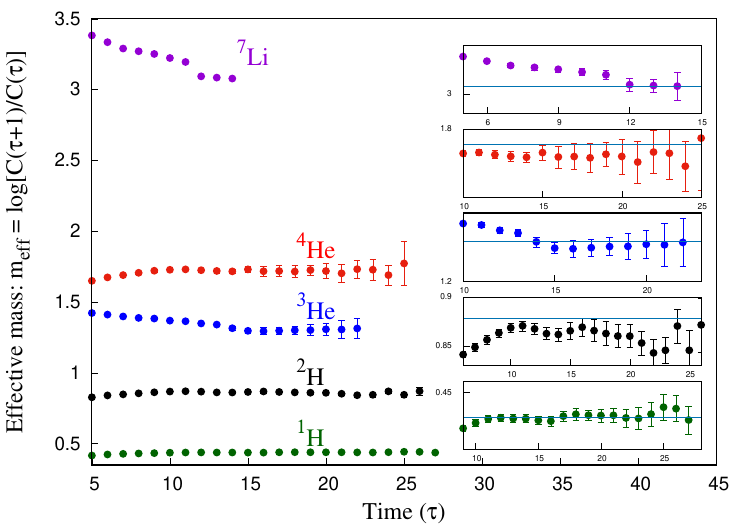}
\includegraphics[width=0.51\textwidth,height= 0.4\textwidth]{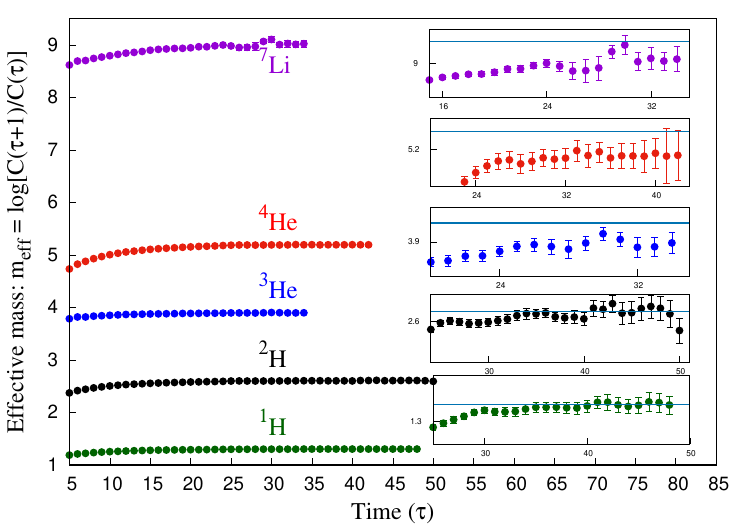}  
    \caption{Effective mass plots for
    $^1$H, $^2$H, $^3$He, $^4$He and $^7$Li (bottom to top) at the strange and charm quark masses (left:  $m_u = m_d = m_s$ and right: $m_u = m_d = m_c$), corresponding to the correlators shown in Fig.~\ref{corf}. The insets are the zoomed-in plots showing 
    possible signal saturation, while the horizontal lines are A $\times$ one nucleon mass, where A is the atomic number.}
    \label{eff_mass_c}
\end{figure}

In Fig.~\ref{eff_mass_c} we show the effective masses obtained from these correlation functions, defined as,
\begin{equation}
    m_{\mathrm{eff}}
    (\tau = t_{\mathrm{sink}} - t_{\mathrm{source}}) = {\mathrm{log}}\left[\frac{C(\tau)}{C(\tau + 1)} \right].
\end{equation}
These effective masses correspond to the correlation functions in Fig.~\ref{corf}. Here again the left pane is for the strange quark mass ($m_u = m_d = m_s$) while the right pane is for the charm quark mass ($m_u = m_d = m_c$). Inside each of these plots, we show a zoomed-in view of the possible signal saturation.  The horizontal lines are the threshold states with mass $\mathrm{A}\times m_{^1{\mathrm{H}}}$, $\mathrm{A}$ being the atomic number and $m_{^1{\mathrm{H}}}$ is the one-nucleon fitted mass. Consequently, these lines can be understood as the threshold, and they aid in guiding  eyes to determine the relative locations of the bound nuclei in relation to these respective thresholds.
The effective masses corresponding to the point-sink are noisier than that of the plane-wave sink, but are consistent to each other.

We fit these correlators with an exponential form ($\sim \exp(-E(L)t)$) to extract their finite volume ($L^3$) ground state energies $E(L)$. For the cases with better SNR, it is even possible to compute the finite volume energy difference,  $\Delta E_N (L) = E_{N}(L) - (nE_p(L) + mE_n(L))$, where $E_N (L)$ is the ground state energy of the nucleus made of $n$ protons and $m$ neutrons. In Table \ref{Table_dE}, we show $\Delta E(L)$ for these nuclei for the case of $m_u = m_d = m_c$, with  $N_\mathrm{conf} = 185$. With a finite volume amplitude analysis of $\Delta E_N (\vec{p})$,  nuclear scattering parameters can be extracted \cite{Luscher:1990ck}, and such studies can be performed in future with the availability of realistic lattices.
\begin{center}
\begin{table}[htbp]
     \centering
     \begin{tabular}{|c|c|}
       \hline
       ~ Nucleus ~ &  ~ $\Delta E(L)  $ 
 MeV ~\\
       \hline
       $^2$H & $-20 \pm 12$\\
       \hline
       $^3$He & $-40 \pm 15$\\
       \hline
       $^4$He & $-72 \pm 20$ \\ 
       \hline      
       $^7$Li & $-400 \pm 70$\\ 
       \hline
     \end{tabular}
     \caption{Finite volume energy difference 
     $\Delta E(L)$ of the ground states of the low-lying nuclei from their respective thresholds,  corresponding to the correlators in Figure \ref{corf}, for the case where  $m_u = m_d = m_c$ ($N_\mathrm{conf} = 185$).} \label{Table_dE}
\end{table}
\end{center}
Of course, these correlators do not correspond to any physical nuclei, for which we need to calculate them with realistic lattices at much smaller quark masses and at large volumes. However, this study demonstrates that it is possible to calculate correlation functions of these nuclei for various set of operators as well as for large volume ensembles, rather easily. The signal-to-noise in their two-point correlation functions are quite good at the charm quark mass, and one can even extract the finite-volume energy differences for these nuclei with good precision. An interesting fact to be noted here is about the presence of energy levels at the charm quark mass which are clearly below their respective thresholds, particularly for $^3$He and $^4$He and  $^7$Li. This suggests the presence of deeply bound states, even though unphysical, for these exotic nuclei, which can be called charmed-Helium and charmed-Lithium. The binding mechanism of these exotic nuclei could be completely different from the normal nuclei and most probably originates from strong color-forces below the confinement length scale. This aspect needs to be investigated in the future.
At the strange quark mass,
there are still signals in each of the correlators up to a time-slice, even for the higher ones, even though we have used only 168 gauge configurations. Use of a large number of configurations (order of thousands and more which are now easily available to any big lattice QCD collaboration), improved interpolating fields and multiple sources on realistic lattices \cite{Yamazaki:2023swq} is expected to provide much better signal-to-noise for these correlators.
With adequate computing resources, we thus expect to achieve precise observables including energy spectra as well as information on the structures of these nuclei in the near future.

For higher nuclei, our proposed method can  also be implemented 
 considering interpolating fields within a cluster structure built with low-lying nuclei. As an example, we consider here the case of $^{12}$C. One possible interpolating field that can easily be constructed is by combining operators for $^{4}$He and  $^{8}$Be nuclei, both separately projected to zero momentum at the sink. The two point correlation functions can then be computed by calculating first the block structures for the constituent $^{4}$He and $^{8}$Be, and then sum over all possible contractions of the blocks. The number of contractions in terms of these blocks will then be of the order $(18!)^2/(6!\times12!)^2\approx 3\times10^8$. With efficient choices of interpolating operators for $^4$He and $^{8}$Be, this computation can indeed be performed. The error probability analysis for implementing the randomized algorithm for $^{12}$C is discussed in Appendix \ref{AppendixC}.  For even higher
nuclei, similar procedures can possibly also be implemented through suitable clustering of lower nuclei.

\subsection{Distributions of terms in determinants and blocks in constructing the correlation functions and their implications}
\label{sec:terms-distribution}
Finally, we would like to discuss
an intriguing feature of the minimal set of determinants, that we discover for these nuclei. 
We calculate the two-point correlation functions using Eq.~(\ref{eq2.9}) with the improved method \ref{Imp_Method}, that we proposed in Sec.~\ref{ss3c}. 
The products of determinants contribute to the 
sum in Eq.~(\ref{eq2.9})
and together provide the total correlation functions. We find that each term in the sum, that comes from the product of determinants, has a characteristic pattern, that we discuss here. More details are provided in the Appendix \ref{distribution}.

\begin{figure} [h!]
    \centering  
    \hspace*{-0.2in}
\includegraphics[width=0.48\textwidth,height=0.39\textwidth]{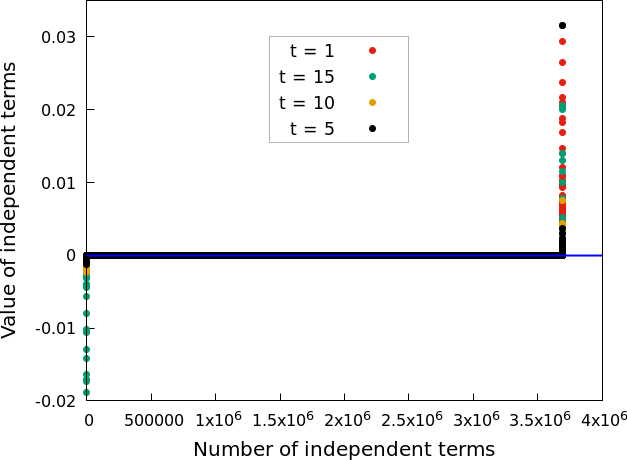}
\includegraphics[width=0.48\textwidth,height=0.39\textwidth]{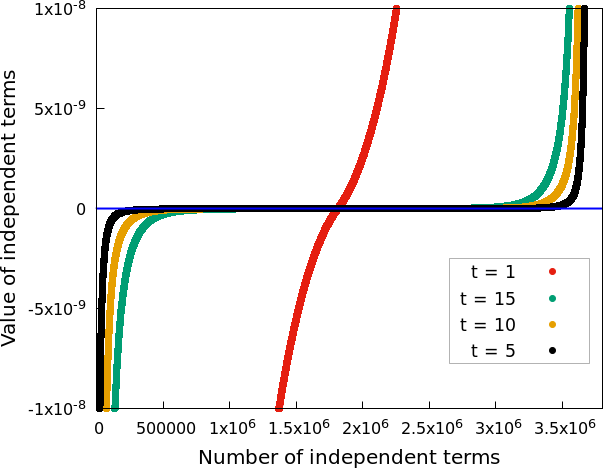}
    \caption{Left: The distributions of non-zero of terms that enters in  Eq.~(\ref{eq2.9}) to construct the two-point correlation functions of $^4$He at various time slices (lattice: $40^3 \times 64$, $a = 0.1189$ fm, and 
    for the case of $m_u = m_d = m_c$). The numbers are ordered from largest negative to largest positive from left to right (see text for details).  
    right: zoomed-in version  showing the high level of symmetry of smaller contributors around zero value. These data are for wall-source and point sink case.}
    \label{distrb_He4}
\end{figure}

We find that most of these terms have near-zero values, and they generally
are distributed around zero (positive and negative), while only a small fraction
of them have large values away from zero. Naturally, the later ones contribute
mostly in building the correlation functions. Moreover, maximum number of
degeneracies comes from the determinants which contribute less to the
correlation functions, {\it i.e.}, maximum contributions to the correlation
functions originates from the tail-end having large values, and they do not have
much degeneracy.  In the left pane of Fig.~\ref{distrb_He4}, we show the
distributions of non-zero of terms for $^4$He nucleus, with the interpolating field given in Eq. (22), computed at various time
slices for a given gauge configuration. We calculate each of these terms at a few time slices, after summing over whole spatial volume. That is, if we sum over all data points at a given time-slice, it will be the value of the two-point correlation function at that time-slice. Though the value of each term changes with the choice of gauge configuration, their distribution does not change, particularly the pattern of top percentage of terms. That is, in each gauge configuration, the indices of the terms that dominate are the same. 
 In Fig.~\ref{distrb_He4}, the numbers for the value of the independent terms are ordered from
largest negative to largest positive from left to right. To show the
distributions for each time slices within one plot, we normalize the
largest positive value of each time slices with the largest positive value at
$t = 1$. The right pane is the zoomed-in version of the left pane showing a
higher level of symmetry for the smaller contributors around zero value. The larger
contributors are only a small fraction of the total number of terms, and they
are much less symmetric with respect to zero-line, suggesting they contribute
mostly in constructing the correlators. We have tested this distribution on
various lattice ensembles and at different time slices and find similar
conclusions. Some of the other results are shown in Appendix \ref{distribution}.

\begin{figure}
    \centering  
    \hspace*{-0.2in}
    \includegraphics[width=0.58\textwidth]{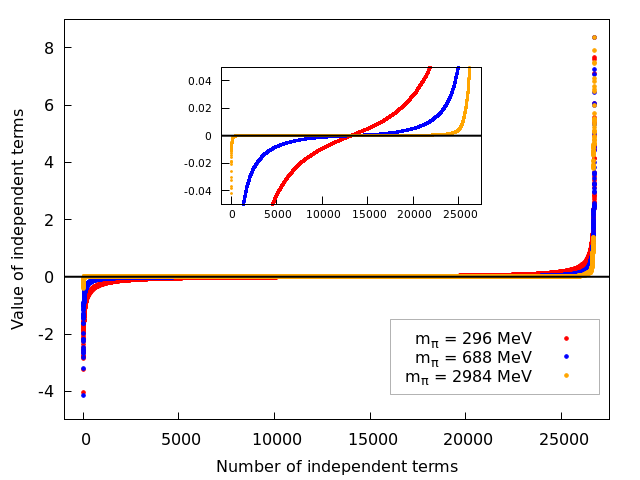}
    \caption{Distribution plots for $^2$He, similar as Fig.~\ref{distrb_He4},
      but at three different quark masses (corresponding pseudoscalar meson
      masses are shown in legends).  Data at different $m_{\pi}$s are normalized
      such that the largest positive values are equal for the three cases. Inset
      in the middle shows a zoomed-in version depicting that the distribution
      becomes less asymmetric near zero as pion mass decreases.}
    \label{dist_all_np_flat}
\end{figure}

\begin{figure}
    \centering  
    \hspace*{-0.2in}
\includegraphics[width=0.5\textwidth,height=0.4\textwidth]{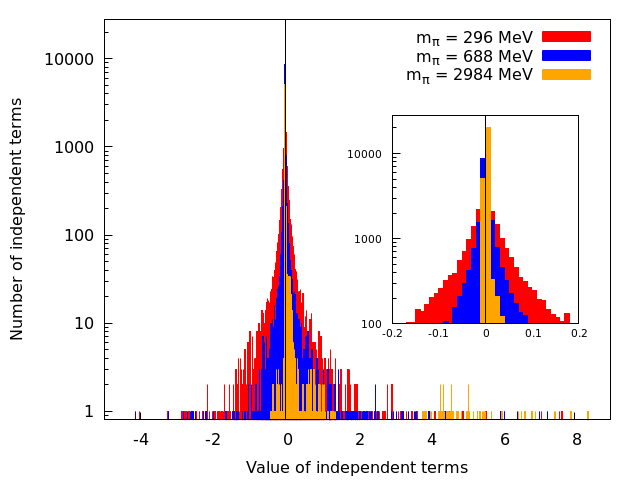}
\includegraphics[width=0.51\textwidth,height=0.4\textwidth]{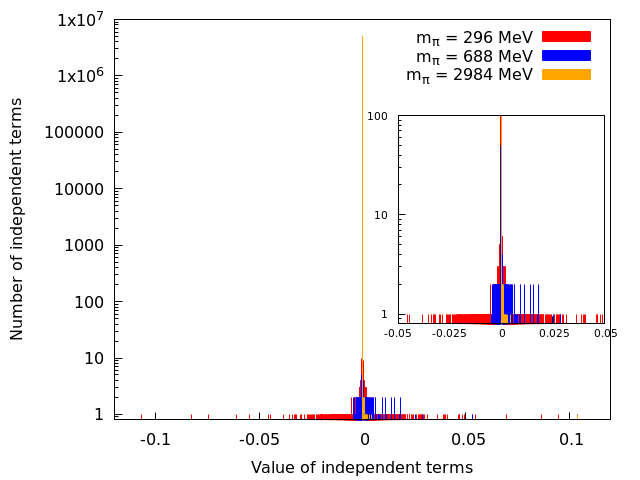}
    \caption{Left: Histogram plots corresponding to Fig.~\ref{dist_all_np_flat} (for $^2$H) with bin-size 0.01. Here again, the inset shows that the asymmetry reduces as pion mass decreases. Right: Similar histogram plot for $^4$He.}
\label{dist_all_np_he4_hist}
\end{figure}

To check the quark mass dependence of the distribution of the minimal set of terms, we carry out the same analysis with propagators generated at various quark masses. In Fig.~\ref{dist_all_np_flat},
we show the results for $^2$H at three different quark masses, corresponding to $m_{\pi} = 296, 688$ and 2984 MeV; inset is the zoomed-in version at the middle.
In the left pane of Fig.~\ref{dist_all_np_he4_hist} we show the same data with a histogram. 
We find that as quark mass decreases  the distribution spreads out and becomes less asymmetric around zero, like a sharp Cauchy-Lorentz distribution but with long tails. It suggests that at lower quark masses, more terms contribute. 
In the right pane of Fig.~\ref{dist_all_np_he4_hist}
we present a similar study for $^4$He, showing that the distribution becomes sharper near zero and the dominance of larger few terms becomes more apparent (see Appendix \ref{distribution} further). Difference in distribution between $^2$H and $^4$He at the same quark mass shows that for higher nuclei, the fraction of required terms would be much smaller compared to the lower ones.

For a plane-wave sink, to compute correlation functions, we utilize Eq.~(\ref{eq2.6}) which originates from a block algorithm. As in the determinant algorithm, we carry out the same exercise to check the distribution of the terms in Eq.~(\ref{eq2.6}). Here again, we find  a similar extreme distribution of the non-zero terms constructing the correlation functions. Since we have already provided a detailed discussion on the distribution for the determinant case, we defer the results for the block-case to Appendix~\ref{distribution}.

\begin{figure}[h!]
    \centering    
  \hspace*{-0.05in} \includegraphics[width=0.49\textwidth,height=0.4\textwidth]{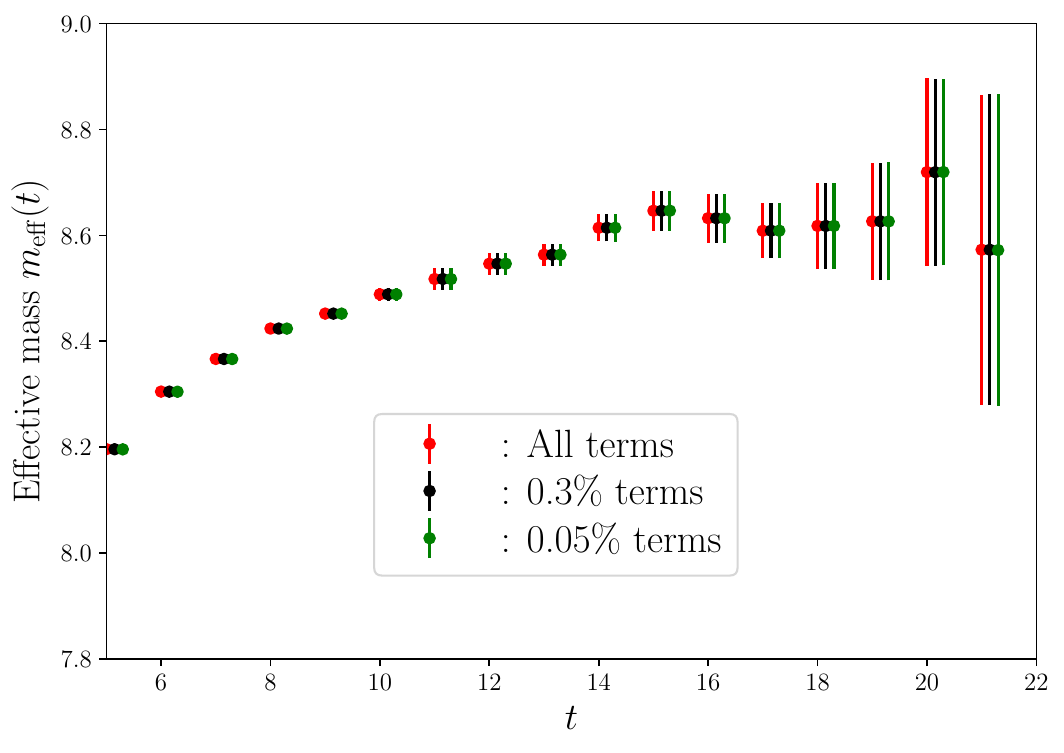}
\includegraphics[width=0.49\textwidth,height=0.4\textwidth]{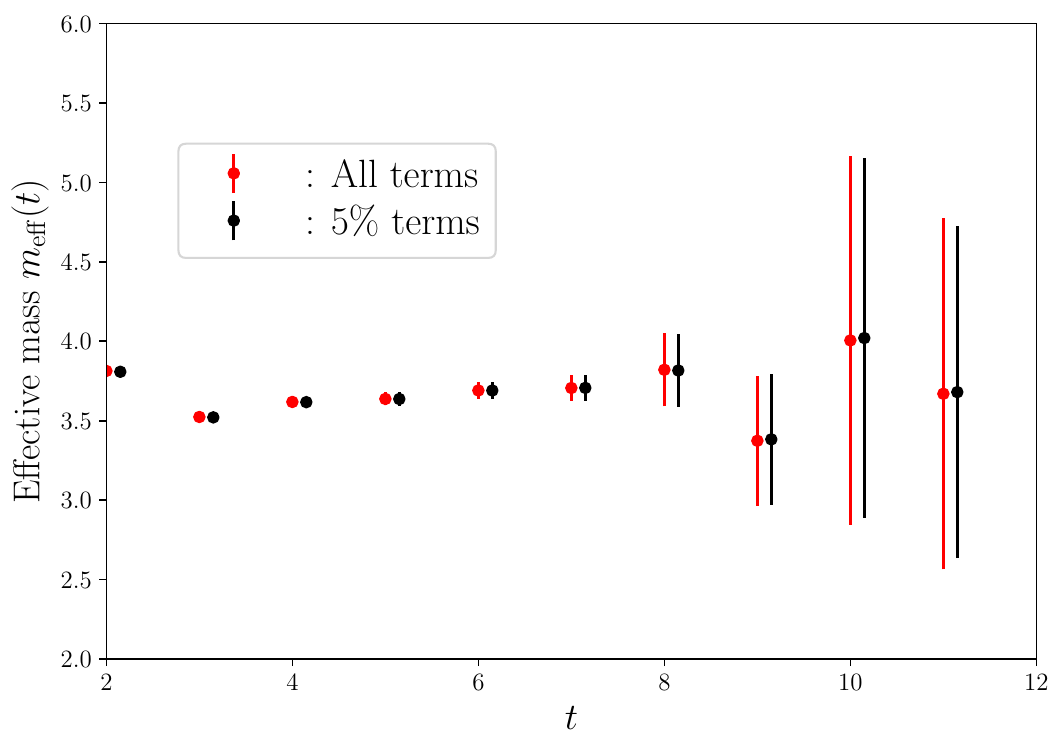}
    \caption{Effective mass of $^4$He obtained using all and only top 0.3\%, and 0.05\% terms (absolute large) of the correlation functions. This result is obtained using a $40^3 \times 64$ ensemble
    for the case of $m_u = m_d = m_c$. Right: Same results for the case of $m_u = m_d = m_s$. Operator set up: Wall source and point sink.}
    \label{em_per_he4}
\end{figure}

To substantiate these observations resolutely, both for the determinant-case and the block-case,  we repeat the full correlator calculations for these nuclei over a set of gauge configurations but only with the top few percent terms. 
In Fig.~\ref{em_per_he4} we show the effective masses of $^4$He obtained with full and smaller fraction of total terms with a point-source setup in determinant case.
We find even if we use only the largest (absolute value) top 0.05\% terms, we can recover almost indistinguishable correlator and resulting effective mass for $^4$He at the charm quark mass. At the strange quark mass required fraction is a bit higher but still less than 5\%.  We repeat this exercise for other nuclei as well, and results for plane-wave cases (block-type) are shown in Appendix \ref{distribution}.
 In Table \ref{tab:dist_summary} we summarize our observation  which led us to conclude: (i) For the operators as in Eqs.~(\ref{NP_op}-\ref{He4_op}),  higher the nuclei lesser and lesser fraction of total terms are needed, (ii) more terms are needed at the lighter quark masses than that at heavier quarks.

The typical shape of the distribution of determinants/blocks are in general
observed to be an extreme statistical one, resembling with a sharp Cauchy-Lorentzian distribution with a long tail, perhaps like the Tracy-Widom distribution \cite{TRACY1993115}. 
 In the future, we will investigate this in detail to find the origin of this statistics, which could well help in understanding the formation of nuclei from a specific color-spin correlation. Moreover, use of specific set of color-spin combinations may help to construct efficient interpolating operators for a nucleus in which more fraction of terms can contribute, yielding better signal-to-noise ratio.

We find these features to be in general true for all nuclei that we have studied. This led us to arrive at the following conjecture: this feature of extreme statistical distribution of the minimal set of determinants in constructing the nuclear correlation functions is true for all nuclei. This could be related to an as-yet-unidentified symmetry which dictates a small number of quark
combinations to be enough in constructing any nuclear correlation functions. Identifying such a symmetry will be quite significant in studying nuclear physics. In the future, we will investigate that in detail to find its origin, and perhaps machine learning techniques can also be employed to analyze the data on independent determinants.
 
 \begin{table}[h!]
    \centering
    \begin{tabular}{c|c|c|c}
         Nucleus &  sink type&quark mass & required fraction \\
         \hline\
          $^2$H, $^3$He & point&  $< m_s$ &$< $ 10\%\\   
        $^2$H, $^3$He & extended& $< m_s$ &$< 20\%$\\
         $^4$He & point & $> m_s$ & $<$ 1\% \\
         $^4$He &  point& $< m_s$ & $\sim$ 5\%\\
         $^4$He & extended &$> m_s$ &$<$  10\%\\
       $^4$He & extended &$< m_s$ &$\sim$  10\%\\
         $^7$Li & cluster &$ > m_s$ &$\sim$  5\%\\
         $^7$Li & cluster &$< m_s$ &$<$   10\%\\         
\hline
    \end{tabular}
    \caption{Summary of observation for the required fraction of terms that are actually needed to compute correlation functions for these nuclei. These fractions of terms are obtained by comparing the effective masses computed from all terms and from fraction of large value terms around the fit (plateau) regions.  
    }
    \label{tab:dist_summary}
\end{table}

Here it is also noteworthy to mention that the use of only a small fraction of terms makes the correlation function computation much cheaper, sometimes more than an order of magnitude than that shown in Tables \ref{tab_results1} and \ref{tab_results2}, computed using our improved determinant/block method. As mentioned earlier, this {\it important} fraction of terms can be fixed on a very small lattice and also only at a spatial point, and hence can be obtained rather cheaply. The proposed use of {\it important} terms will make it possible to compute correlation functions for more higher nuclei.

\section{Summary and outlook}\label{sec_vi}
The study of nuclear physics using quarks and gluons, the fundamental degrees of freedom of the strong force, is a very challenging problem. Lattice QCD calculations are, in principle, most suitable for this task. However, in practice, the two main issues that impede such attempts are the uncontrollable growth of quark-combinatorics in a correlation functions that are needed to compute physical observables, and the control of errors in such calculations. In this work, we present a state-of-the-art solution to the first problem, the so-called Wick-contraction problem of nuclei.
We propose two novel methods: the first one is based on the determinant algorithm, which arises from the symmetry of the tensors involved in Wick-contractions. Therein, an algorithm is proposed to identify a ``minimal'' set of independent determinants or block terms exploiting their degeneracies, that are 
sufficient to compute a given correlation function.  Being dependent only on the color-spin symmetry, this minimal set, which is unique to a given interpolating field, needs to be determined only once and that can be searched on a small lattice and only at one space-time point. Once this construction is done, that minimal set can be employed across any lattice ensembles, irrespective of their sizes. This renders the algorithm highly efficient, robust, and hence state-of-the-art.
The second method highlights the potential advantages of using tensor-based
programming models, using \tensorflow{} as an example.  Use of  \tensorflow{} allows one to write a high-level ``code'' automatically for a given interpolating operator of a nucleus, which will be very useful and less error-prone for exploring optimized operators for performing calculations in the framework of generalized eigenvalue problems. Moreover, since these tools  can automatically perform optimizations in tensor contractions and take advantage of the parallelism afforded by hardware accelerators, we find it to be more efficient than the traditional approaches till it gets constrained by the hardware memory limit. We encourage LQCD
practitioners to adapt such advantages in computing correlation functions in lattice QCD calculations. In our study, we have successfully demonstrated the efficacy of our proposed methods for nuclei ranging from A = 1 to 7, surpassing the efficiency of existing algorithms at least by an order of magnitude which becomes even higher as the atomic number increases. The wall clock timings, illustrating the effectiveness of our methods, are provided in  Tables \ref{tab_results1} and \ref{tab_results2}. With contemporary hardware we find \tensorflow{} to be more efficient for lighter nuclei (till A $< 5$), and after that our proposed optimized-determinant algorithm emerges as the more efficient choice. With the efficiency of our proposed methods it would be possible to compute
correlation functions for nuclei even with complicated interpolating fields and
smearings, at least for the lighter ones (A $\sim$ 12) as discussed in the results section. 
In addition, we uncover an intriguing feature common to all nuclei that we studied: the distribution of determinants and terms in blocks constructing a correlation function follows an extreme statistics pattern, and the heavier the nuclei the more extreme is the distribution. This phenomenon is related to an inherent spin-color correlation, which dominates in the nuclear correlation functions. Perhaps it hints to a spin-color symmetry, warranting further future exploration. Due to the extreme statistical distributions, we find only a small fraction of determinants and blocks are actually needed to construct the nuclear correlation functions, and those can be obtained rather cheaply on a smaller lattice. Utilization of this important small set of determinants makes the correlator computation even more economical, potentially by another order of magnitude, compared to the results presented  in Tables \ref{tab_results1} and \ref{tab_results2}. This enhances the possibility of studying  nuclei with much higher atomic number. Given this advance, subject to the
control over signal-to-noise problem in correlation functions, it would be
possible to extract nuclear properties, for example energy levels and
structures. Taking advantage of our proposed methods and incorporating finite volume
techniques of lattice QCD
\cite{Luscher:1990ck,Luscher:1990ux,RUMMUKAINEN1995397,Hansen:2014eka,Briceno:2017max,Hansen:2019nir}
one can envision extracting excited states of nuclei, scattering
parameters and even delve into the study of nuclear reactions
\cite{Beane:2015yha,Savage:2016kon,Tiburzi:2017iux,Kronfeld:2019nfb},
information on which are difficult to obtain from experiments. Similarly,
the study of nuclear three-point functions, which are also possible to compute through our
proposed methods, can help us in investigating the intricate structures of these nuclei.
Admittedly, the next challenge in such endeavors lies in controlling errors, particularly addressing signal-to-noise issues. However, in the ongoing exascale era of high-performance computing, lattices with bigger volumes and smaller lattice spacings will increasingly be available \cite{Yamazaki:2023swq}. 
Moreover, usage of GEVP with a variety of operators including, local, non-local, and multiple source sink setup, can help to find optimized operators leading to better overlap to individual close-by energy levels.  It is expected that controlling of systematics in the near future will be feasible with the aid of those lattices and appropriate judicious operator choices for nuclei. As emphasized in the introduction, such calculations possess profound significance as their outcomes create a synergy across nuclear physics, astrophysics, cosmology, high-energy physics, and many-body physics. We envision that an amalgamation of lattice QCD input for the low-lying nuclei, possibly till A  $\sim$ 12, and the subsequent use of effective field theories, including DFT/QMC calculations, could be the way forward for a comprehensive understanding of nuclei. This work provides an important milestone on this exciting journey.

\begin{acknowledgments}
\noindent This work is supported by the Department of Atomic Energy, Government
of India, under Project Identification Numbers RTI 4001 and RTI 4002. Computations
were carried out on the Cray-XC30 of ILGTI, TIFR, Guppy and Grouse
clusters of the Department of Theoretical Physics, TIFR, and the
Airawat system of the National Supercomputing Mission, CDAC, India. We
would also like to thank Rishi Pathak, Ajay Salve and Kapil Ghadiali
for computational supports. N. M. would like to thank 
 Vivek Datar and R. Palit for discussions.
P. S. acknowledges support from Adobe Systems Incorporated via a gift to TIFR, from DST, India under project number MTR/2023/001547, from the
Infosys-Chandrasekharan Virtual Centre for Random Geometry at TIFR,
and computational facilities
provided by the TIFR CC HPC Facility.
A. K. acknowledges the visitor program of the Department of Theoretical Physics, TIFR.
\end{acknowledgments}

\clearpage
\appendix

\section{GPU implementation with CUDA}\label{GPU}
Here, we provide the details of implementation procedure of the determinant algorithm, as discussed in Sec.~\ref{sec:determ-based-algor}, using CUDA-C. In the initialization step, we fetch the determinant algorithm parameters from the index data files and transfer them to the device memory. We proceed by declaring and allocating memory to the propagator, correlator, and several hyper-parameters on the device side. Additionally, we read the first temporal slice of the propagator and store it on the device for further computations. The propagator is denoted by $S^{ab}_{\alpha\beta}(t,\vec{\bf x})$, with color indices ($a, b$) and spinor indices ($\alpha , \beta$).

\begin{figure}[htbp]
	\centering
\includegraphics[width=0.8\textwidth,height = 0.6\textwidth]{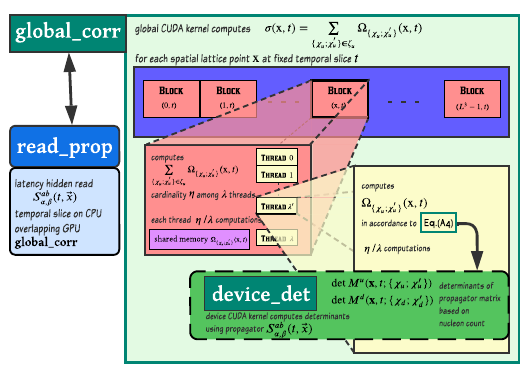}
	\caption{Schematic diagram showing CUDA-C implementation of the determinant algorithm for deuteron.}
	\label{fig:deuteron_algo}
\end{figure}

The next step involves iteration over the temporal slices of the propagator, represented as $t \in \left[0, T-1\right]$, where $T$ denotes the temporal lattice size. For each spatial lattice point $x$ at a specific temporal slice $t$, we calculate one-point correlations. These calculations are executed by the global CUDA kernel, named \pyth|global_corr|, and are performed on the device side. The flowchart shown in Fig.~\ref{fig:deuteron_algo} provides a simplified illustration of the intricate details. 
On the device side, we consider a CUDA grid where each block is assigned to a spatial lattice point. Within each grid-block, independent computations are performed for a specific spatial lattice point, denoted as $x$, at a fixed temporal slice $t$. As explained in Sec.~\ref{impl}, we need to sum over a product of finite number of distinct non-zero determinants. In Sec.~\ref{impl}, we discussed that it is advantageous to find out the minimal set of spin-color indices $\{\{\chi_f;\chi^{\prime}_f\}\}|_{min}$ for each flavor $f\in\{u,d\}$.\footnote{We are only including up and down flavors as we are discussing the algorithm in the context of usual nuclei which only have valence quarks of these two flavors.} Let us define the minimal set of indices, $\{\{\chi_u;\chi^{\prime}_u\}\}|_{min}$ and $\{\{\chi_d;\chi^{\prime}_d\}\}|_{min}$ as $\boldsymbol{\zeta}_u$ and $\boldsymbol{\zeta}_d$ respectively. Then we can rewrite Eq.~(\ref{eq2.9}) for the case of two flavors as in the following,
\begin{equation}
    G(t)=\sum_{\mathbf{x}}\sum_{\{\chi_u;\chi^{\prime}_u\}\in\boldsymbol{\zeta}_u}\sum_{\{\chi_d;\chi^{\prime}_d\}\in\boldsymbol{\zeta}_d}\Theta(\{\chi_u;\chi^{\prime}_u\};\{\chi_d;\chi^{\prime}_d\})\det M^u(\{\chi_u;\chi^{\prime}_u\}) \det M^d(\{\chi_d;\chi^{\prime}_d\}).\label{eqa1}
\end{equation}
Here $\Theta(\{\chi_u;\chi^{\prime}_u\};\{\chi_d;\chi^{\prime}_d\})$ represents the weights of the particular product of the determinants, and it is determined taking into account the degeneracy of the product of the determinants and the factors of coefficient matrices,  $C^{\prime\chi_{1}^{\prime}\chi_{2}^{\prime}\cdots \chi_{n}^{\prime}}_{\alpha}$ and $C^{\chi_1\chi_2\cdots \chi_{n}}_{\alpha}$, as in Eq.~(\ref{eq2.9}). We can then express  Eq.~(\ref{eqa1}) as below,
\begin{eqnarray}
G(t)&=&\sum_{\mathbf{x}}\sigma(\mathbf{x},t) , \\
    \sigma(\mathbf{x},t)&=&\sum_{\{\chi_u;\chi^{\prime}_u\}\in\boldsymbol{\zeta}_u}\Omega_{\{\chi_u;\chi^{\prime}_u\}}(\mathbf{x},t) ,\\
    \Omega_{\{\chi_u;\chi^{\prime}_u\}}(\mathbf{x},t)&=& \sum_{\{\chi_d;\chi^{\prime}_d\}\in\boldsymbol{\zeta}_d}\Theta(\{\chi_u;\chi^{\prime}_u\};\{\chi_d;\chi^{\prime}_d\})\det M^u(\mathbf{x},t;\{\chi_u;\chi^{\prime}_u\})\nonumber\\&& \hspace{3cm}\times\, \det M^d(\mathbf{x},t;\{\chi_d;\chi^{\prime}_d\}) \label{eqa4}.
\end{eqnarray}
Within each grid-block, independent computations of $\sigma(\mathbf{x},t)$ are performed for a specific spatial lattice point, at a fixed temporal slice $t$. Consequently, each block requires executing $\eta$ computations where $\eta$ is the cardinality of set $\boldsymbol{\zeta}_u$. These terms are then summed together, that is,
\begin{equation}
	\sigma (x,t) = \sum_{\{\chi_u;\chi^{\prime}_u\}\in\boldsymbol{\zeta}_u}\Omega_{\{\chi_u;\chi^{\prime}_u\}}(\mathbf{x},t).
\end{equation}
For an efficient implementation, we can divide the $\eta$ computations among $\lambda$ threads, ensuring that each block-thread performs $\eta/\lambda$ computations. To optimize the performance, we store these intermediate results, denoted as $\Omega_{\{\chi_u;\chi^{\prime}_u\}}$, in the shared memory cache. Then, within each thread, we calculate final $\Omega_{\{\chi_u;\chi^{\prime}_u\}}$ by evaluating the expression in Eq.~(\ref{eqa4}). The determinants of matrices $M^u(\mathbf{x},t;\{\chi_u;\chi^{\prime}_u\})$ and $M^d(\mathbf{x},t;\{\chi_d;\chi^{\prime}_d\})$ are evaluated in a CUDA device kernel, \pyth|device_det|. The matrices $M^u$ and $M^d$ are defined using the propagator matrix as discussed in Sec. 3. For example, for $^2$D-nucleus, $\Delta$-matrices will be  $3\times 3$ matrices (corresponding to three $u$ and three $d$ quarks) of the following form:
\begin{equation}
M^{u/d}(x,t;\{\chi_{u/d};\chi_{u/d}^{\prime}\})
= \begin{vmatrix}
S^{a_{1},b_{1}}_{\alpha_{1},\beta_{1}}(x,t) & \,\,
S^{a_{1},b_{2}}_{\alpha_{1},\beta_{2}}(x,t)
& \,\,
S^{a_{1},b_{3}}_{\alpha_{1},\beta_{3}}(x,t)\\
S^{a_{2},b_{1}}_{\alpha_{2},\beta_{1}}(x,t) & \,\,
S^{a_{2},b_{2}}_{\alpha_{2},\beta_{2}}(x,t)
& \,\,
S^{a_{2},b_{3}}_{\alpha_{2},\beta_{3}}(x,t)\\
S^{a_{3},b_{1}}_{\alpha_{3},\beta_{1}}(x,t) & \,\,
S^{a_{3},b_{2}}_{\alpha_{3},\beta_{2}}(x,t)
& \,\,
S^{a_{3},b_{3}}_{\alpha_{3},\beta_{3}}(x,t) \\
\end{vmatrix} ,
\end{equation}
with $\chi_{u/d}$ and $\chi_{u/d}^{\prime}$ are the combined spin-color indices defined as $\chi_{u/d}=\{a_1,\alpha_1,a_2,\alpha_2,a_3,\alpha_3\}$ and $\chi_{u/d}^{\prime}=\{b_1,\beta_1,b_2,\beta_2,b_3,\beta_3\}$. 

After the execution of the global CUDA kernel, \pyth|global_corr| at a specific time slice $t$, we are left with $\sigma(x, t)$ for each spatial lattice point on the device side. Subsequently, we transfer this data to the host side and perform the summation over the spatial lattice points to derive the final one-point correlation function at the fixed temporal slice $t$, that is 
\begin{equation}
	\Sigma(t) =  \sum_{x} \sigma(x,t).
\end{equation}
 It is important to note that while the \pyth|global_corr| computation takes place on the GPU side, we read the necessary propagator slice for the subsequent iteration concurrently. By doing this, we effectively overlap the reading time of the propagator, thus hiding the latency and maximizing overall resource utilization. These calculations are repeated over all temporal slices. At the end free all the allocated memory on the device and host side.

\section{Tensor Flow}\label{tf}
As discussed in the main paper, the \tensorflow{} framework (combined with \opteinsum{}) offers two type of
advantages.
The first is syntactic: the \emph{syntax} for writing the code computing a given
LQCD expression is compact and human readable, and often has one-to-one
correspondence with the corresponding mathematical equations.
It is clear that this makes the code much easier to both write and verify, as
compared to equivalent hand-written code in \clang{} or \fortran{}.\footnote{Of
  course, here one has to assume that the \tensorflow{} and \opteinsum{} systems
  are implemented correctly.  This, however, is analogous to assuming that
  \clang{} or \fortran{} compilers are implemented correctly.}
However, this syntactic compactness has another advantage: it makes it much
easier to write programs that will \emph{generate} the
\tensorflow{}/\opteinsum{} code for correlation functions in lattice QCD.
We discuss this in more detail below: the need for such code-generating programs
arises because even for systems with a moderate number of quarks, there may be
too many terms for one to reliably generate even the corresponding mathematical
equations by hand.

The second advantage of frameworks like \tensorflow{} is their ability to take
advantage of specialized hardware like GPUs without requiring significant change
in the code.  In principle, it may be possible that for a specific LQCD
computation, one can produce hand-optimized code using lower-level languages and
libraries such as \clang{} and CUDA.  However, as pointed out above, this is
both much slower to write and can also be error-prone.  Frameworks like
\tensorflow{} therefore allow one to quickly compare higher-level algorithms for
carrying out the computation before significant hand-optimization effort is
expended for any one algorithm.
\subsection*{Programatically generating \tensorflow{} code for computing correlation functions in LQCD}
Since systems like \tensorflow{} offer high-level constructs and operations such
as tensors and contraction as primitives, code written in such systems can often
closely mirror the mathematical expressions being translated.  This
``declarative'' nature of the code also makes it easier to write programs (in
traditional programming languages such as \texttt{python} or \texttt{c++}) to
automatically generate the \tensorflow{} code for computing correlation
functions, given a description of the nucleus to be studied.  In this work, we
exploited this to write a program which, given the number of protons and neutron
in the nucleus, automatically generates both (i) a description of the terms in
the expression for computing correlation functions, and (ii) the corresponding
\tensorflow{} code for evaluating these terms.

We present here examples of the output of such a program, for the simple case of
a single proton system.  The program generates both the \LaTeX{} code for the
corresponding expression (Eq.~\eqref{code-1p-eq}, with the \LaTeX{} code\footnote{For
  reasons of presentation, the spacing in the \LaTeX{} code generated by the
  program has been slightly edited to produce Eq.~\eqref{code-1p-eq}.} generated by
the program given in Fig.~\ref{fig:code-1p-tex}) as well as the corresponding code
using the \texttt{opt\_einsum} and \tensorflow{} packages (Fig.~\ref{fig:code-1p}).
\begin{align}
  +\,\epsilon_{i_0 j_0 k_0 }  C_{\beta_0 \gamma_0 }  \epsilon_{i_0' j_0' k_0' }  C_{\beta_0' \gamma_0' } 
  S(x_0)_{i_0 \alpha_0 i_0' \alpha_0' } S(x_0)_{j_0 \beta_0 j_0' \beta_0' } S(x_0)_{k_0 \gamma_0 k_0' \gamma_0' } \nonumber
  \\
  -\,\epsilon_{i_0 j_0 k_0 }  C_{\beta_0 \gamma_0 }  \epsilon_{i_0' j_0' k_0' }  C_{\beta_0' \gamma_0' } 
  S(x_0)_{i_0 \alpha_0 j_0' \beta_0' } S(x_0)_{j_0 \beta_0 i_0' \alpha_0' } S(x_0)_{k_0 \gamma_0 k_0' \gamma_0' }.
 \label{code-1p-eq}
\end{align}
Here the subscript ``0" refers to the indices of the source point.

\begin{figure}[h]
	\centering
\begin{lstlisting}[language=Python]
 + \
oe.contract(eps, ['i_0', 'j_0', 'k_0'], C, ['beta_0', 'gamma_0'], eps, ['i_0_prime', 'j_0_prime', 'k_0_prime'], C, ['beta_0_prime', 'gamma_0_prime'], S, ['x_0', 'i_0', 'alpha_0', 'i_0_prime', 'alpha_0_prime'], S, ['x_0', 'j_0', 'beta_0', 'j_0_prime', 'beta_0_prime'], S, ['x_0', 'k_0', 'gamma_0', 'k_0_prime', 'gamma_0_prime'], ['alpha_0', 'alpha_0_prime']) - \
oe.contract(eps, ['i_0', 'j_0', 'k_0'], C, ['beta_0', 'gamma_0'], eps, ['i_0_prime', 'j_0_prime', 'k_0_prime'], C, ['beta_0_prime', 'gamma_0_prime'], S, ['x_0', 'i_0', 'alpha_0', 'j_0_prime', 'beta_0_prime'], S, ['x_0', 'j_0', 'beta_0', 'i_0_prime', 'alpha_0_prime'], S, ['x_0', 'k_0', 'gamma_0', 'k_0_prime', 'gamma_0_prime'], ['alpha_0', 'alpha_0_prime'])
\end{lstlisting}
  \caption{Automatic code generation (using the \texttt{opt\_einsum} package along with
    \tensorflow{}) for a single proton.\label{fig:code-1p}}
\end{figure}

\begin{figure}[h]
  \centering
\begin{lstlisting}[language={[latex]tex}]
\documentclass[a4papper,11pt]{article}
\usepackage{amsmath}
\begin{document}
\begin{multline}
 + \epsilon_{i_0 j_0 k_0 }  C_{\beta_0 \gamma_0 }  \epsilon_{i_0' j_0' k_0' }  C_{\beta_0' \gamma_0' }  \\
S(x_0)_{i_0 \alpha_0 i_0' \alpha_0' } S(x_0)_{j_0 \beta_0 j_0' \beta_0' } S(x_0)_{k_0 \gamma_0 k_0' \gamma_0' } \end{multline}
\begin{multline}
 - \epsilon_{i_0 j_0 k_0 }  C_{\beta_0 \gamma_0 }  \epsilon_{i_0' j_0' k_0' }  C_{\beta_0' \gamma_0' }  \\
S(x_0)_{i_0 \alpha_0 j_0' \beta_0' } S(x_0)_{j_0 \beta_0 i_0' \alpha_0' } S(x_0)_{k_0 \gamma_0 k_0' \gamma_0' } \end{multline}
\end{document}
\end{lstlisting}
  \caption{\LaTeX{} code for the expression in
    \cref{code-1p-eq}\label{fig:code-1p-tex}}
\end{figure}

Similarly, for other nuclei we can print the optimized codes automatically. However, those will be much longer in number of lines and hence are not possible to include here.

\section{Identity testing for determinants}\label{sec:append-ident-test}
\label{AppendixC} Our method of detecting degeneracy for determinants is based on the idea of
randomized identity testing, which in turn is based on the following simple but
powerful result.  In this form, the result is often attributed to
Schwarz~\cite{schwartz_fast_1980} and Zippel~\cite{zippel79:_probab}, but the
ideas underlying it seem to have been rediscovered many
times~\cite{demillo_probabilistic_1978,erickson_counting_1974,ore22:_uber}.
\begin{theorem}[\textbf{Identity testing
    lemma}~\cite{schwartz_fast_1980,zippel79:_probab}]
  \label{thm:sch-zip}
  Let $p(x_1, x_2, \dots x_m) \in \F[x_1, x_2, \dots, x_m]$ be a polynomial of
  total degree at most $d$ in $m$ variables, with coefficients in the field
  $\F$, such that not all coefficients of $p$ are zero.  Let $C$ be a finite
  subset of $\F$, and let $r_1, r_2, \dots, r_n$ be random elements of $C$,
  chosen uniformly at random from $C$, independently of each other.  Then
  \begin{equation}
    \label{eq:1}
    \Pr[p(r_1, r_2, \dots r_m) = 0] \leq \frac{d}{\abs{C}}.
  \end{equation}
\end{theorem}

Consider now the setting of Eq.~\eqref{eq2.10}.  We can now view the
determinants $\det\mathbf{M}^{f}(\{\chi;\chi'\})$ as polynomials of degree $n_f$
(the number of quarks) with coefficients in the field $\mathbb{Q}$, with the
variables being the entries of the {propagator} $S$. Thus, if we want to test if the \emph{polynomials}
$\det\mathbf{M}(\{\xi; \xi'\})$ and $\det\mathbf{M}(\{\chi;\chi'\})$ are
identical (where $\{\chi;\chi'\}$ and $\{\xi;\xi'\}$ are distinct sets of
indices), \cref{thm:sch-zip} suggests doing so by checking that they evaluate to
the same value when evaluated at random instantiations of the matrix entries of
$S$.  In particular, if we perform the check with $T$ independent realizations
of the $S$ matrix, each with entries uniformly and independently at random from
a set $C$, then the probability of a given pair of determinants being
erroneously declared identical, when they are not, is at most $(n_f/\abs{C})^T$
(note that if the test in \cref{thm:sch-zip} says that two determinants are
\emph{not} identical, than that conclusion is correct: the probability of error
arises only when a test yields the same value for both the determinants for a
given instantiation of $S$).  Thus, even in the worst case, the probability of
making even one error is no more than
$\binom{N\cdot N'}{2}\cdot(n_f/\abs{C})^T$, where $N\cdot N'$  represents an upper bound on the number of terms to be considered, as in
\cref{ss3c}.  In practice, not all the $\binom{N\cdot N'}{2}$ comparisons will
need to be made in each trial (since many of the pairs would have been
distinguished in earlier trials), and the error probability bound is thus likely
to be even smaller.

We consider here an approximate numerical estimate of the above error
probability bound, particularly considering calculations involving higher
nuclei.  Considering the carbon nucleus $\prescript{12}{6}{\mathrm{C}}$, for
which $n_f = 18$ for $f \in \{u, d\}$, we take $C = \{1, 2, \dots, 360\}$, and
$T = 40$.  This leads to an approximate upper bound on $N \times N'$ of $24^{12}$ if we consider the optimized estimate given in the paragraph below Eq.~\eqref{eq2.6},
and that gives an approximate upper bound on the probability of error of about
$6.1\times 10^{-20}$.  If we instead use the naive upper bound  of $24^{24}$ described there, then increasing the value of $T$ to $65$ gives an error probability of a similar order of magnitude.  As mentioned earlier, the actual probability of error
would be even smaller. This suggests that it may be feasible to compute carbon
correlation functions as well using the determinant method we propose.

We note that the tests above needs to be performed only \emph{once} for each
nucleus under study.  Once we know the pairs of indices corresponding to which
the determinants are identical, we can simply store this information in a
look-up table, and use it for any input instantiation of the propagator, $S$.

\subsection*{Implementation Procedures} We note here some points regarding
the implementation of the above strategy.
\begin{enumerate}
\item Note that the \emph{coefficients} of the determinants above are integers
  (even though the matrix entries in $S$ may be 
  {arbitrary complex numbers}).
  Thus, for greater accuracy and simplicity, it is better to take the set $C$
  from which evaluation points are drawn to be a set of integers, say from $1$
  to $\abs{C}$.
  This ensures that all the arithmetic performed in the tests is exact integer
  arithmetic, and any spurious conclusions due to floating point round-off are
  avoided.
  Exact polynomial time rational algorithms for the evaluation of the
  determinant can be used for large $n_f$.
\item However, since we would typically like $\abs{C}$ to be about ten times
  $n_f$, a potential problem with large $n_f$ (roughly when $n_{f} \geq 7$) may
  be the possibility of integer overflow (if machine integer types are used) or
  slow speed and large memory use (if exact rational arithmetic is used).
  In particular, if one uses machine integers, one may suspect that some of the
  test results above may be spurious.

  If desired, this can be fixed using another standard but powerful method from
  computational number theory and algebra: the Chinese Remainder Theorem.  For
  our purposes, we need only the following simple special case: if integers $a$
  and $b$ satisfy $a \equiv b \mod p_i$ for distinct primes
  $p_1, p_2, \dots, p_k$, then they also satisfy
  $a \equiv b \mod \prod_{i=1}^kp_i$.  In particular, if we know further that
  $\abs{a}, \abs{b} < (1/2)\prod_{i=1}^kp_i$, then they must in fact satisfy
  $a = b$.  Note that the determinants we evaluate can be at most
  $(n_f\abs{C})^{n_f}$ in absolute value.  Thus, if we pick at least
  $n_f \log (n_f\abs{C}) + 1$ primes, we can conclude that the determinants are
  equal if they are equal modulo each of the primes.  However, modulo each such
  prime $p$, the determinant can be seen as a determinant in the finite field
  $\F_p$, and polynomial time methods such as Gaussian elimination can then be
  applied using standard 64-bit unsigned integers (even using the
  unrealistically high values $n_f = 1000$ and $\abs{C} = 10000$, we would not
  need primes bigger than $2^{20}$ in the above scheme).
\end{enumerate}

In passing, we remark that while the ideas underlying the Chinese remainder
theorem and the identity testing lemma are quite simple, they are powerful
enough to underlie some of the most important results in computational number
theory and algebra, including deterministic and randomized primality
testing~(see, e.g.,
\cite{miller_riemanns_1976,rabin_probabilistic_1980,agrawal_primality_2003,agrawal_primes_2004}).

\section{Distribution of independent terms in correlation functions}\label{distribution}
In Sec.~\ref{sec_v}, we presented the characteristics of the minimal set of determinants that comprise the major contribution in the computation of correlation functions. Here we elaborate and discuss that further with more such results.
We calculate the two-point correlation functions using Eq.~(\ref{eq2.9}) with the improved method \ref{Imp_Method}, that we proposed in \ref{ss3c}. 
We then analyze the non-zero independent terms that enters Eq.~(\ref{eq2.9}) after removing the zero-contributors. 

\begin{figure}[h!]
    \centering    
\includegraphics[width=0.48\textwidth,height=0.38\textwidth]{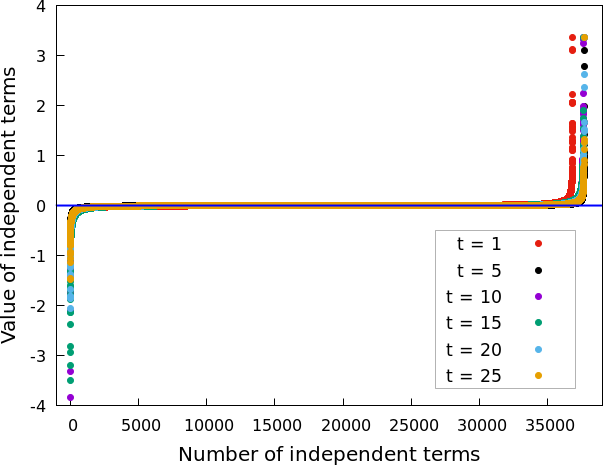}
\includegraphics[width=0.48\textwidth,height=0.38\textwidth]{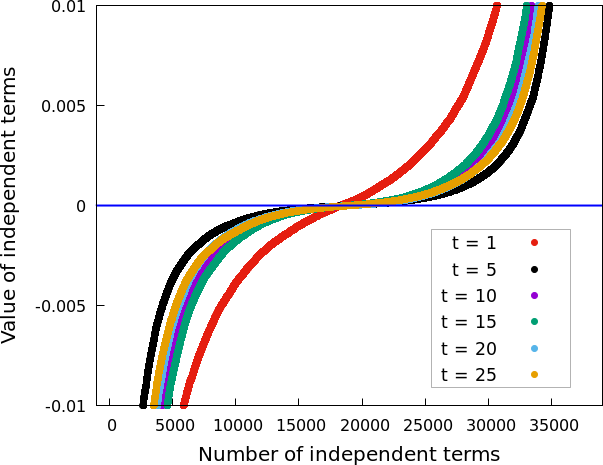}
    \caption{The distributions of non-zero of terms that enters in  Eq.~(\ref{eq2.9}) to construct the two-point correlation functions of deuteron at various time slices (lattice: $40^3 \times 64$, $a = 0.1189$ fm, and 
    for $m_u = m_d = m_c$). The numbers are ordered from the largest negative to largest positive (left to right). To show the distributions for each time slices inside the same plot, the largest positive values for each time slices are normalized with the largest positive value at $t = 1$.    
    The right pane is the zoomed-in version of the left pane showing the high level of symmetry (of positive and negative values) of smaller contributors around zero value. These data are for wall-source and point sink case.}
    \label{distrb_NP}
\end{figure}

In Fig.~\ref{distrb_He4} we have  shown the distributions of non-zero  terms at various time slices
 for $^4$He. Here we show that for $^2$H in Fig.~\ref{distrb_NP}. Here again, we  normalize the largest positive value of each time slices with the largest positive value at $t = 1$. The right pane is the zoomed-in version of the left pane, showing the higher level of symmetry of smaller contributors around zero value.

\begin{figure}[h!]
    \centering    
\includegraphics[width=0.45\textwidth]{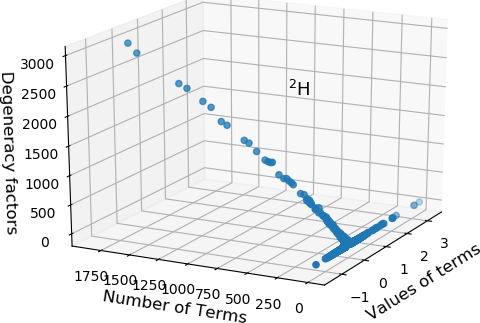}
\includegraphics[width=0.45\textwidth]{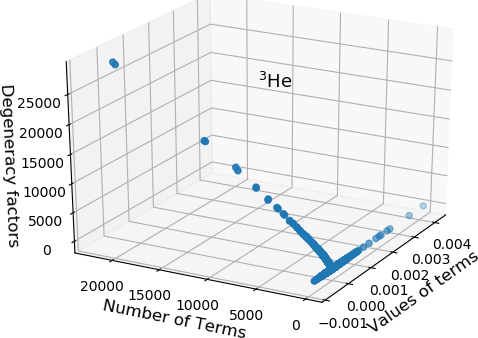} 

\includegraphics[width=0.45\textwidth]{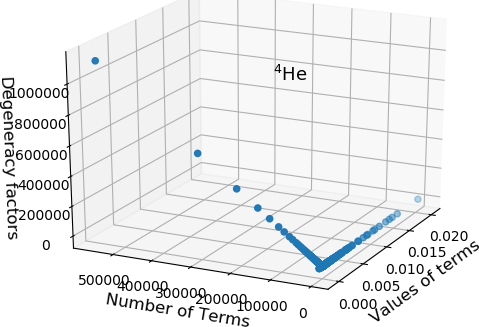}  
\includegraphics[width=0.45\textwidth]{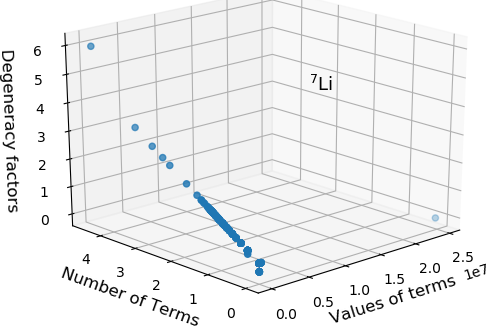} 
\caption{The distributions of non-zero of terms that enters in  Eqs.~(\ref{eq2.9}) and (\ref{eq2.6}) to construct the two-point correlation functions (at $t = 1$, and for the case of $m_u = m_d = m_c$, and for one gauge configuration) for different nuclei.  
For $^7$Li, lattice size is $24^3 \times 64$ and for others it is $40^3 \times 64$.
The $x$-axis represents non-zero terms, $y$-axis are their values while the $z$-axis shows their degeneracies (for $^7$Li, $y$ and $z$-axes are in terms of ${\mathrm{log}}_{10}$).}
    \label{distrb_bin}
\end{figure}

Next, we investigate whether there is any degeneracy in these non-zero terms. In Fig.~\ref{distrb_bin} we show these terms and their degeneracies for $^2$H and $^3$He, $^4$He and $^7$Li (due to large number of terms,  data have been binned for presentation purpose). It is quite clear from these plots that the distribution peaks near zero, showing that degeneracy is maximum for the terms with near-zero value. Only a small fraction have larger values with a rather extended asymmetric tail. Naturally, the contributors from the tails mainly construct the correlation functions. They also do not have much degeneracy. It is also evident that the distributions become more extreme as the atomic number increases suggesting only a fraction of terms are actually needed to calculate the full correlation functions for higher nuclei.

\begin{figure}[t!]
    \centering    
    \hspace*{-0.3in}
\includegraphics[width=0.33\textwidth, height=0.3\textwidth]{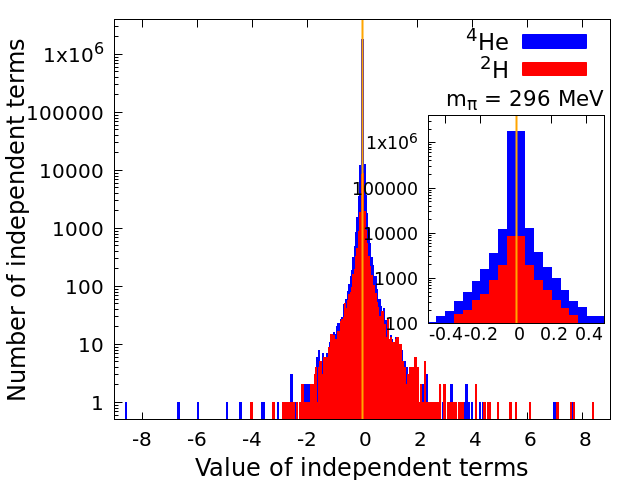}
\includegraphics[width=0.33\textwidth, height=0.3\textwidth]{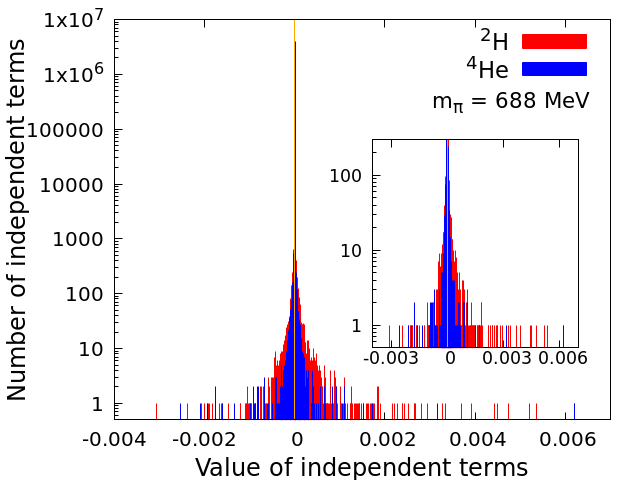}
\includegraphics[width=0.33\textwidth, height=0.3\textwidth]{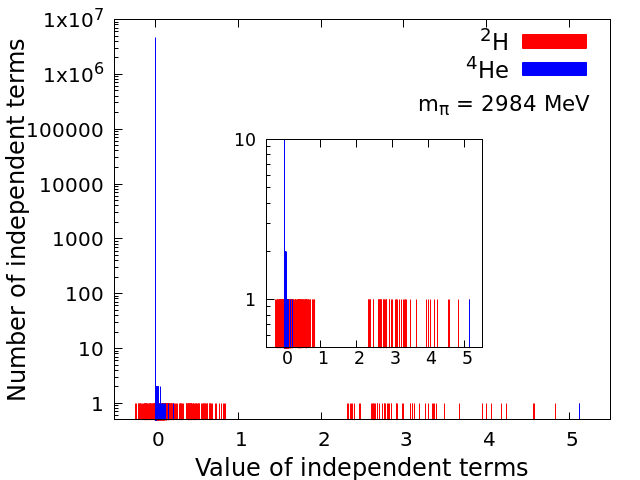}
    \caption{Comparison of distribution data of $^2$H and $^4$He at three different quark masses. Distribution is more extreme for $^4$He than that of $^2$H and that pattern also increases as the quark mass increases.}
    \label{hist_np_he4_comp}
\end{figure}

In Figs.~\ref{dist_all_np_flat} and  \ref{dist_all_np_he4_hist} we showed the quark mass dependence of this distribution. We compare below the quark mass dependence across two nuclei, $^2$H and $^4$He, and results are shown in Fig.~\ref{hist_np_he4_comp}. It shows that the distribution peaks much more around zero for $^4$He, that is, a larger fraction of terms becomes less relevant to the correlator. This becomes more and more prominent as quark mass increases. It demonstrates that in constructing these correlation functions for higher nuclei one needs substantially a small fraction of total terms. This procedure of using only
the {\it important} terms will definitely help to compute correlation functions for much higher nuclei.

To demonstrate that indeed a small fraction of terms are sufficient to construct the correlation function we use the probe of effective mass. 
In Fig.~\ref{em_per_he4} we showed effective mass for $^4$He with different fraction of terms. Below in Fig.~\ref{em_per20_np}, we show such results for $^2$H at two different quark masses: $m_u = m_d = m_c$ (left pane), and $m_u = m_d = m_s$ (right pane).
\begin{figure}
    \centering    
    \hspace*{-0.05in}
\includegraphics[width=0.48\textwidth]{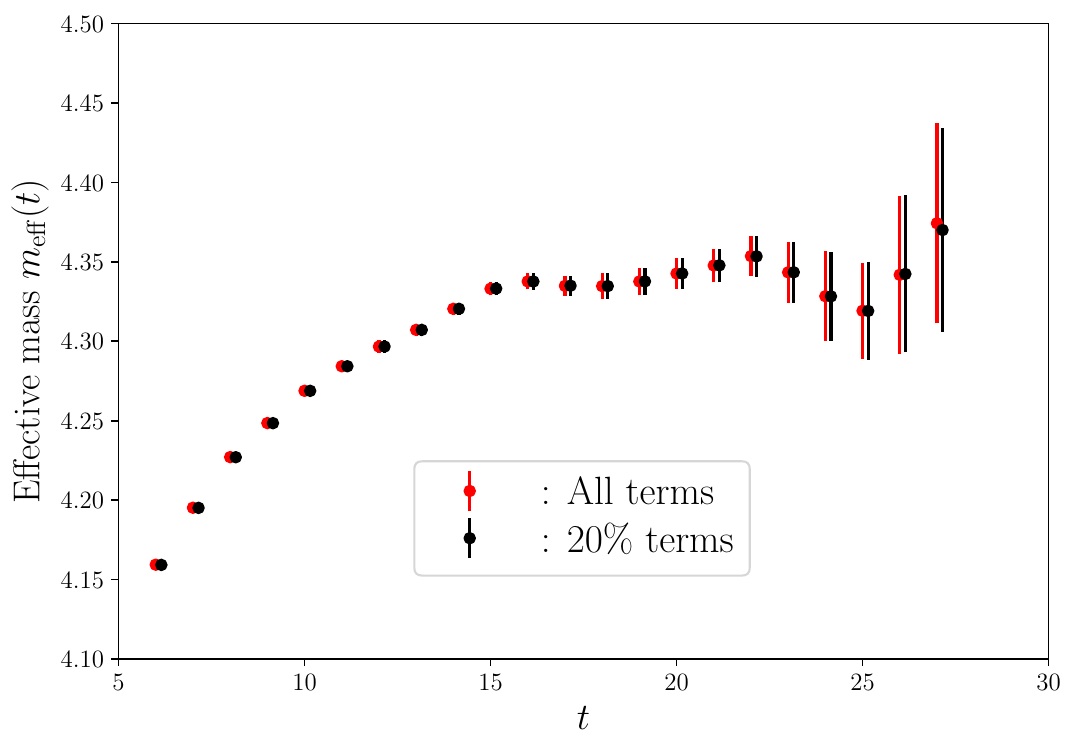}
\includegraphics[width=0.48\textwidth]{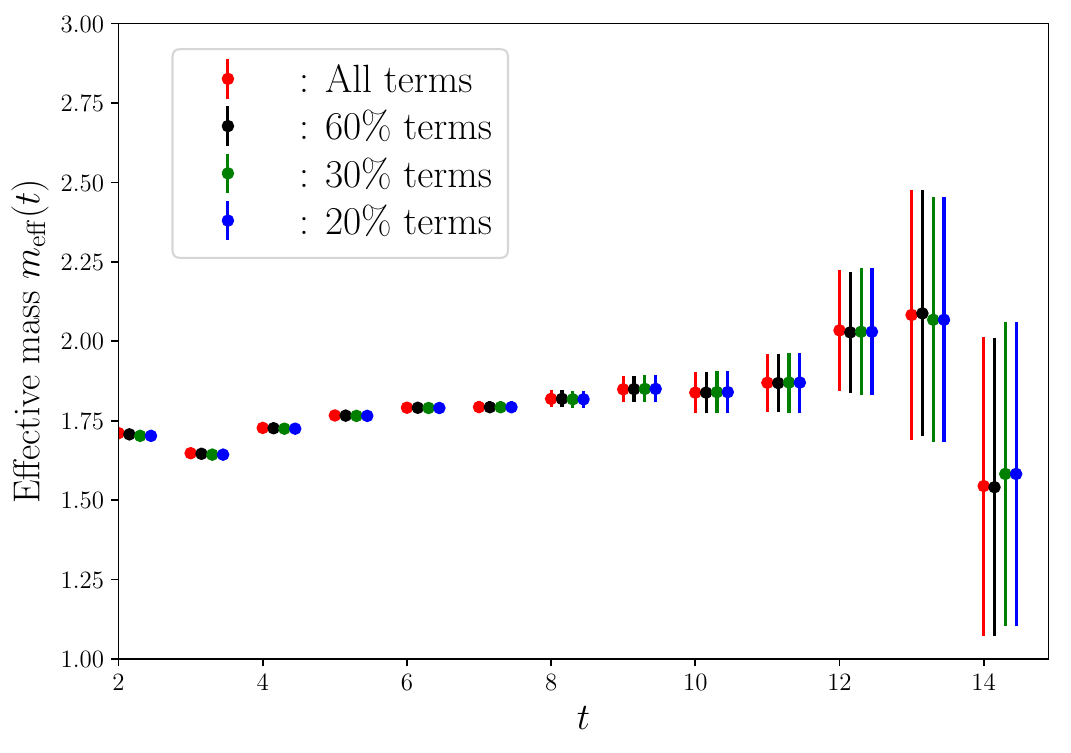}
    \caption{Effective mass of $^2$H obtained using all and  top certain fraction of terms (absolute large) of the correlation functions. Left: 
    For the case of $m_u = m_d = m_c$ on a $40^3 \times 64$ lattice ensemble with $N_{\mathrm{config}} = 100$. 
    Data for 20\% terms are shifted slightly to the right to avoid overlapping. Right: For the case of $m_u = m_d = m_s$ on a  $24^3 \times 64$ lattice ensemble with $N_{\mathrm{config}} = 156$. Operator set up: Wall source and point sink for both cases.}
    \label{em_per20_np}
\end{figure}

\begin{figure}[h!]
    \centering    
\includegraphics[width=0.48\textwidth]{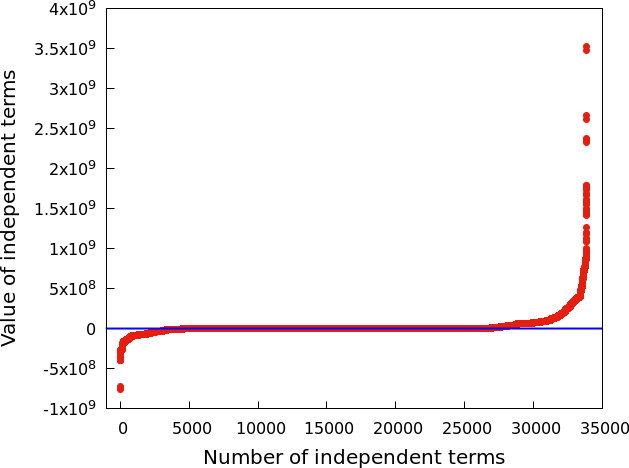}
\includegraphics[width=0.48\textwidth]{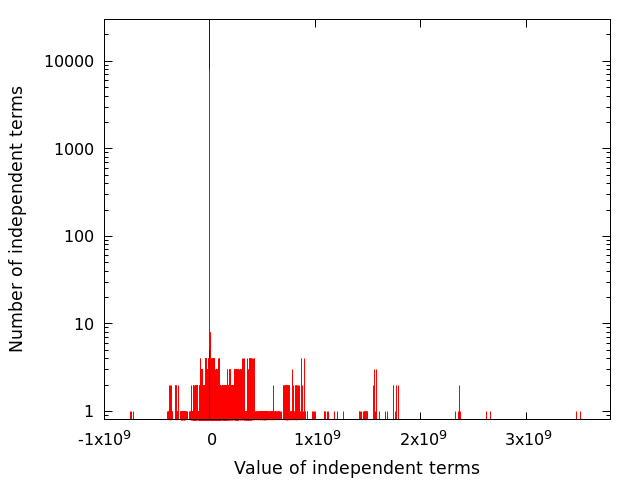}
    \caption{Left: same as Fig.~\ref{distrb_NP} but for $^3$He on the same lattice (at $t = 10$). 
    Right: Corresponding histogram plot (in log-scale) with bin size 1. The blue lines show the position of zero value.
    For this case we choose a sink where individual nucleons are projected to zero momentum.}
    \label{distrb_He3}
\end{figure}

For plane-wave sink  operators, we utilize  Eq.~(\ref{eq2.6}) within 
block algorithm formalism, to compute the 2-point correlation functions for $^2$H, $^3$He and $^4$He nuclei. For this case, a correlation function is calculated by contracting the block tensors and summing over all possible permutations of indices of the tensors. A generalization of Eq.~(\ref{eq2.6}) for a nuclei with mass number $A$ leads to the following expression for a zero-momentum projected correlation function,
\begin{eqnarray}
    G_{A}(t)&=&\sum_{\boldsymbol{\chi},\alpha_1,\cdots\alpha_n}C(\chi_1,\chi_2,\cdots,\chi_{3A};\alpha_1,\cdots,\alpha_{A})\sum_{\sigma}sgn(\sigma)B(t,\alpha_1,\chi_{\sigma(1)},\chi_{\sigma(2)}, \chi_{\sigma(3)}) \nonumber \\
    &&\hspace*{0.3in} \times \quad B(t,\alpha_2,\chi_{\sigma(4)},\chi_{\sigma(5)}, \chi_{\sigma(6)})\cdots B(t, \alpha_A,\chi_{\sigma(3A-2)},\chi_{\sigma(3A-1)}, \chi_{\sigma(3A)})
\end{eqnarray}
Here $\boldsymbol{\chi}=\{\chi_1,\chi_2,\cdots,\chi_{3A}\}$ with $\chi_i$'s as the combined spinor-color indices of quark fields at the source. Assuming isospin symmetry we have omitted the label for proton and neutron in the above equation. 
The above correlator gets constructed with the 
 $\chi_i$'s and $\alpha_i$'s 
for which the coefficient tensor $C$ is non-zero. We perform the same analysis as the determinant case and find the non-zero contributors.
In Fig.~\ref{distrb_He3} we show the distribution of $^3$He with plane-wave sink
 where individual nucleons of $^3$He are projected to zero-momentum separately. The right pane is the histogram plot of the same data. Both of this plot show that as in point-sink case, for an extended sink the typical characteristic of distribution showing the extremeness of the distribution and the importance of only a smaller number of terms. It shows that this characteristic is independent of the choice of sink.

\begin{figure}[h!]
    \centering    
\includegraphics[width=0.6\textwidth]{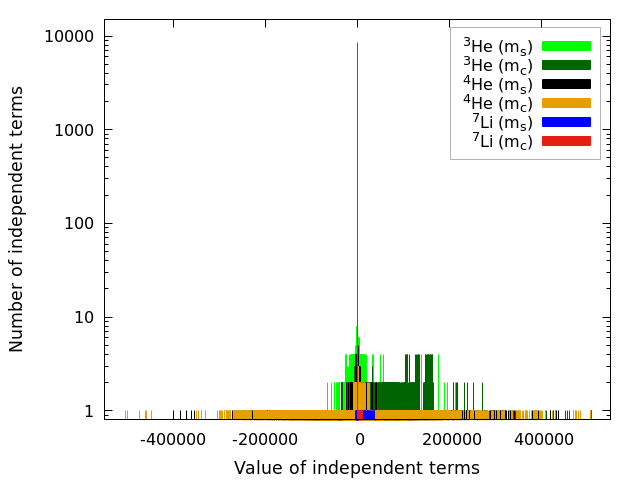}     
    \caption{Histogram plot (bin size 0.01) showing the distribution of terms (normalized with the largest values) constructing two-point correlators for various nuclei at $m_u = m_d = m_c$ and $m_u = m_d = m_s$,  with plane-wave sinks. The distribution of non-zero terms becomes sharper around zero as the quark mass increases, as well as for nuclei with higher mass numbers.
    }
    \label{hist_all}
\end{figure}

In Fig.~\ref{hist_all} we present a histogram plot  showing the distribution of terms (normalized with the largest values) constructing two-point correlators for various nuclei at two different quark masses with plane-wave sink. The distribution of non-zero terms becomes sharper around zero, leading to the requirement of smaller fraction of important terms in constructing the correlation functions as the quark mass increases, as well as for nuclei with higher mass numbers.

\begin{figure}[h!]
    \centering    
   \hspace*{-0.05in}  
\includegraphics[width=0.48\textwidth, height=0.35\textwidth]{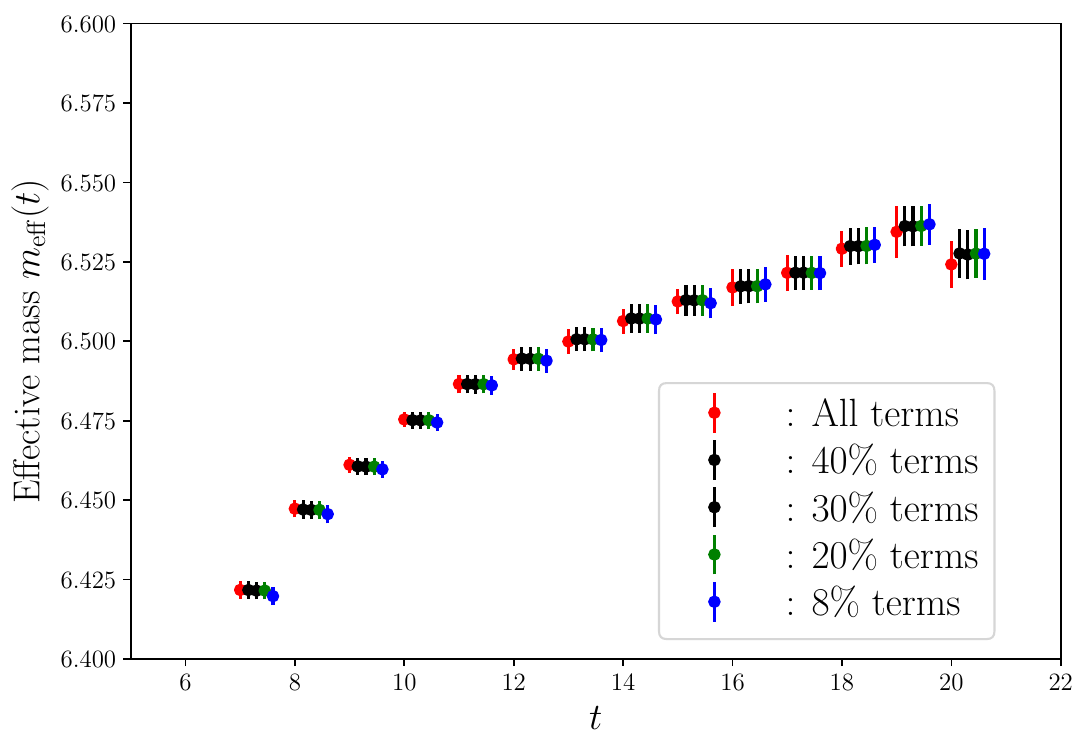}
\includegraphics[width=0.48\textwidth, height=0.35\textwidth]{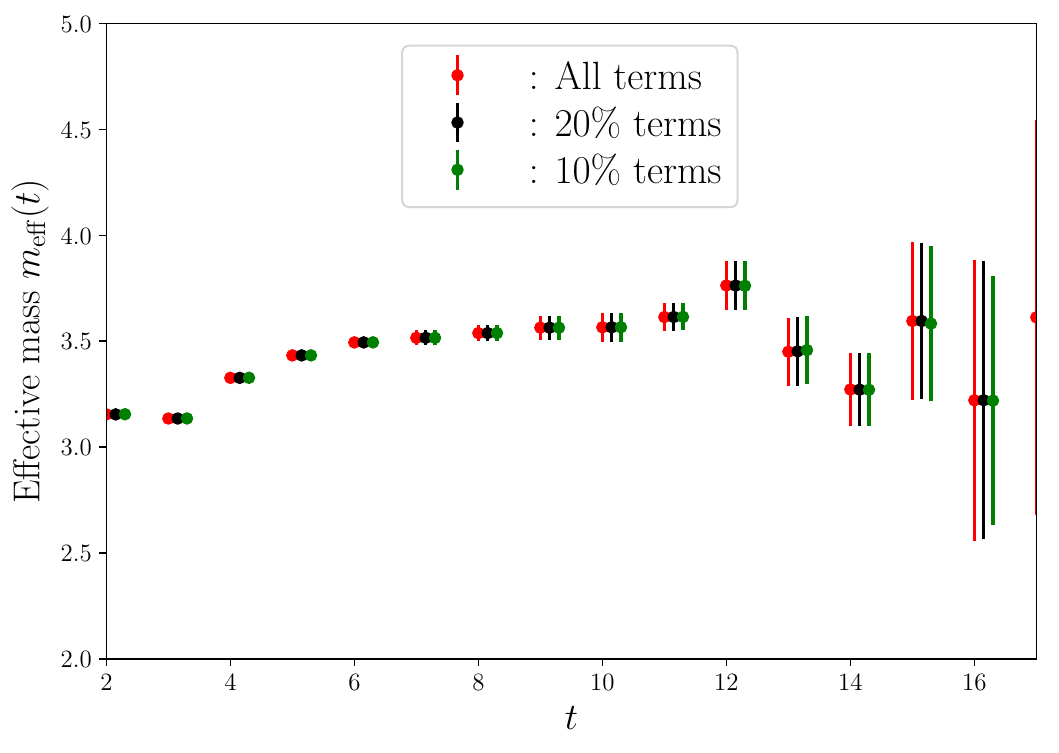}      
    \caption{
    Effective masses of $^3$He (left side) and $^4$He (right side), obtained using all and top several other percentage of terms (absolute large) of the correlation functions. Results for $^3$He (left) is
    for the case of $m_u = m_d = m_c$, on a $40^3\times 64$ ensemble with $N_{conf} = 100$, while for $^4$He (right), 
     $m_u = m_d = m_s$, lattice size is $24^3\times 64$ and $N_{conf} = 137$.  Operator set up: Wall source and plane-wave sink (see Fig.~\ref{corf_fig}).}
    \label{em_per_he3_4}
\end{figure}

In Fig.~\ref{em_per_he3_4} we present effective masses for $^3$He (left pane) and $^4$He (right pane), with a plane-wave sink setup, showing again that a small fraction of terms are sufficient to compute the physical results irrespective of the type of sink.  The right pane corresponds to the bottom right plot of Fig.~\ref{corf_fig}.

\begin{figure}[h!]
    \centering    
   \hspace*{-0.05in}  
\includegraphics[width=0.48\textwidth, height=0.35\textwidth]{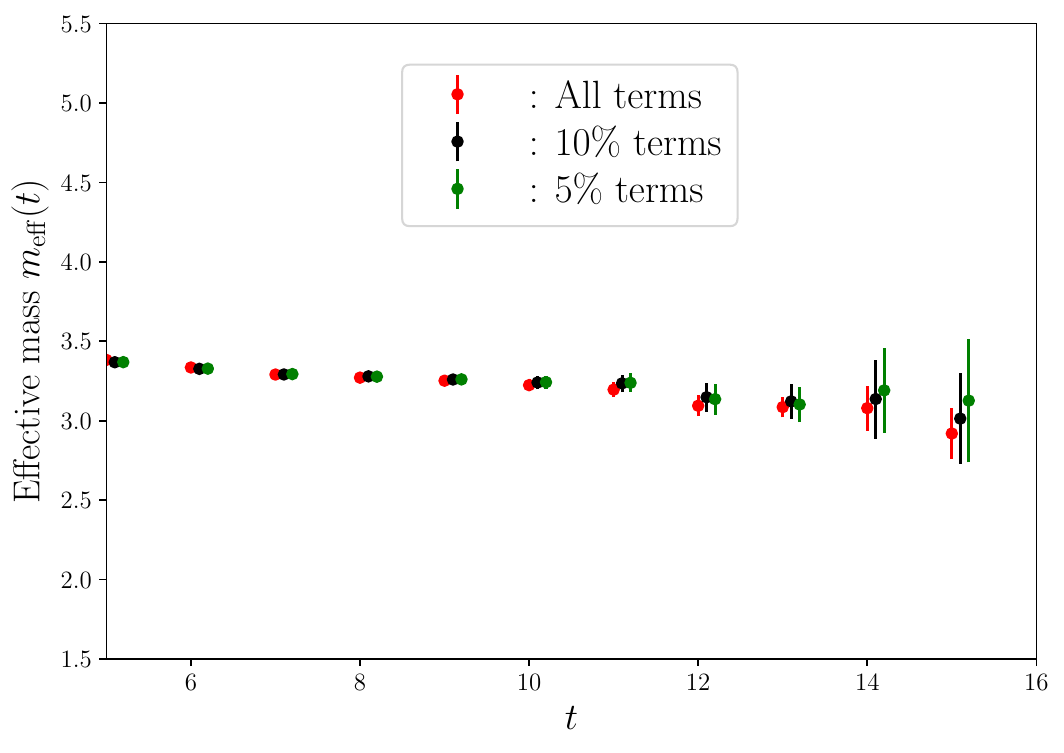}
\includegraphics[width=0.48\textwidth, height=0.35\textwidth]{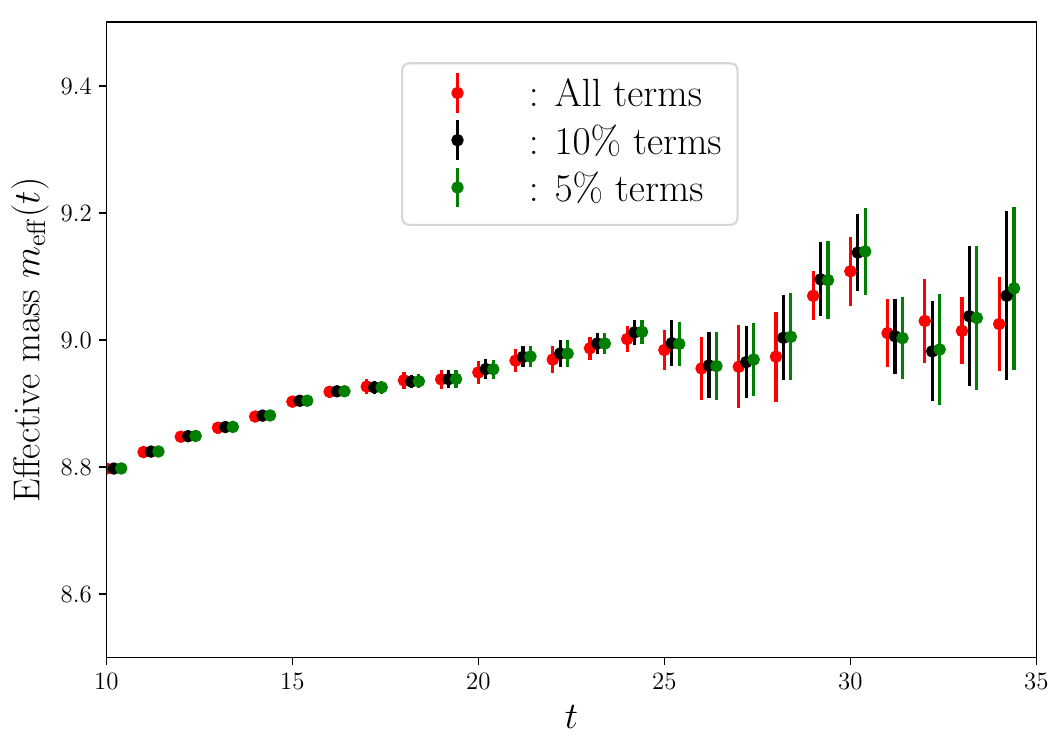}         
    \caption{
    Effective masses of $^7$Li 
    obtained using all and top several other percentage of terms (absolute large) of the correlation functions. The left pane corresponds to the  case of $m_u = m_d = m_s$, while the right pane is for  the  case of $m_u = m_d = m_c$. Lattice and operator set ups are the same as in Fig.~\ref{eff_mass_c}.}
    \label{em_per_Li}
\end{figure}

In Fig.~\ref{em_per_Li} we present effective masses for $^7$Li on the same lattice set up as in Fig.~\ref{eff_mass_c}, with  left pane corresponding to  $m_u = m_d = m_s$, and right pane corresponding to $m_u = m_d = m_c$). Here we use a cluster set up with Tritium and $^4$He. These results show that with 5-10\% of the total terms one can get the similar effective mass as obtained with all terms.

\begin{figure}[h!]
    \centering    
\includegraphics[width=0.49\textwidth, height=0.45\textwidth]{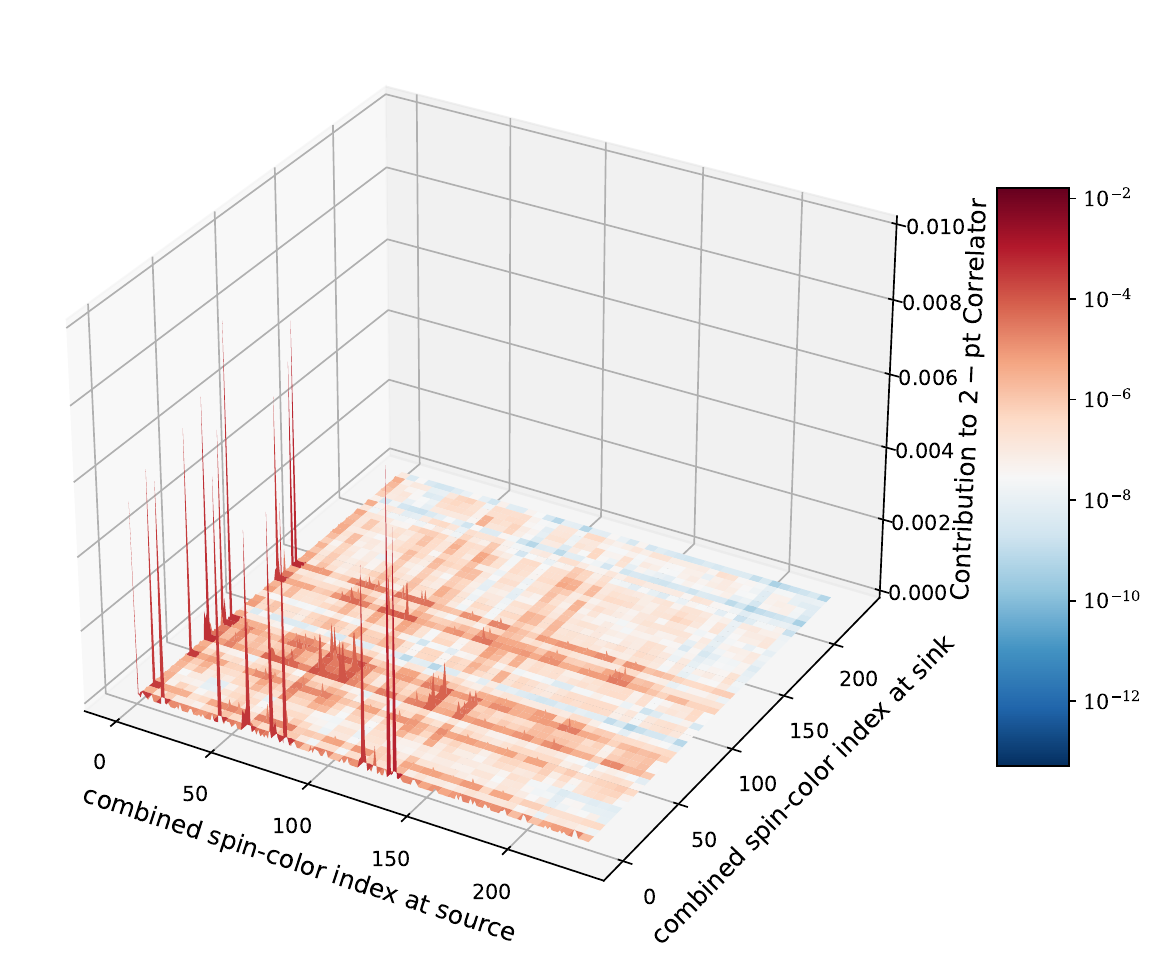}
\includegraphics[width=0.49\textwidth, height=0.45\textwidth]{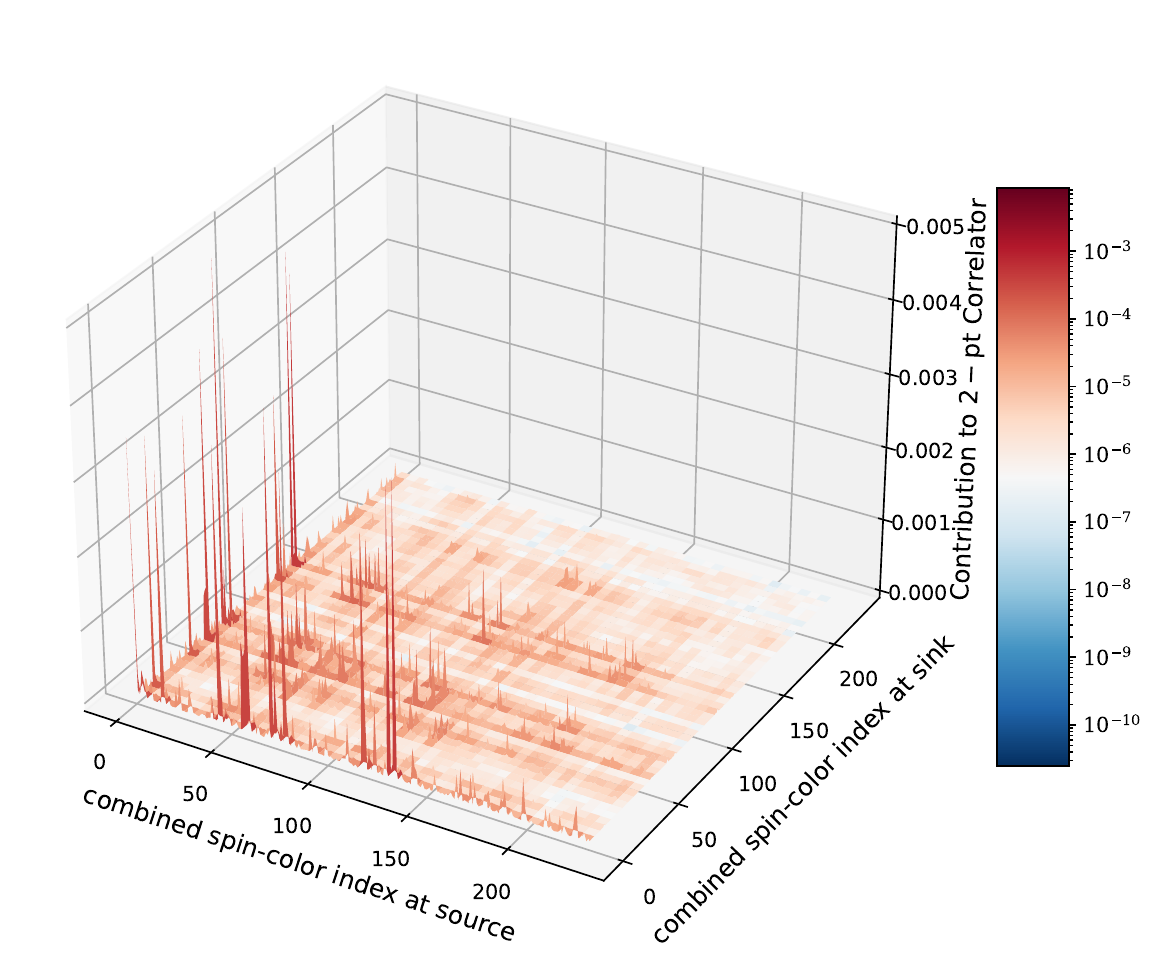}
    \caption{Three-dimensional plots showing the distribution of terms constructing two-point correlators for $^2$H at $m_u=m_d=m_c$ (left panel) and $m_u=m_d=m_s$ (right panel) on a $24^3\times64$ lattice. The $x$ and $y$ axis represents serial numbers of spin-color indices (see discussion above) at source and sink respectively for which tensor $C$ in Eq.~(\ref{eq2.7}) is non-vanishing. The stripes with deeper colors (towards red) are the dominant contributors.}

    \label{3d_NP}
\end{figure}

\begin{figure}[htb]
    \centering    
\includegraphics[width=0.49\textwidth, height=0.42\textwidth]{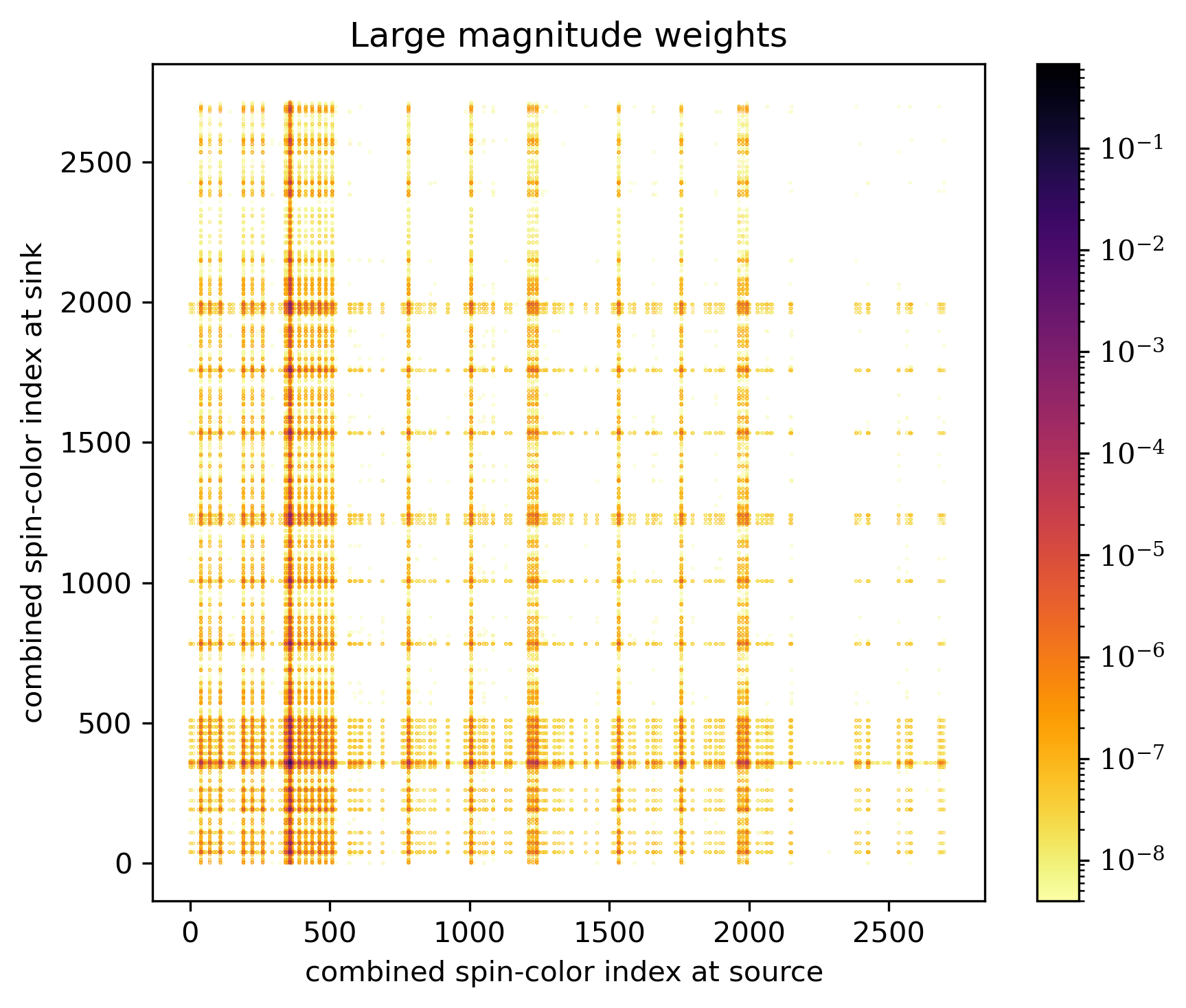}
\includegraphics[width=0.49\textwidth, height=0.42\textwidth]{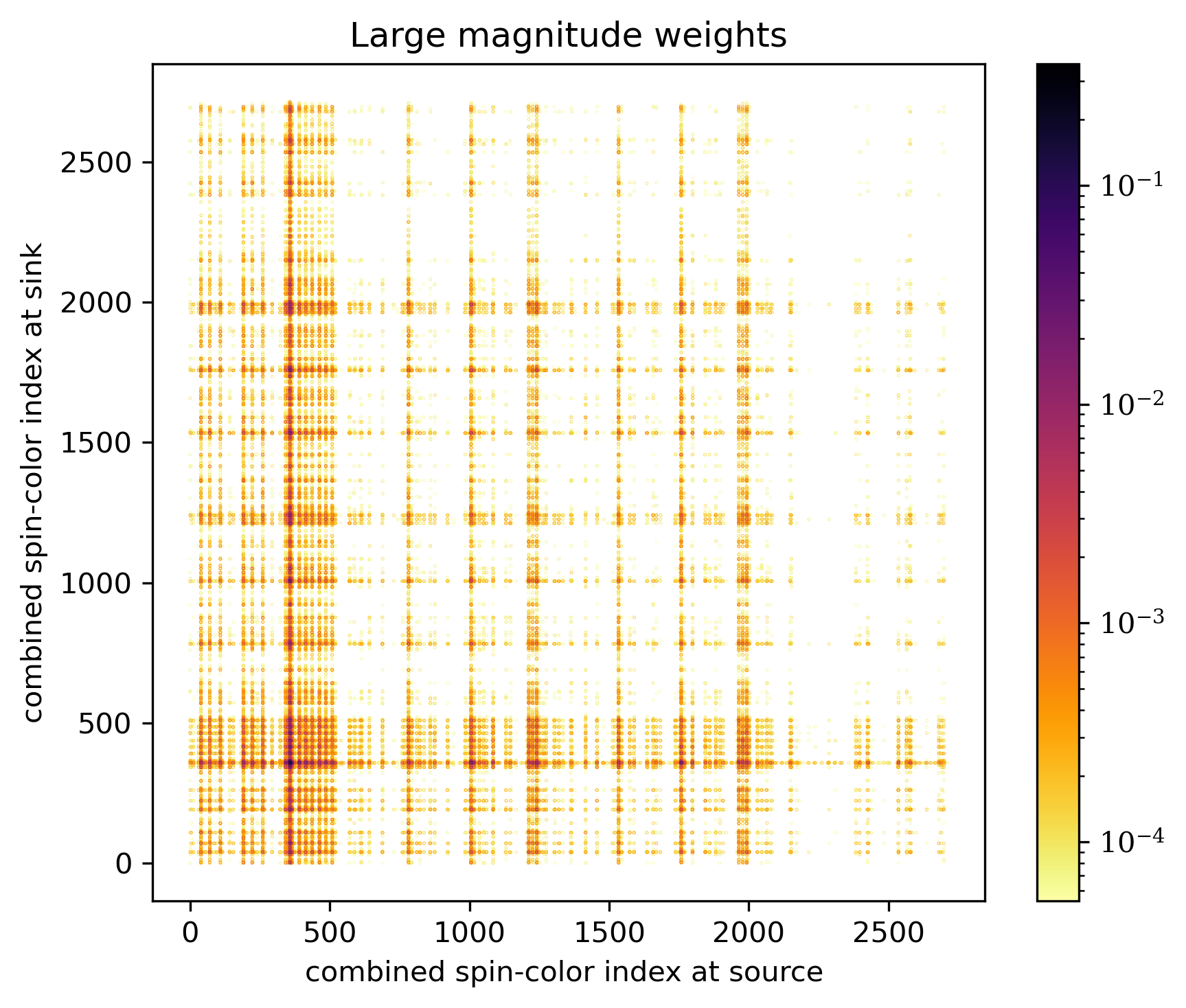}
    \caption{Colormap plots corresponding to Fig \ref{3d_NP}. The stripes with deeper colors are the dominant contributors.}
 \label{colormap_NP}
\end{figure}

\begin{figure}[htb]
    \centering    
\includegraphics[width=0.49\textwidth, height=0.4\textwidth]{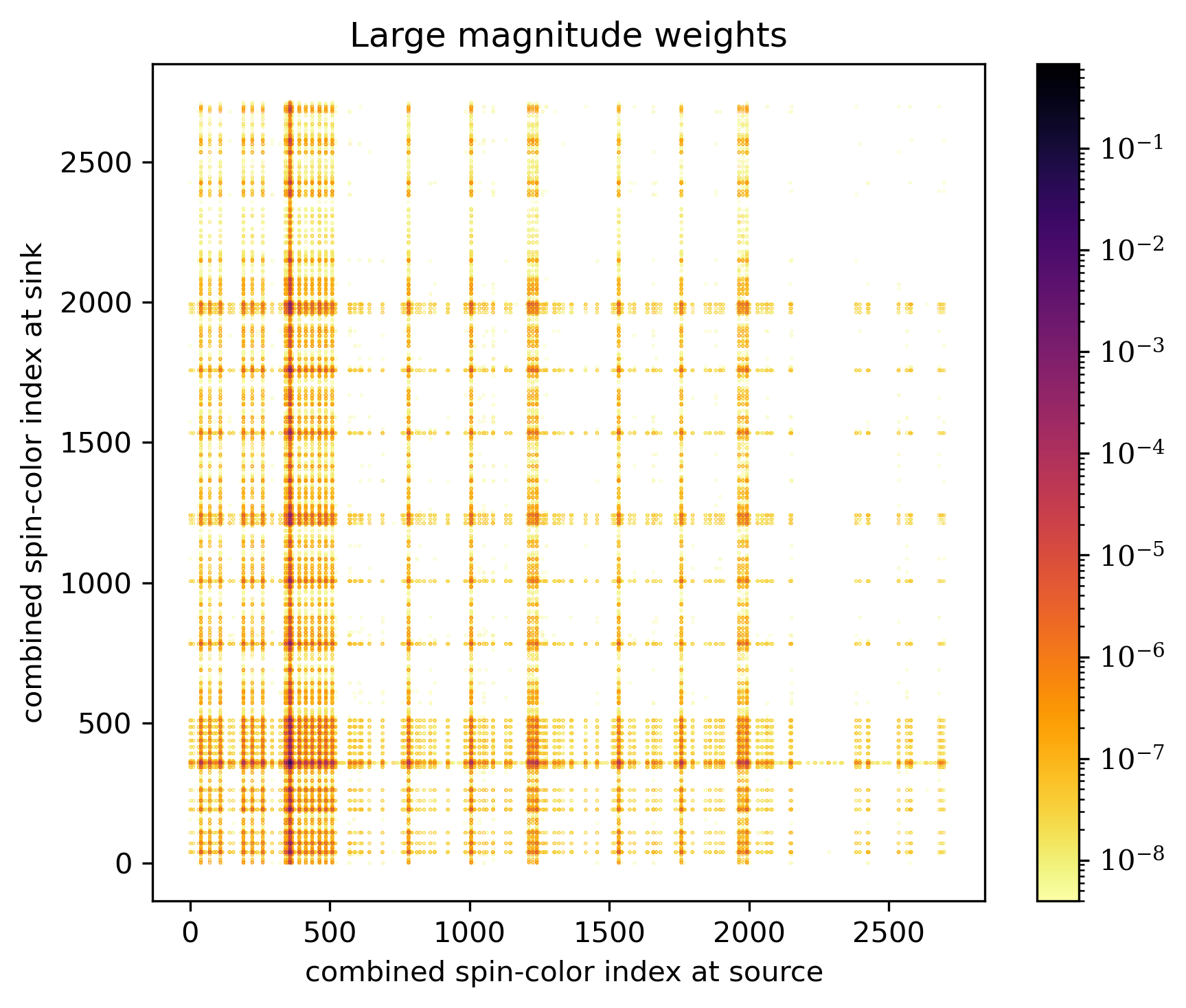}
\includegraphics[width=0.49\textwidth, height=0.4\textwidth]{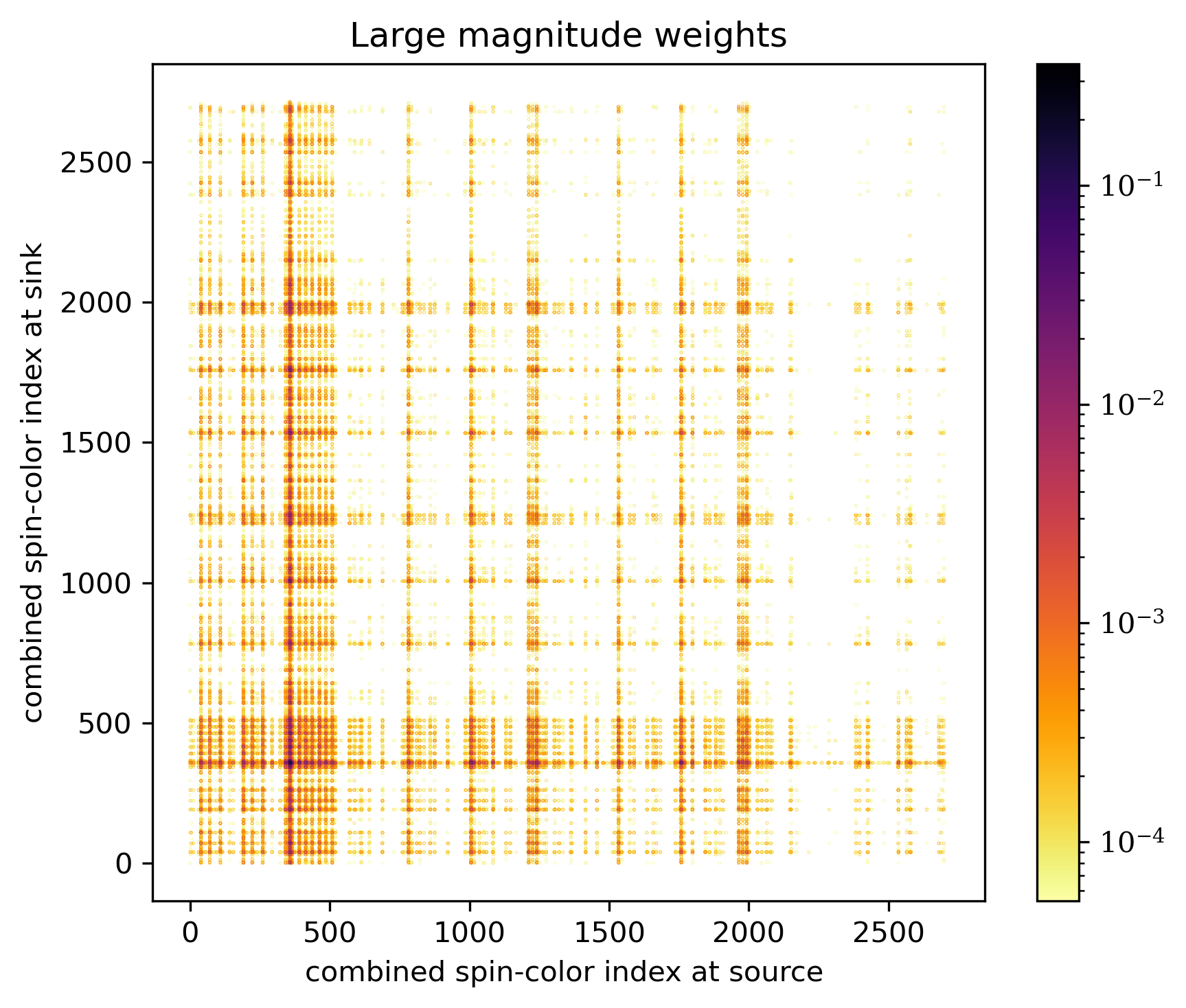}
    \caption{Colormap plots showing the distribution of top $1\%$ terms constructing two-point correlation functions for $^4$He. Data are obtained at $m_u=m_d=m_c$ (left panel) and $m_u=m_d=m_s$ (right panel) on a $24^3\times64$ lattice. The stripes with deeper color are the dominant contributors.
    }

    \label{colormap_He4}
\end{figure}

To understand the dominance of certain spin-color combinations in the nuclear correlation functions
we employ three-dimensional plots and colored heat-map plots showing  how the distribution of terms %
varies with spin-color indices at the source and sink. 
In Figs.~\ref{3d_NP} and \ref{colormap_NP}, 
we present such plots for the correlation function of $^2$H at $m_u=m_d=m_c$ (left panel) and $m_u=m_d=m_s$ (right panel) on a $24^3\times64$ lattice.
As we have discussed in Appendix \ref{GPU}, the two-point correlator of a nuclei can be expressed as (for ease of reading we reproduce that here),
\begin{equation*}
    G(t)=\sum_{\mathbf{x}}\sum_{\{\chi_u;\chi^{\prime}_u\}\in\boldsymbol{\zeta}_u}\sum_{\{\chi_d;\chi^{\prime}_d\}\in\boldsymbol{\zeta}_d}\Theta(\{\chi_u;\chi_d\};\{\chi^{\prime}_u;\chi^{\prime}_d\})\det M^u(\{\chi_u;\chi^{\prime}_u\}) \det M^d(\{\chi_d;\chi^{\prime}_d\}), %
\end{equation*}
and the details about the notations can be found around Eq.~(\ref{eqa1}). For ease of plotting with many indices, we arrange the spin-color indices at the source and sink i.e. $\{\chi_u,\chi_d\}$ and $\{\chi^{\prime}_u, \chi^{\prime}_d\}$ in increasing order and label them by a serial number to identify them. The $x$ and $y$ axes in Figs.~\ref{colormap_NP} and \ref{colormap_He4} represent these spin-color serial numbers indices at source and sink, respectively. The value of the contributions to the correlation functions at a given source-sink pair, $(x,y)$, is represented by the weight of the colormap.  While 
Fig.~\ref{colormap_NP} shows this color-spin correlation for $^2$H for $m_u=m_d=m_c$ and $m_u=m_d=m_s$, Fig.~\ref{colormap_He4} represents the same for $^4$He. However, due to the large number of terms for $^4$He, most of which are very small, we plot only the top 1\% terms which are sufficient to construct more than 99\% of the full correlator. From the color of stripes it is quite apparent that certain spin-color indices dominate to a very large extent compared to others, and interestingly in most large contributors once the spin-color is fixed at the source the domination does not depend on the sink spin-color indices, and vice versa.

\clearpage

\bibliography{nuclei_lattice}

\end{document}